\documentclass[aps,prd,10pt,twocolumn,showpacs,preprintnumbers,amsmath,amssymb,floatfix,nofootinbib]{revtex4}

\usepackage{natbib} 

\usepackage{color}

\usepackage{amsmath}
\usepackage{epsfig}
\usepackage{array}
\usepackage[hyperfootnotes=true]{hyperref} 
\usepackage{afterpage}
\usepackage{color}

\begin{document}

\title{A Radiation-Hydrodynamics Code Comparison for Laser-Produced Plasmas: \\ FLASH versus HYDRA and the Results of Validation Experiments}

\author{  Chris Orban$^{1}$, Milad Fatenejad$^{2}$, Sugreev Chawla$^{3,4}$, Scott Wilks$^{4}$ and Donald Lamb$^{2}$}

\affiliation{
\vspace{0.2cm}
(1) Department of Physics, The Ohio State University, Columbus, OH 43210  \\
(2) ASC Flash Center for Computational Science, University of Chicago, Chicago, IL \\
(3) Center for Energy Research, University of California San Diego, La Jolla, CA \\
(4) Lawrence Livermore National Laboratory, Livermore, CA, 94551
}

\email{orban@physics.osu.edu}

\date{\today}

\begin{abstract}
The potential for laser-produced plasmas to yield fundamental insights into high energy density physics (HEDP) and deliver other useful applications can sometimes be frustrated by uncertainties in modeling the properties and expansion of these plasmas using radiation-hydrodynamics codes. In an effort to overcome this and to corroborate the accuracy of the HEDP capabilities recently added to the publicly available FLASH radiation-hydrodynamics code, we present detailed comparisons of FLASH results to new and previously published results from the HYDRA code used extensively at Lawrence Livermore National Laboratory. We focus on two very different problems of interest: (1) an Aluminum slab irradiated by 15.3 and 76.7~mJ of ``pre-pulse'' laser energy and (2) a mm-long triangular groove cut in an Aluminum target irradiated by a rectangular laser beam. Because this latter problem bears a resemblance to astrophysical jets, Grava et al., Phys. Rev. E, 78, (2008) performed this experiment and compared detailed x-ray interferometric measurements of electron number densities to HYDRA simulations. Thus, the former problem provides an opportunity for code-to-code comparison, while the latter provides an opportunity for both code-to-code comparison and validation. Despite radically different schemes for determining the computational mesh, and different equation of state and opacity models, the HYDRA and FLASH codes yield results that are in excellent agreement for both problems and with the experimental data for the latter. Having validated the FLASH code in this way, we use the code to further investigate the formation of the jet seen in the Grava et al. (2008) experiment and discuss its relation to the Wan et al. (1997) experiment at the NOVA laser. \\

{\bf IM Release Number:} LLNL-JRNL-636375 (Distribution Unlimited)
\end{abstract}


\maketitle

\section{Introduction}
  \label{sec:intro}

The potential for laser experiments to yield fundamental insights into High-Energy-Density Physics
(HEDP) is in many ways limited by the sophistication and accuracy of current-generation 
``three-temperature'' (3T)\footnote{ We use the term ``three-temperature'' (or ``3T'') to denote the approximation
that electrons and ions move together as a single fluid but with two different temperatures, and that this fluid 
can emit or absorb radiation. In the 3T simulations presented throughout this paper each cell has an electron temperature, an ion temperature, and radiation energy densities in a number of photon energy bins.} radiation-hydrodynamics codes that simulate the heating, conduction and radiation of laser-irradiated fluids. In deconstructing the results from ultra-high intensity, 
short-pulse laser experiments, for example, Particle-In-Cell (PIC) simulations of the ultra-intense pulse interaction
with the target may depend sensitively on radiation-hydrodynamics simulations of the heating and
ionizing effect of stray ``pre-pulse'' laser energy in the nanoseconds before the arrival of
the main pulse. Interferometric instruments to measure electron number densities in the ``pre-plasma'' created by
this pre-pulse are often unavailable or, in some cases, the target geometry disallows probe beams from directly accessing
the pre-plasma (e.g. in cone targets as in \cite{AkliOrban_etal2012}), and so the pre-plasma properties must be predicted from
a radiation-hydrodynamics code. The uncertainties in these simulations may frustrate efforts to gain 
a better understanding of ion acceleration or electron transport that could prove to be valuable for a variety of
applications, such as radiation therapy, x-ray generation or the activation and detection of fissile materials.

Another important use of these codes is in modeling inertial confinement fusion experiments at facilities 
like Omega and the National Ignition Facility (NIF) \cite{Boehly_etal1997,Moses_etal2009,Lindl_etal2011}. Rosen et al. \cite{Rosen_etal2011} describe some of the
subtleties encountered in understanding indirect-drive experiments and a recent panel report
by Lamb \& Marinak et al. \cite{LambMarinak2012} outlines a number of remaining uncertainties in simulating ignition-relevant 
experiments on NIF. Lamb \& Marinak et al. \cite{LambMarinak2012} emphasize the need for code-to-code comparisons and validation
in a wider effort to reproduce the diagnostics of NIF implosions. Although there have been some recent investigations
with other codes \cite{Bellei_etal2013,RAGE}, the HYDRA code  
\cite{Marinak_etal1996,Marinak_etal1998,Marinak_etal2001}, developed at Lawrence
Livermore National Laboratory (LLNL), is the canonical choice for radiation-hydrodynamics modeling 
of these experiments. Simulations using the HYDRA code were integral to achieving orders-of-magnitude
larger fusion yields than initially produced on NIF \cite{Callahan_etal2012}.

Uncertainties and inaccuracies in radiation-hydrodynamics modeling can also frustrate the design and
interpretation of experiments to investigate fundamental plasma properties (e.g., opacities, equation of state, 
hydrodynamic instabilities) at HEDP-relevant densities and temperatures 
\cite{Harding_etal2009,vanderHolst_etal2012,Taccetti_etal2012}. 
Both Omega and NIF, among other facilities, have completed a number of experiments in this category,
and will continue to do so in the future \cite{NIF}. 



With these concerns in mind, and in an effort to confirm the accuracy of the HEDP capabilities recently added 
to the FLASH radiation-hydrodynamics code, we compare the predictions of 
FLASH to previously-published and new results from the HYDRA code. FLASH is a finite-volume 
Eulerian code that operates on a block-structured mesh using Adaptive Mesh Refinement (AMR) \cite{PARAMESH},
whereas the HYDRA code uses an Arbitrary-Lagrangian-Eulerian (ALE) scheme to determine the 
computational grid \cite{Hirt_etal1974,Castor2004,Kucharik2006},
which can deform and stretch in response to the movement and heating of the fluid.
We show results for the 15.3-76.7 mJ (a.k.a. ``pre-pulse'') irradiation of Aluminum slab targets over 
a period of 1.4 ns. This problem tests the codes in 2D cylindrical geometry. We also show results for a mm-long
triangular groove cut in an Aluminum target irradiated by a rectangular beam. The results of this experiment, which 
maintains translational symmetry along the grove as a test of plasma expansion in 2D cartesian geometry, were 
reported and modeled with HYDRA simulations in Grava et al. 2008 \cite{Grava_etal2008}. They 
investigated the problem for its resemblance to astrophysical jets where radiative cooling plays an important 
dynamical role, and as a miniature version of
similarly-motivated experiments at the Nova laser \cite{Wan_etal1997,Stone_etal2000}. The experiment was 
performed at Colorado State University and, importantly, they present x-ray
interferometric measurements of electron number density from 1-20 ns after the target irradiation that afford a 
powerful validation test. We therefore compare the results of FLASH to both the HYDRA simulations they present
and to the experimental data. Using publicly and commercially-available equation of state (EOS) 
and opacity models with FLASH, we find excellent agreement between FLASH and HYDRA, and confirm 
 the resemblance of the FLASH results to the experimental measurements presented in Grava et al. \cite{Grava_etal2008}.


In the next section we provide an overview of the physics FLASH and HYDRA have in common in modeling the problems just 
described. \S~\ref{sec:diff} emphasizes differences between the physics in the two codes and how the equations
of radiation-hydrodynamics are solved. \S~\ref{sec:prepulse} describes and presents detailed 
results from two pre-pulse problems. \S~\ref{sec:grava} 
presents comparisons of FLASH predictions to the experiments and modeling in Grava et al. \cite{Grava_etal2008}.
\S~\ref{sec:jet} uses FLASH simulations to investigate the formation of the jet in the Grava et al. experiment.
\S~\ref{sec:disc} discusses and synthesizes the results from both 
code-to-code comparisons and comments on some of the minor discrepancies that arose. 
\S~\ref{sec:concl} recapitulates our findings, and points to future code 
development and validation efforts with FLASH.

\section{Requisite Physics in Modeling Laser-Produced Plasmas}
\label{sec:req}

This section gives an overview of the physics required to model the laser-produced plasmas 
we are concerned with here and describes the commonalities of the FLASH and HYDRA codes. 
Both FLASH and HYDRA are built on the premise that, except at ultra-high intensities 
($\gtrsim 10^{18} \,  \rm{W}/\rm{cm}^2$), laser interactions with solid density targets can be treated 
as a hydrodynamic problem with the laser rays acting as a source of energy on the grid. This laser energy 
is absorbed by {\it electrons} at a rate specified by the inverse bremsstrahlung approximation,
after which this energy can be transferred to ions so that, in the absence of other heating (or cooling),
the ion and electron temperatures will equilibrate on the electron-ion equilibration 
timescale, $\tau_{e,i}$ \cite{Spitzer1963,LeeMore1984}. In 3T treatments (as in HYDRA and FLASH), this fluid is 
allowed to radiate and the diffusion of this radiation can be modeled by tracking the radiation energy densities 
in a fixed number of photon energy groups\footnote{Both the FLASH and HYDRA simulations presented in this paper use the standard multi-group diffusion approximation to model radiation transport. HYDRA now includes a more-sophisticated algorithm for radiation transport \cite{Marinak_etal2009}.}.
The equations of energy conservation for this 3T
fluid (with hydrodynamic terms dropped for simplicity) are then
\begin{equation}
 \frac{\partial (\rho  e_{\rm{ele}} )}{ \partial t} = \rho \frac{c_{v,\rm{ele}}}{\tau_{e,i}} (T_{\rm{ion}} - T_{\rm{ele}}) - \nabla \cdot \vec{q}_{\rm{ele}} + Q_{\rm{las}}  + Q_{\rm{abs}} - Q_{\rm{emis}} \label{eq:eele}
\end{equation}
and
\begin{equation}
\frac{\partial (\rho e_{\rm{ion}})}{\partial t} = \rho \frac{c_{v,\rm{ele}}}{\tau_{e,i}} (T_{\rm{ele}} - T_{\rm{ion}}), \label{eq:eion}
\end{equation}
where $e_{\rm{ele}}$ and $e_{\rm{ion}}$ are the electron and ion specific internal energies, $T_{\rm{ele}}$ and $T_{\rm{ion}}$ 
are the electron and ion temperatures, $c_{v,\rm{ele}}$ is the electron heat capacity computed from the EOS, $\vec{q}_{\rm{ele}}$ is the heat flux from electron thermal conduction, $Q_{\rm{las}}$ is the laser heating and
\begin{equation}
\frac{\partial u_g }{\partial t} = \nabla \cdot \vec{q}_{{\rm rad},g} - Q_{{\rm abs},g} + Q_{{\rm emis},g}; \, \, g = 1, ..., N_g \label{eq:erad}
\end{equation}
and
\begin{equation}
Q_{\rm abs} = \sum_{g=1}^{N_g} Q_{{\rm abs},g} , \, \, \, Q_{\rm emis} = \sum_{g=1}^{N_g} Q_{{\rm emis},g} \label{eq:qrad}
\end{equation}
are the equations of radiative transfer for $N_g$ energy groups. $Q_{\rm abs}$ and $Q_{\rm emis}$ represent the
total radiation energy absorbed or emitted. For the present discussion we omit the equations of 
mass and momentum conservation, as well as equations connecting $Q_{{\rm abs},g}$ and $Q_{{\rm emis},g}$
to the opacity and temperature of the plasma, all of
 which can be found in a number of more-complete presentations of
the equations of radiation-hydrodynamics \cite{MihalasMihalas1984,Castor2004,Flash2012}.

Implicit in Eqs.~\ref{eq:eele}-\ref{eq:qrad}, as already mentioned, is an EOS model 
that is needed to determine the 
electron heat capacity as well as the relevant pressure for the fluid in whatever density and temperature
state it may achieve during the course of the simulation. Also important is a model for the electron 
thermal heat flux into or out of a cell relative to the surrounding temperature gradient,
\begin{equation}
\vec{q}_{\rm ele} = - K_{\rm ele} \nabla T_{\rm ele}. \label{eq:qele}
\end{equation}
$K_{\rm ele}$ can be a complicated function of density, temperature and the material properties,
such as the Lee \& More 1984 \cite{LeeMore1984} model used in the FLASH \& HYDRA simulations presented here, and some codes implement 
$K_{\rm ele}$ as a tabulated quantity. However, near or significantly above 100~eV, $K_{\rm ele}$ typically 
asymptotes to the classical Spitzer formula \cite{Spitzer1963}. In cases where a large value of $|\nabla T_{\rm ele}|$ 
would give rise to unphysically 
large heat fluxes, $\vec{q}_{\rm ele}$ is capped to some fraction of the maximum physically-allowable heat flux, e.g.,
\begin{equation}
\vec{q}_{\rm max, ele} = \alpha_{\rm ele} n_{\rm ele} k_{\rm B} T_{\rm ele} \sqrt \frac{k_{\rm B} T_{\rm ele}}{m_{\rm ele}},
\end{equation}
where $n_{\rm ele}$ is the electron number density, $k_{\rm B}$ is Boltzmann's constant, $m_{\rm ele}$ is the mass of
the electron, and $\alpha_{\rm ele}$ is a dimensionless number much less than one \cite{Atzeni2004}. 
For sake of comparison the HYDRA and FLASH simulations presented here both assume $\alpha_{\rm ele} = 0.05$ 
for the flux limiter. 

Both FLASH and HYDRA incorporate the aforementioned equations, coupling them with a hydrodynamics
solver and a ray-trace algorithm to model the propagation and deflection of laser rays through the 
grid. This coupling is sufficient to model a wide range of laser-produced plasmas, including 
 both the pre-plasma test problems and the validation experiments considered in 
Grava et al. \cite{Grava_etal2008} that we investigate here. In the next section we consider the subtle 
and not-so-subtle differences in the way that FLASH and HYDRA solve these equations.

\section{Differences between FLASH \& HYDRA implementations}
\label{sec:diff}

\begin{figure*}
\includegraphics[angle=0,width=7in]{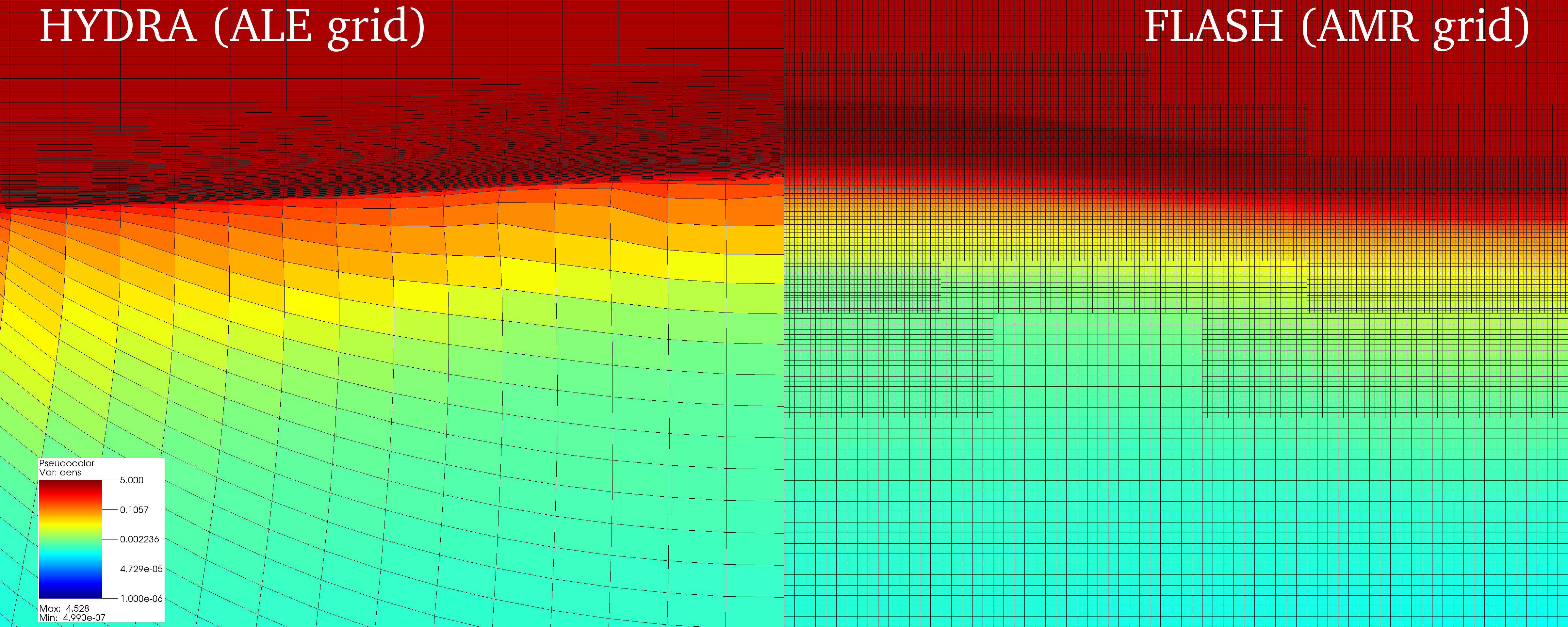}
\caption{ Snapshots of the density from a HYDRA (left) and a FLASH (right) pre-pulse simulation with the computational mesh overlayed. Results shown are for the 15.3~mJ pre-pulse at $t = 1.0$~ns; density units are in g cm$^{-3}$ and total grid dimensions are 30 $\mu$m x 75 $\mu$m. Laser rays travel from the bottom to the top, irradiating target material at the top center of the plot.}\label{fig:grid}
\end{figure*}

\subsection{ALE versus Eulerian AMR}
\label{sec:aleamr}

Apart from the subtle differences in the \emph{physics} implemented in HYDRA
and FLASH, by design the codes have radically different approaches to defining and
evolving the computational mesh and solving the equations of hydrodynamics. Whereas FLASH performs 
its hydrodynamic calculations on a fixed, finite volume Eulerian grid that can be refined (or de-refined) on the fly to maintain high
resolution in interesting areas using AMR, by contrast the 
ALE approach used by HYDRA allows the grid to distort and move
with the fluid flow with (preferably) minor deviations from this Lagrangian behavior to prevent severe
tangling of the mesh. As an illustration of these two methods, Fig.~\ref{fig:grid}
shows a density snapshot from the 15.3~mJ pre-pulse test (described in \S~\ref{sec:prepulse})
at $t = 1$~ns with the grid overlayed. 

Choosing one grid scheme over another also substantially changes the \emph{forms} of the differential equations solved.\footnote{The reader can find more complete discussions of grid schemes and the forms of the differential 
equations solved, including the equations of momentum and energy conservation, in 
\cite{Knupp_etal2002,Castor2004,Kucharik2006} and other sources.}
We highlight the ramifications of choosing one scheme over another using the 
continuity equation; i.e., 
\begin{equation}
\frac{\partial \rho }{\partial t} + \nabla \cdot (\rho \vec{v}) = 0, \label{eq:continuity}
\end{equation}
where $\rho$ is the mass density of the fluid and $\vec{v}$ is the vector field of the fluid velocity.
An AMR grid scheme is an essentially Eulerian approach; therefore at each cell the density
is represented at a fixed point in space in a fixed volume\footnote{For the present discussion we ignore the detail that $V_i$ may change when the mesh is refined or de-refined.}, $V_i$,
so that the total cell mass, $m_i$ increases or decreases with the cell density, $\rho_i = m_i / V_i$, 
as fluid moves into or out of the cell. With this in mind the continuity equation at a cell
can be rephrased as
\begin{equation}
\frac{\partial \rho_i }{\partial t} = \frac{\partial}{\partial t} \left( \frac{m_i}{V_i} \right) = \frac{1}{V_i}\frac{\partial m_i }{\partial t} = - \left[ \nabla \cdot (\rho \vec{v}) \right]_i ,
\end{equation}
where the right hand side is the divergence of the mass flux into the cell.

This result can be compared to an ALE grid scheme, which is an essentially Lagrangian approach that instead keeps 
the \emph{mass} in each cell, $m_i$, (at least approximately) fixed as the mesh expands and flows with the 
fluid. In examining the consequences of this approach, it is helpful to recast Eq.~\ref{eq:continuity} in a more suggestively Lagrangian form,
\begin{equation}
\left( \frac{\partial}{\partial t} + \vec{v} \cdot \nabla \right) \rho + \rho \nabla \cdot \vec{v} = 0.
\end{equation}
The term in the parenthesis is readily identified as the time derivative in the frame of the 
moving fluid $(\equiv D / Dt)$. The continuity equation for a cell moving with the fluid 
flow becomes
\begin{equation}
\frac{D \rho_i}{Dt} = -\rho_i \left[ \nabla \cdot \vec{v} \right]_i  , \label{eq:convder1}
\end{equation}
and once again, $\rho_i = m_i / V_i$, so that the left hand side of this expression becomes
\begin{equation}
\frac{D \rho_i}{Dt} = \frac{D}{Dt} \left(\frac{m_i}{V_i}   \right) = m_i \frac{D}{Dt} \left(\frac{1}{V_i}  \right) .\label{eq:convder2}
\end{equation}
Using Eq.~\ref{eq:convder2} in Eq.~\ref{eq:convder1},
\begin{equation}
\frac{1}{V_i} \frac{D V_i}{Dt} = \left[ \nabla \cdot \vec{v} \right]_i . \label{eq:vol}
\end{equation}
This result has important consequences because it means that a strong divergence of the 
fluid flow, as is generic in laser ablation, will expand the cells and consequently
lower the resolution near the critical density where most of the laser energy
is deposited. Conventionally, and, for example, in the HYDRA simulation presented in Fig.~\ref{fig:grid}, 
users of purely-Lagrangian and ALE codes 
will initialize a simulation with significantly higher resolution in the solid or near-critical-density
regions that will absorb most of the laser energy in order to maintain modest resolution throughout the 
grid after these cells have become stretched by the ablative flow. 

To be absolutely precise, Eqs.~\ref{eq:convder1}-\ref{eq:vol} 
are exact for purely-Lagrangian codes but are only approximately true for 
ALE codes such as HYDRA.  A fuller presentation would describe how ALE codes can apply a ``re-zoning''
operation to prevent severe tangling of the mesh, which will result in a
(preferably small) change to the mass in each cell \cite[][and references therein]{Knupp_etal2002,Kucharik2006}.

The loss of resolution in purely Lagrangian and ALE meshes does not automatically mean
that the AMR approach is better suited to computationally efficient modeling of
ablation-driven plasmas. For example, the literature has documented problems 
where the AMR scheme can yield poor results \cite{Quirk1994,Tasker_etal2008,Springel2010}. Suffice it to say that 
neither the ALE nor the AMR scheme has been definitively proven to be better-suited 
for modeling laser ablation. In this paper we hope to gain some insight into 
this question; however, as discussed in the next section, the physics in the two codes
is not {\it exactly} the same. Also, any discrepancies stemming directly 
from the choice of grid scheme should disappear at sufficiently high resolution 
\cite{Robertson_etal2010}. It may be that the fine meshes 
shown in Fig.~\ref{fig:grid} are already in this regime. The principal 
difference between the two methods may simply be in making different decisions
for the spatially-varying and time-varying resolution of the computational grid. 
(The FLASH simulations presented here, for example, determine the refinement level based on 
changes in mass density and electron temperature.) As in a recent astrophysical code 
comparison for modeling galaxy formation (\cite{Scannapieco_etal2012} and commentary in \cite{Hopkins2012}), 
we expect that the hydrodynamic behavior is well-captured by both schemes but that other 
assumptions -- such as those discussed in the next section -- are where more important differences can emerge.

\subsection{EOS \& Opacity Models}

An obvious place where discrepancies can arise is in the different EOS models used by the two codes.
Without any special reasons to prefer one EOS model for Aluminum over another,
 we present FLASH simulations using the SESAME model \cite{SESAME1}, and the commercially-available
PROPACEOS model \cite{Macfarlane_etal2006}.\footnote{We also experimented with the BADGER EOS model \cite{BADGER} developed at the  University of Wisconsin, but were unable to resolve some questions regarding the publicly-available version of 
the code in time for publication.} These
well-respected models can at least highlight the EOS dependence of the FLASH 
results relative to HYDRA, which uses either the QEOS \cite{QEOS} or LEOS\footnote{\url{https://www-pls.llnl.gov/?url=about_pls-condensed_matter_and_materials_division-eos_materials_theory}} model. (The Al slab pre-pulse tests were performed
with the LEOS model while the V-shaped groove validation experiment was modeled using QEOS.) Intuitively,
it is important to note that the EOS model controls not only the pressure of the 
plasma, given a density and temperature, but also affects the efficiency of 
heat conduction by determining the mean ionization fraction, $\bar{Z}$, 
as well as the specific heat in Eqs.~\ref{eq:eele}~\&~\ref{eq:eion}.  Aside from subtleties in interpolation, 
these models can disagree on the quantities of interest by up to factors of a few
in some density-temperature regimes \cite{BADGER}.

Discrepancies may also arise from differences in opacity models. The FLASH
results presented here use only the PROPACEOS model for the opacity of Aluminum, 
while the HYDRA results discussed here use an average-atom opacity model developed at LLNL.  However, the opacities can only
be a source of discrepancy in cases where radiation-energy loss plays an important
role. Since the atomic number of Al is still relatively low, the usual channel for
energy loss through Bremsstrahlung \cite[][Ch. V.]{ZeldovichRaizer} is not necessarily a large effect.
So although the opacity models from different research groups
can disagree at the level of a few to tens of percent\footnote{Our own comparisons of 
the all-frequency-integrated rossland means for Al from PROPACEOS and publicly-available 
Opacity Project \cite{Seaton_etal1994,Mendoza_etal2007} data, for example, are consistent with this conclusion.}, 
we regard the opacity model to be of 
secondary importance when searching for the cause of differing results.

\subsection{Treatment of the Laser}
\label{sec:laser}

For both of the problems investigated here, the laser model in FLASH
differs from HYDRA in that HYDRA includes the ponderomotive force 
(a.k.a. light pressure) of the laser as an extra term in the momentum
equation \cite{Berger_etal1998}. However, the ponderomotive force for 
the most intense laser pulse we consider here ($10^{13}$ W/cm$^2$)
is smaller than atmospheric pressure and should be overwhelmed by
the plasma pressure for any solid-density plasma hotter than
{\it room temperature}. Therefore the lack of ponderomotive force
in FLASH should be a negligible source of error at these intensities.

Both FLASH and HYDRA deposit the laser energy on the 
grid using the inverse bremsstrahlung approximation in the sub- to near-critical
density regions where the laser rays propagate. 
 Both codes use the Kaiser \cite{Kaiser2000} algorithm to calculate the trajectories
of the rays and interpolate the deposited energy to the computational mesh at each timestep. 
 For the 2D cartesian geometry of the V-shaped groove problem the setup and refraction 
of these rays should be nearly identical in the two codes.
For the 2D cylindrical Aluminum slab simulations described in the next section,
there are some differences in the simulation setup. In FLASH, a few thousand
laser rays, each weighted by the spatial profile of the laser spot, enter the 
grid parallel with each other and are not actually permitted to refract 
laterally in order to ensure that the laser energy is deposited
as smoothly as possible. (Earlier tests in which the rays were allowed to refract laterally gave
very similar results but with somewhat more noise in the energy deposition.)
By contrast, the rays in HYDRA converge with an f/3 ratio and refraction away
from the laser axis is allowed. Essentially, the FLASH code is set up with
the aim of maximizing the smoothness of the energy deposition (as the Flash
Center currently encourages its users to do) while HYDRA is set up to maximize the 
correspondence between simulated rays and real laser rays, as is the customary
way of modeling pre-pulses at LLNL.

\subsection{Remaining Minor Differences}

While the HYDRA simulations discussed here include heat conduction via both electrons (Eq.~\ref{eq:qele})
and ions, the FLASH simulations neglect the contribution from ions. Heat conduction is proportional
to the thermal velocity of the species, therefore the contribution from ions is suppressed relative to that of
electrons by a factor of $\sqrt{m_{\rm ele} / m_{\rm ion}} \sim 0.0045$ for Al. Although the next
FLASH release will include this effect, we neglect it here.

Finally, as a stand-in for vacuum, the FLASH \& HYDRA simulations include very low density Helium 
($5 \cdot 10^{-7}$ g cm$^{-3}$ unless noted otherwise) outside the target. There may be subtle differences in the way
that FLASH \& HYDRA treat the Al/He interface as the ablated aluminum expands into the low density He.
As the ablated Al moves away from the target, some of the kinetic energy of the high-velocity Al is transferred to the 
He ions, which, being low density, are super-heated, creating an out-of-equilibrium (a.k.a. 3T)
shock \cite{Shafranov1957,MihalasMihalas1984,Fatenejad_etal2011}. But since the super-heated Helium is 
very low density the properties of the expanding Al further upstream from this interface are, in practice, 
insensitive to the specific value for the density of the vacuum-like He. Nevertheless, for completeness, it bears mentioning
that the treatment of this shock is another place where the codes may yield different behavior.

\section{Code-to-Code Comparisons for an Aluminum Slab Target Irradiated by a Laser Pre-Pulse}
\label{sec:prepulse}

\begin{figure}
\includegraphics[angle=0,width=3.0in]{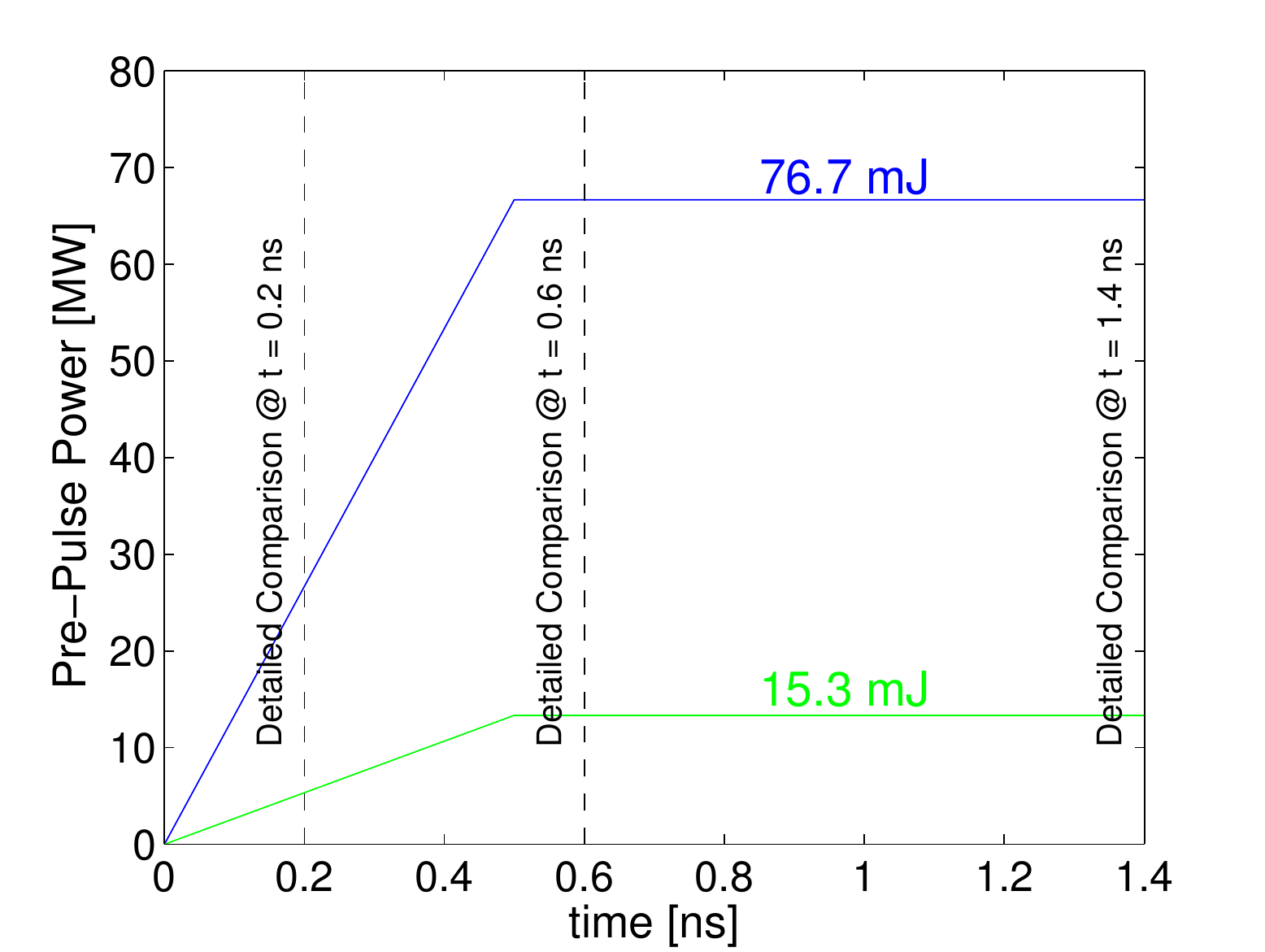}
\caption{ Total laser power versus time for the pre-pulse tests. We present detailed comparisons of the results at $t = $0.2 ns, 0.6~ns, and 1.4~ns for the 15.3~mJ pre-pulse in Fig.~\ref{fig:20mJresults} and for the 76.7~mJ pre-pulse in Fig.~\ref{fig:100mJresults}.
}\label{fig:power}
\end{figure}

\begin{figure}
\includegraphics[angle=0,width=3.0in]{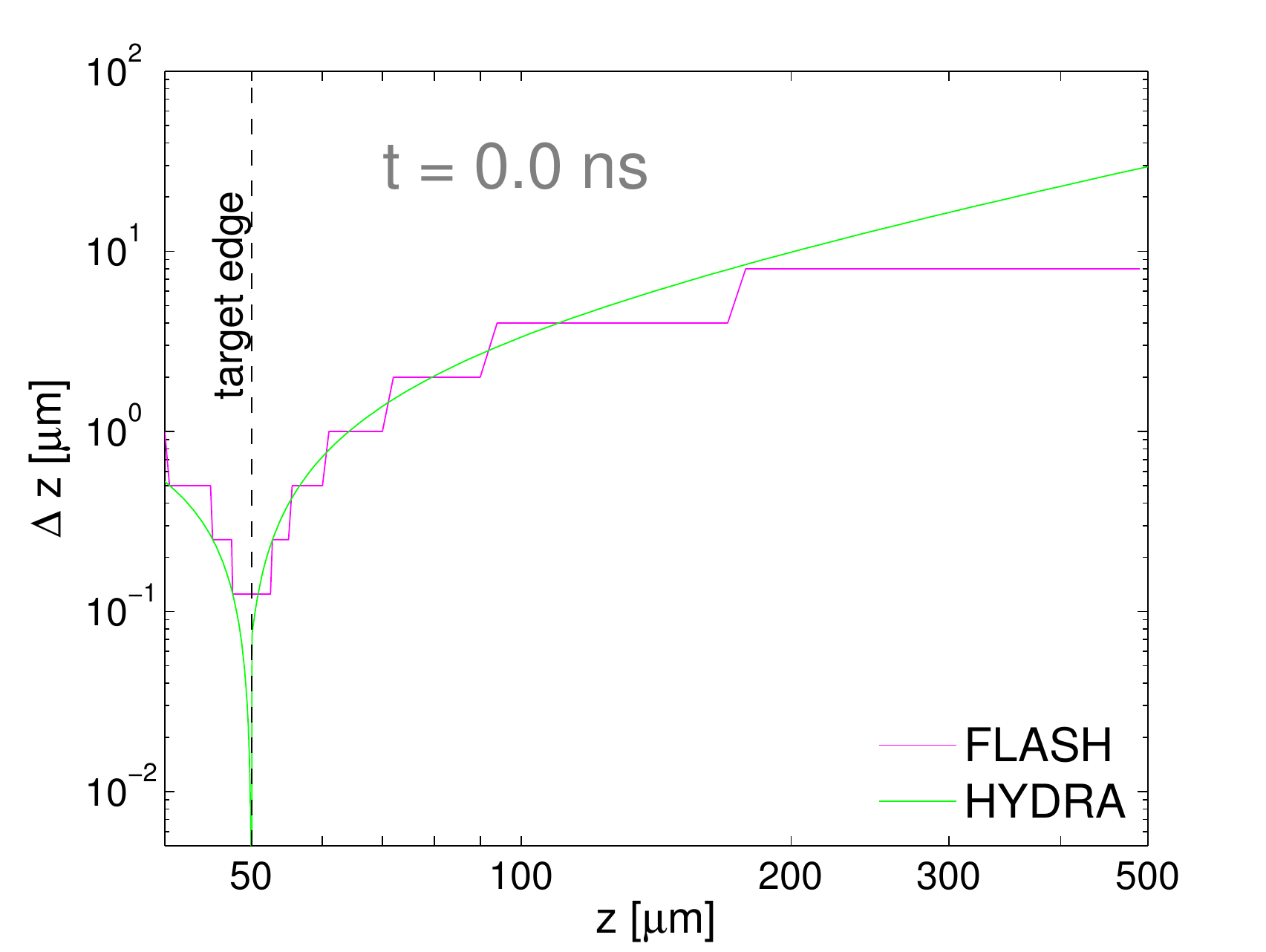}
\caption{ Comparing the initial spatial resolution through the laser 
axis ($r = 0, z$). The Aluminum slab target edge is at 50~$\mu$m and 
extends leftward of this mark.
}\label{fig:res0}
\end{figure}

In this section, we describe code-to-code comparisons between FLASH and HYDRA for a problem in which an Al slab is irradiated by a laser pre-pulse.

\subsection{Design and Simulation Setup}

We chose to compare the two codes for two different pre-pulse energies, 15.3~mJ \& 76.7~mJ, 
using a 0.5~ns linear ramp up and 0.9~ns of constant power as in the temporal profile shown in 
Fig.~\ref{fig:power}. During the first few tens of femtoseconds, the laser energy deposition will heat and 
ionize the target, taking it from room temperature all the way to the plasma regime. The fluid approximation 
is at least partially invalid during these first few moments, but since ionization and heating happens 
quickly this (possibly inaccurate) phase only occurs for a brief interval after which the familiar ablation 
front and hot corona have formed, which can adequately be described by 
the approximations discussed earlier (\S~\ref{sec:req}). Conceivably, then, the two 
codes may differ in these first few instants but agree well for much of the later evolution.
We highlight the results at $t = 0.2$~ns as a 
gauge of how various quantities appear early on but after this transition has taken place. The choice 
of highlighting $t = 1.4$~ns, by contrast, is motivated to show the results after 0.9~ns of steady laser power,
while the $t = 0.6$~ns comparison shows the plasma properties fairly soon after the ramp-up has reached the plateau. 
We show results at these three times for both the 15.3~mJ and 76.7~mJ pre-pulse levels assuming a 20~$\mu$m full width at 
half maximum (FWHM) gaussian beam. For reference, the pre-pulse of the Titan laser (LLNL) is estimated to be 8.5~mJ 
deposited in 2.3~ns, with most of the energy deposition occurring in the last $\sim$1.0~ns \cite{Ma_etal2012} while the 
76.7~mJ pre-pulse is somewhat more akin to deliberate pre-pulses or intrinsic pre-pulses at large laser facilities 
\cite{MacPhee_etal2010}. Note that Akli \& Orban et al. \cite{AkliOrban_etal2012} used FLASH to successfully 
model the pre-pulse of the Titan laser interacting with an Aluminum cone target in order 
to create initial conditions for Particle-in-Cell modeling of the ultra-intense pulse interaction. 
The FLASH simulation in \cite{AkliOrban_etal2012} was run in an analogous way to the FLASH simulations presented
 in this section, with 2D cylindrical geometry, similar laser intensities and pulse durations 
and the same version of the FLASH code.

Already, at $t = 0$, before any evolution has been computed, the codes are set 
up in different ways to handle the problem. Fig.~\ref{fig:res0} compares
the resolutions of the two codes through the laser axis at $r = 0$. The 
AMR scheme in FLASH gives rise to the jagged appearance of the line showing
the FLASH resolution. These simulations employ six levels of mesh refinement 
to span resolutions from 0.125~$\mu$m to 8~$\mu$m. In HYDRA the resolution 
changes smoothly except close to the edge of the target at $z = 50$~$\mu$m
where it becomes extremely fine, $\approx 10^{-3} \mu$m for the smallest cell.
Although this might seem excessive, it is common practice to set up ALE grids
 in this way, anticipating that the ablation will coarsen the grid because of 
the many orders of magnitude change in density that will occur. 

\begin{figure*}
\includegraphics[angle=0,width=2.5in]{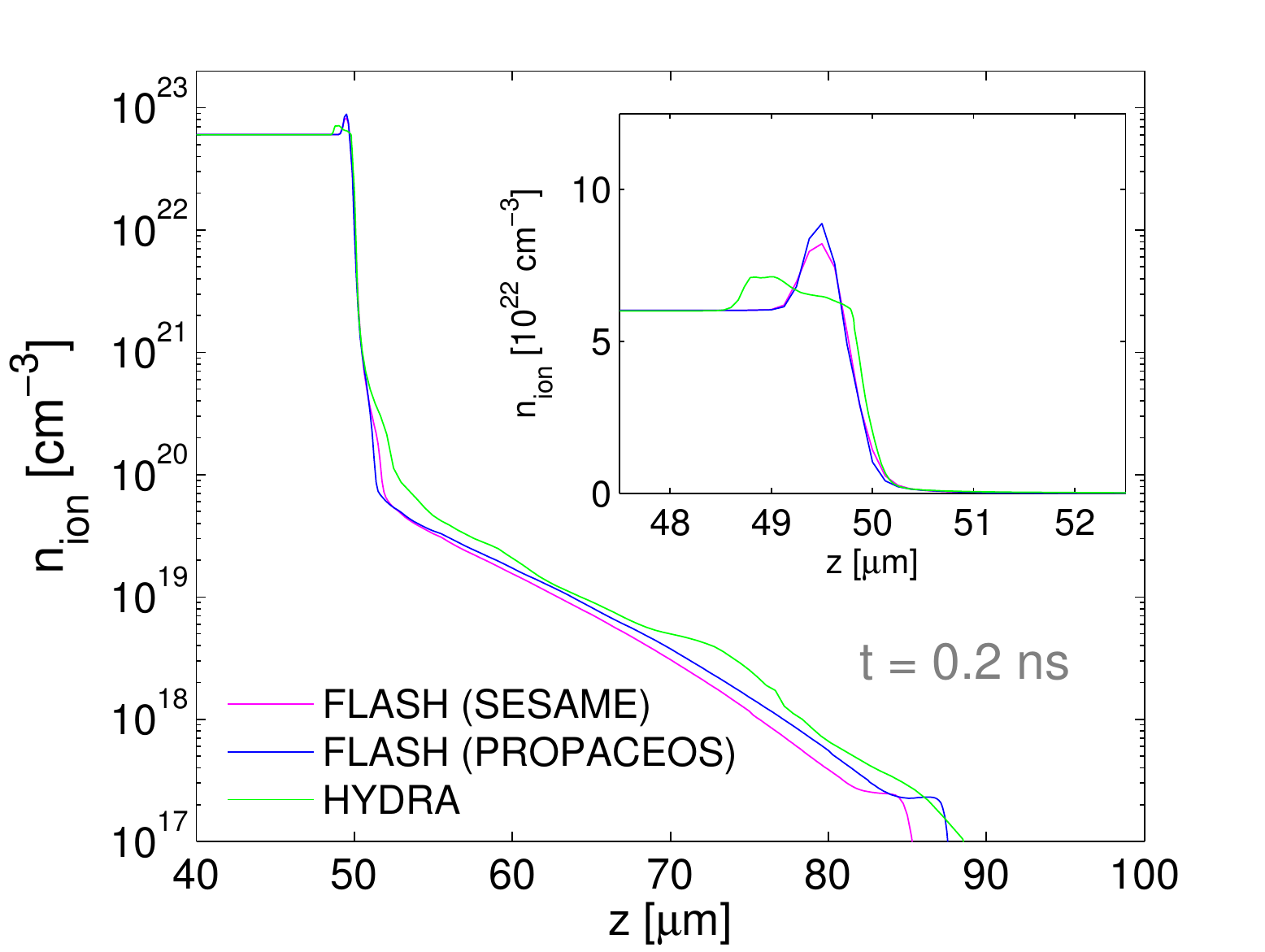}\includegraphics[angle=0,width=2.5in]{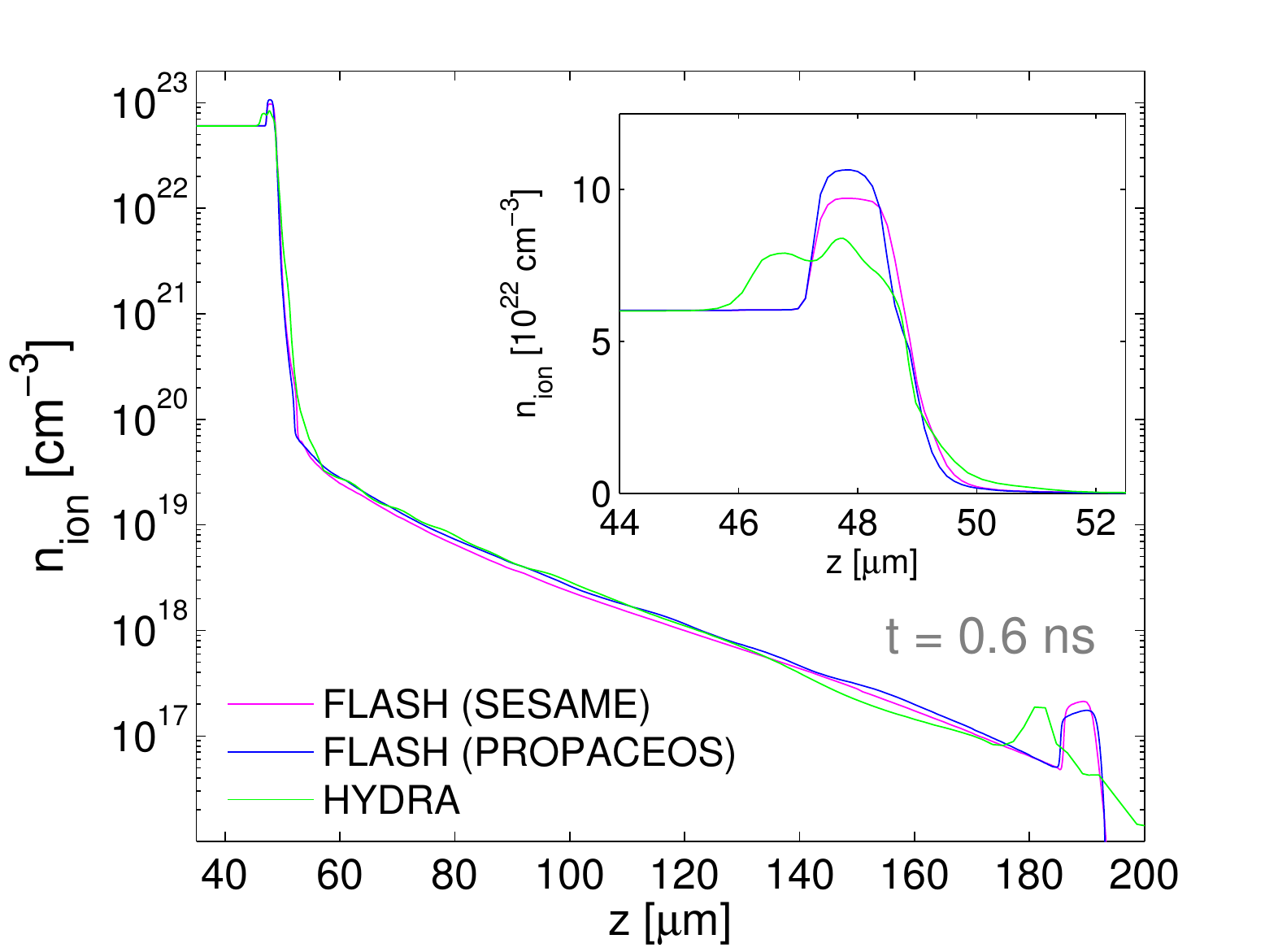}\includegraphics[angle=0,width=2.5in]{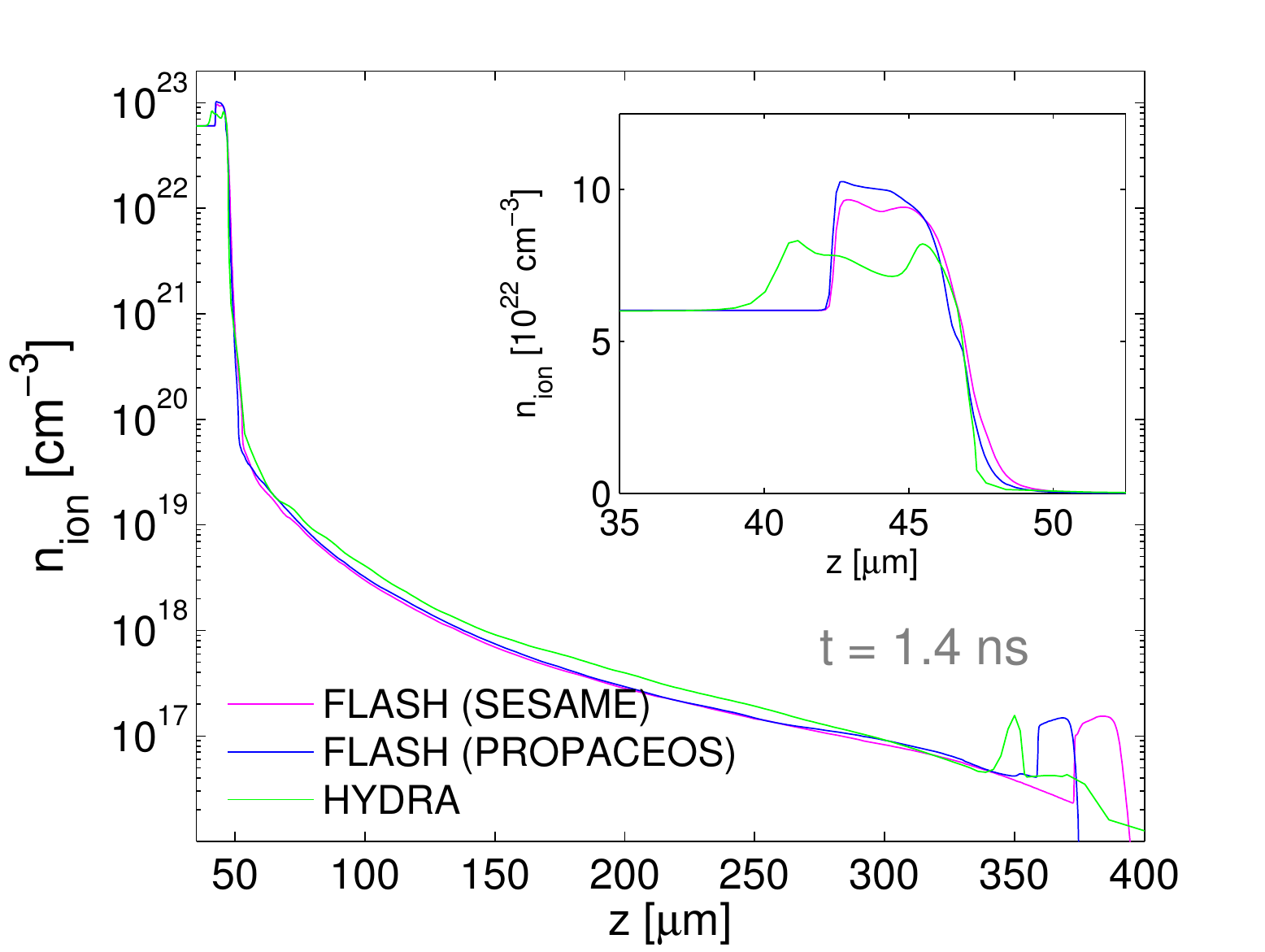}
\includegraphics[angle=0,width=2.5in]{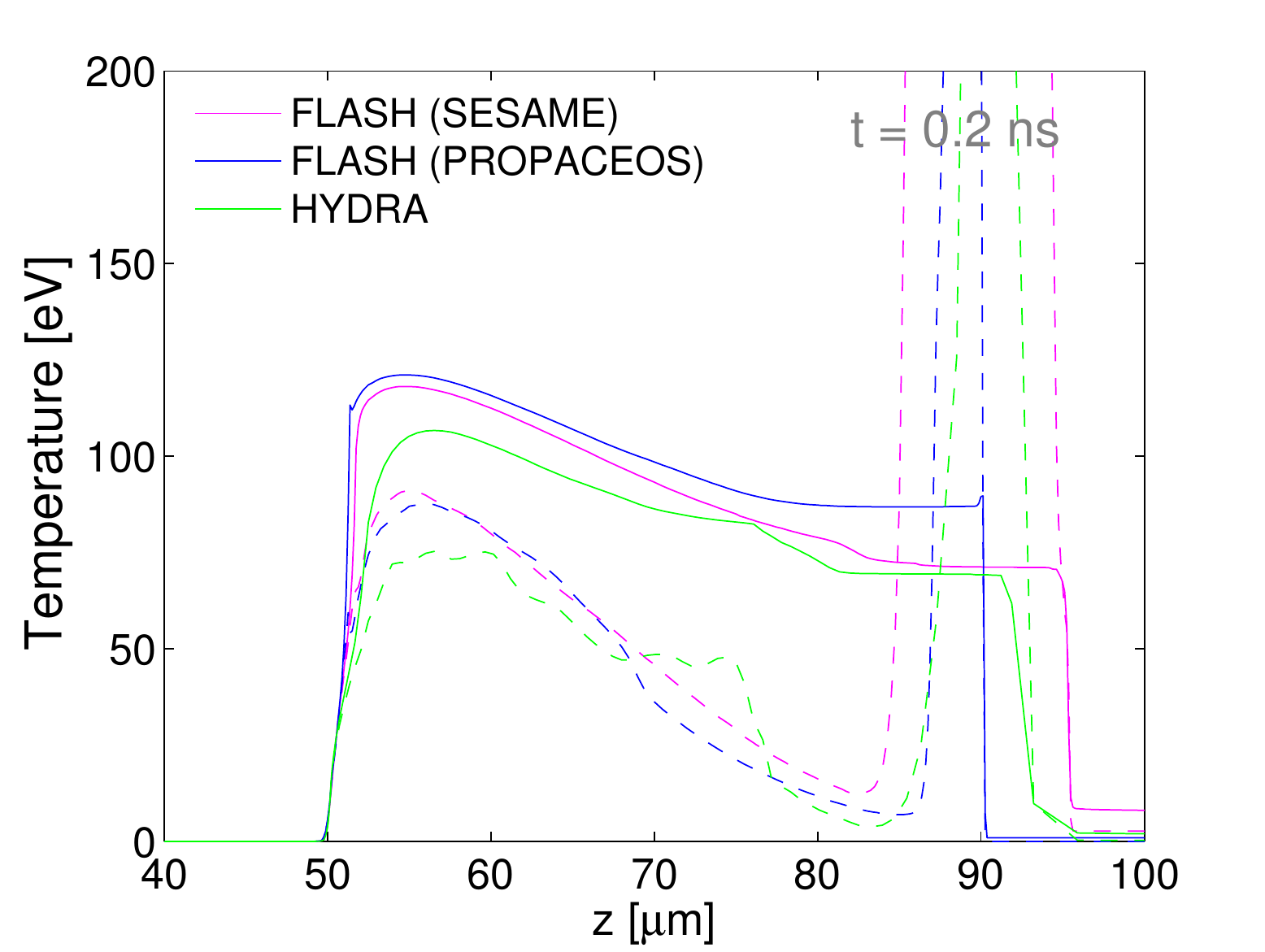}\includegraphics[angle=0,width=2.5in]{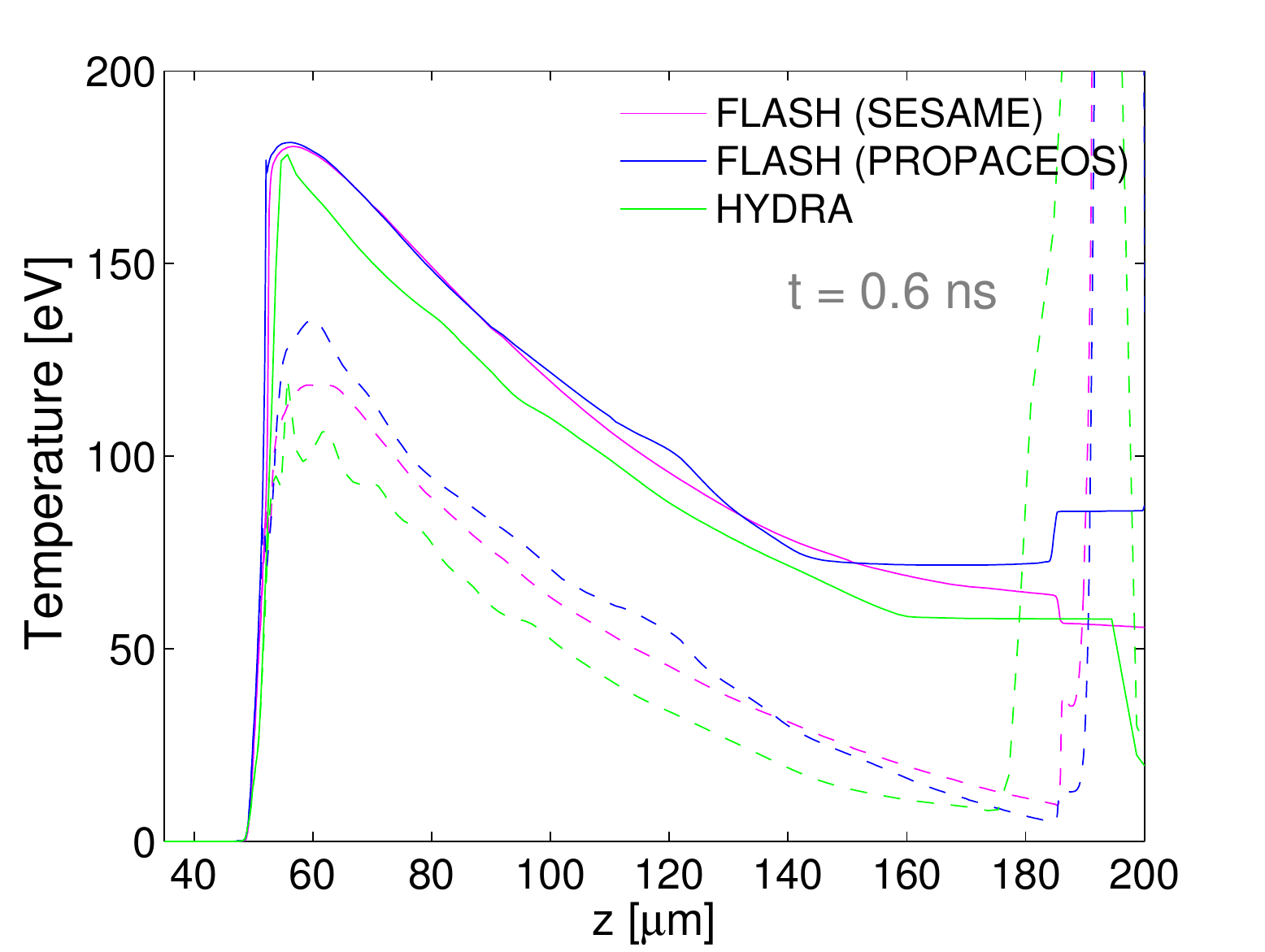}\includegraphics[angle=0,width=2.5in]{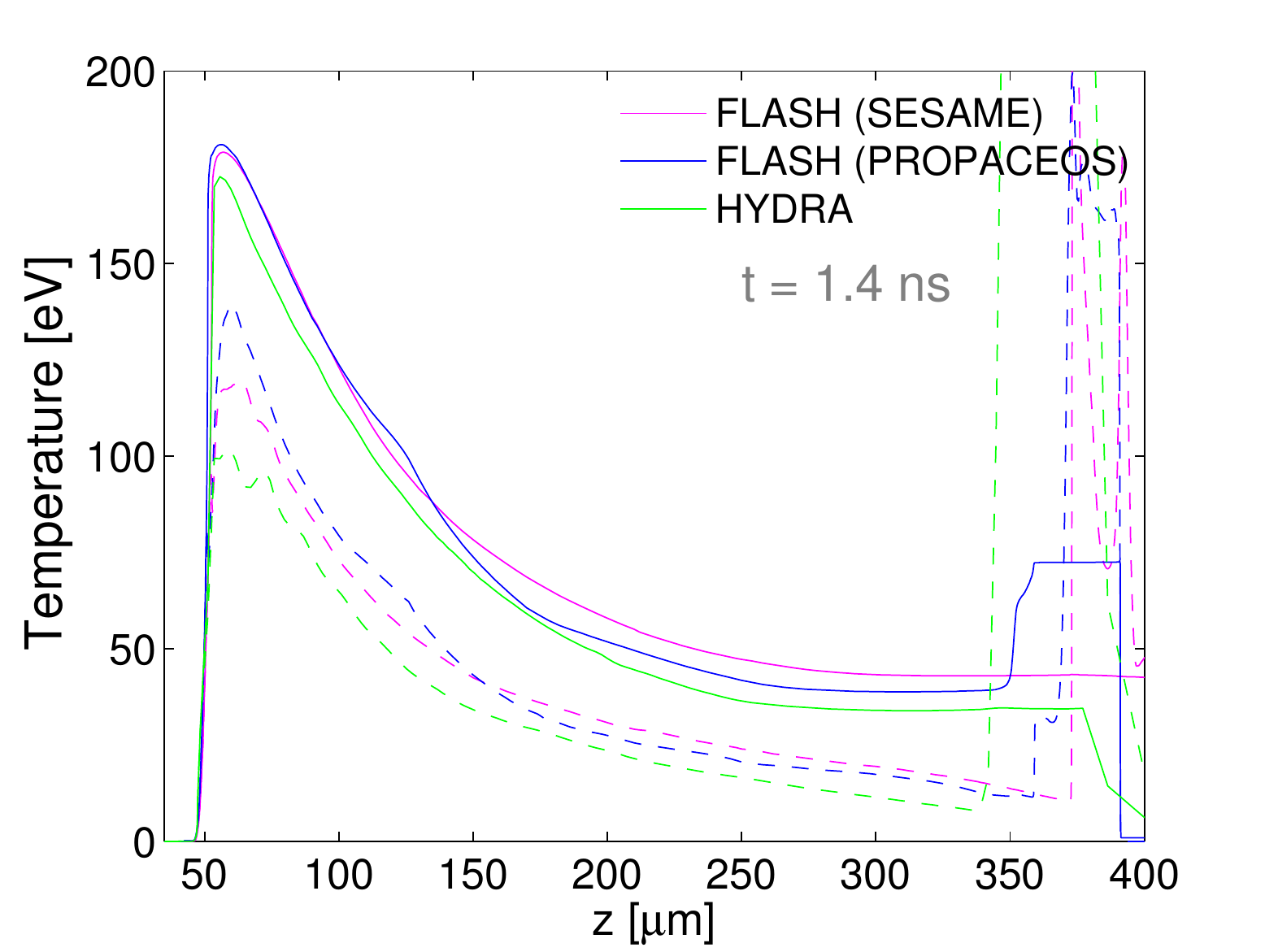}
\includegraphics[angle=0,width=2.5in]{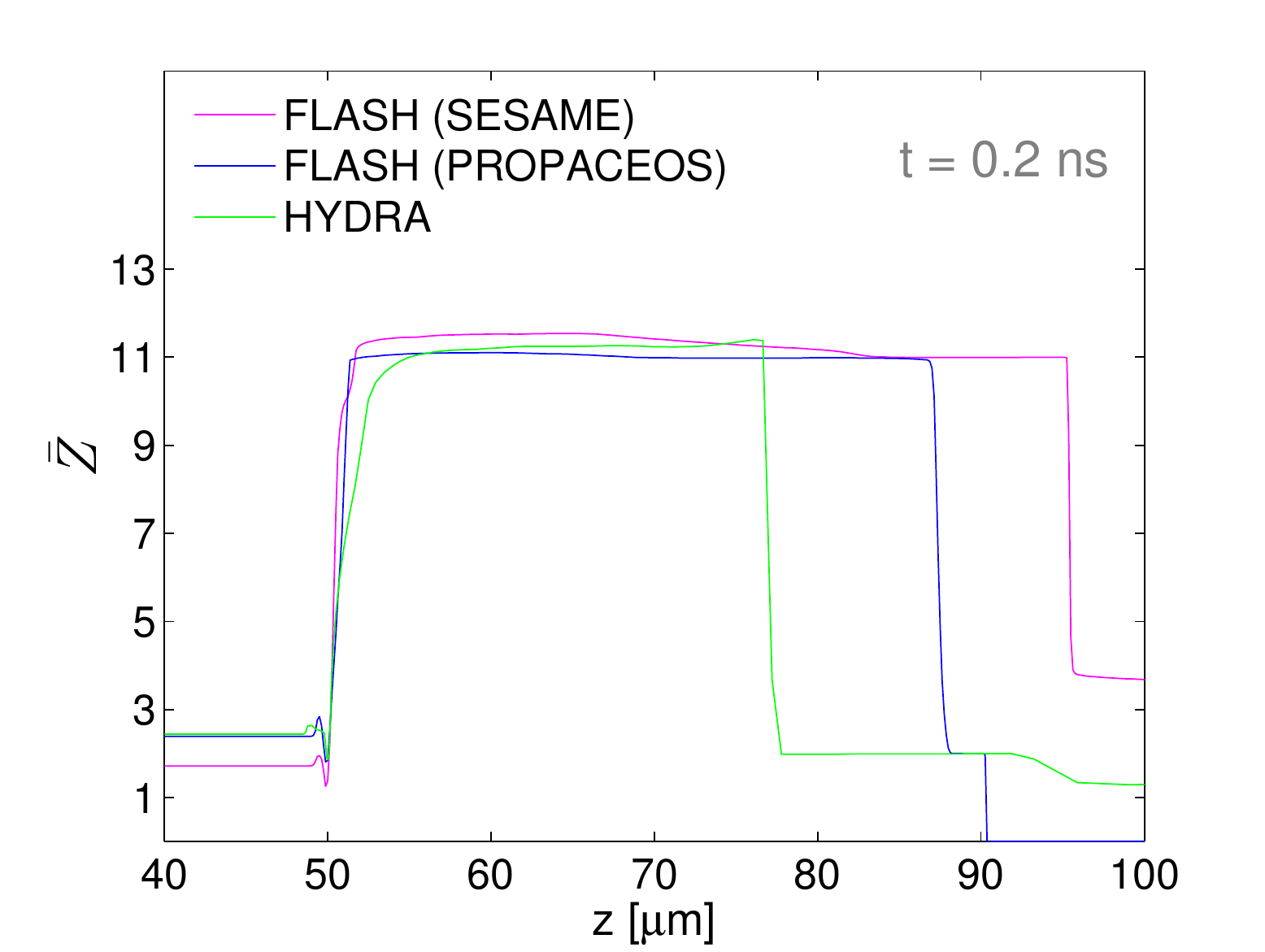}\includegraphics[angle=0,width=2.5in]{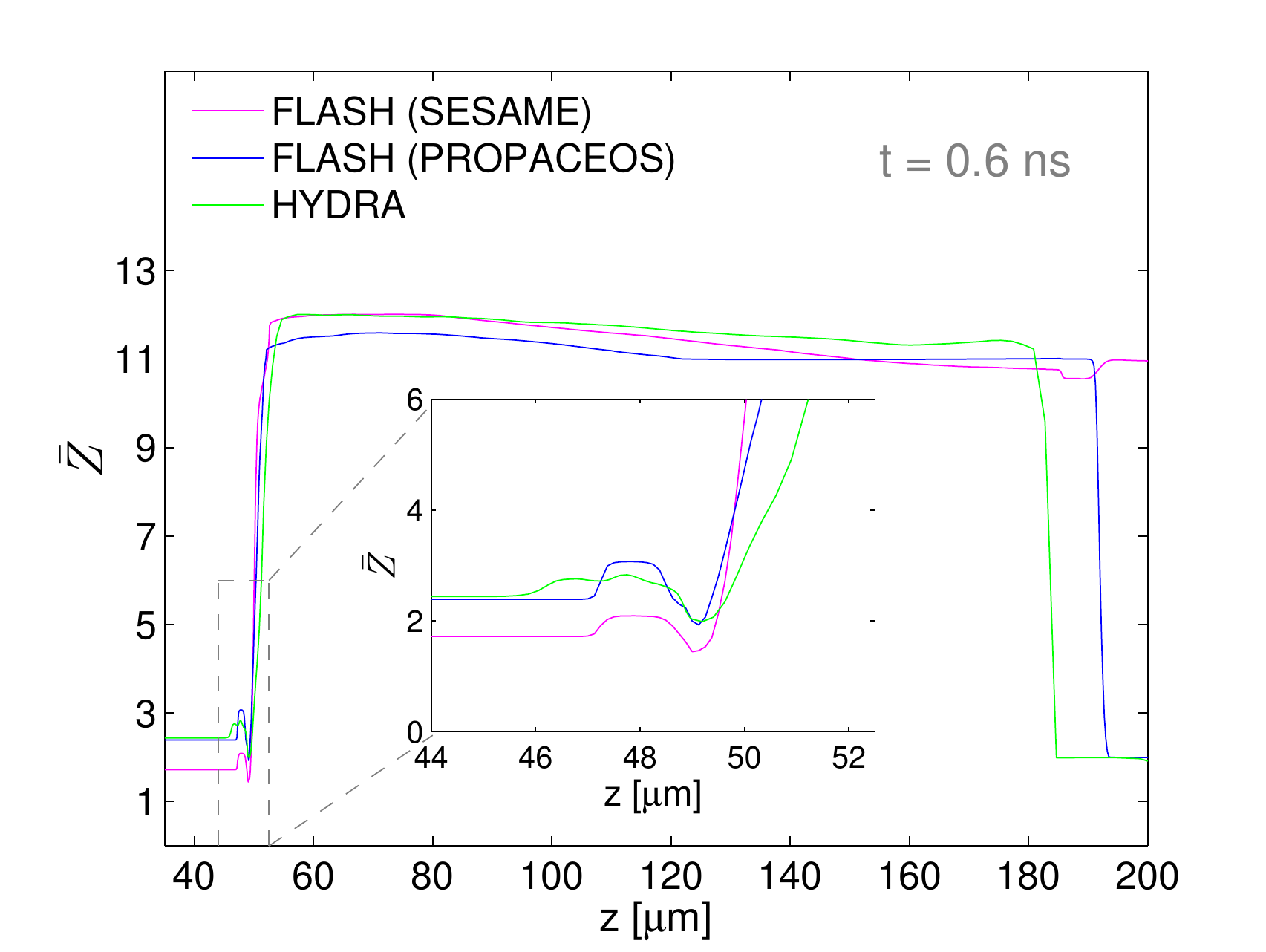}\includegraphics[angle=0,width=2.5in]{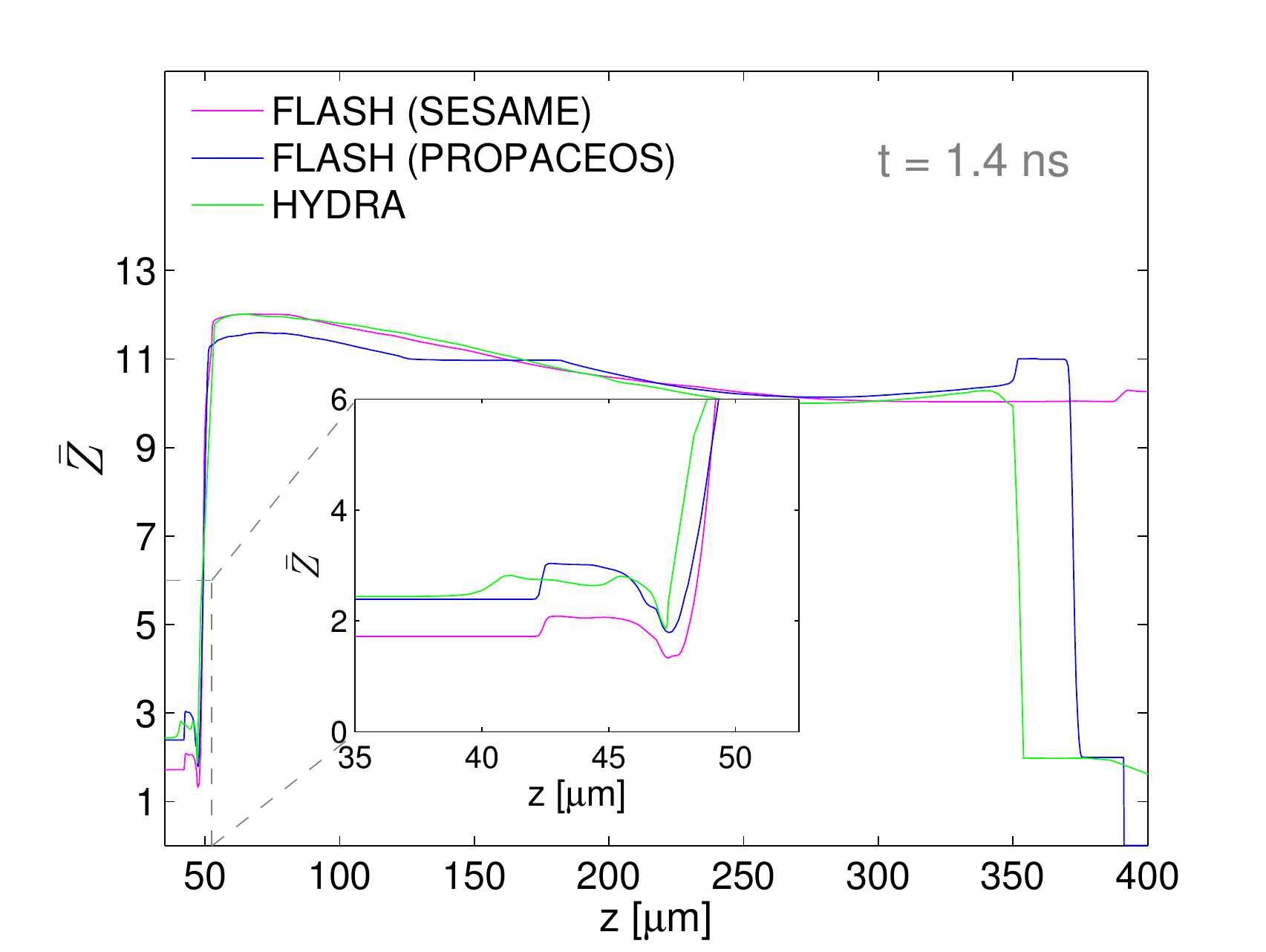}
\includegraphics[angle=0,width=2.5in]{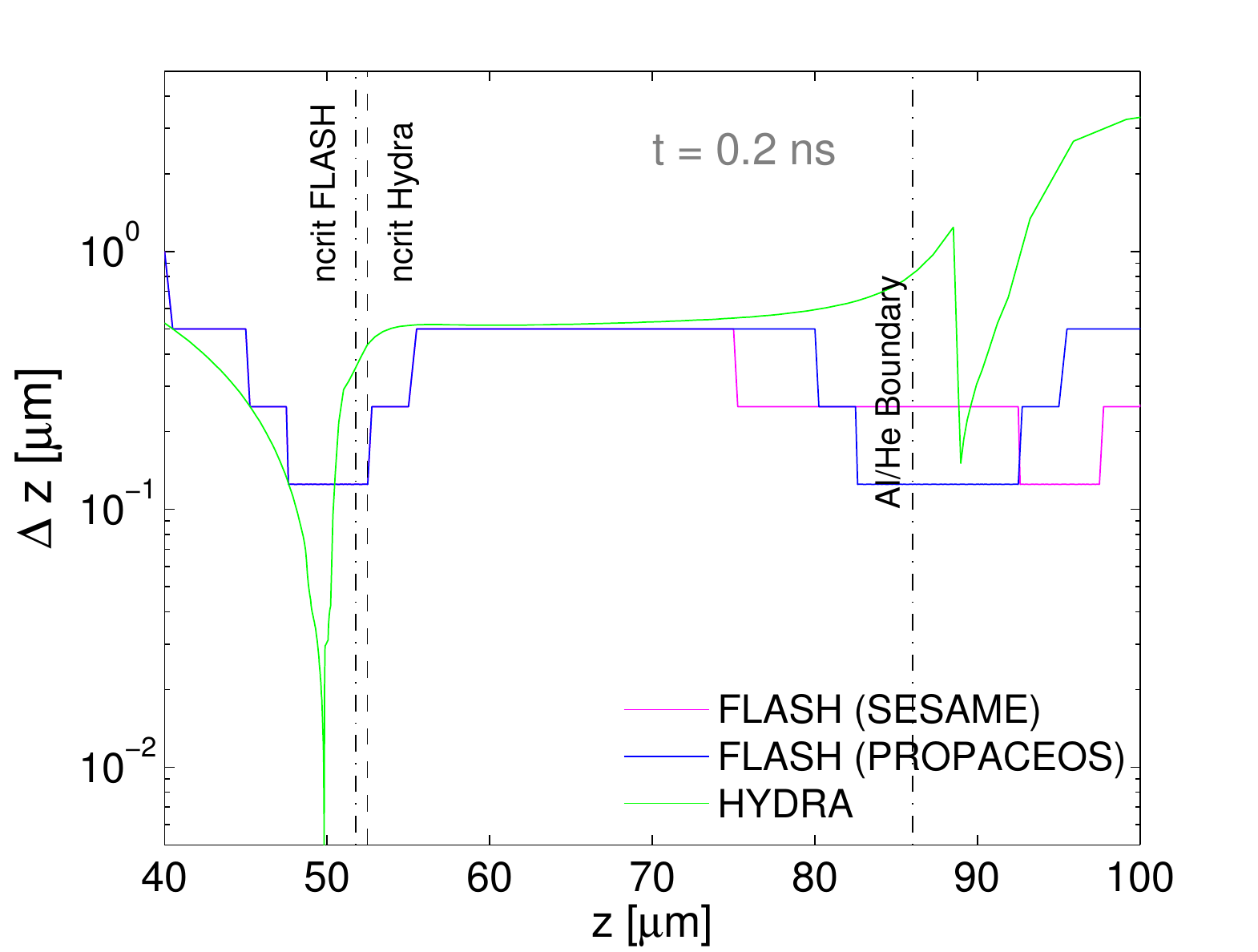}\includegraphics[angle=0,width=2.5in]{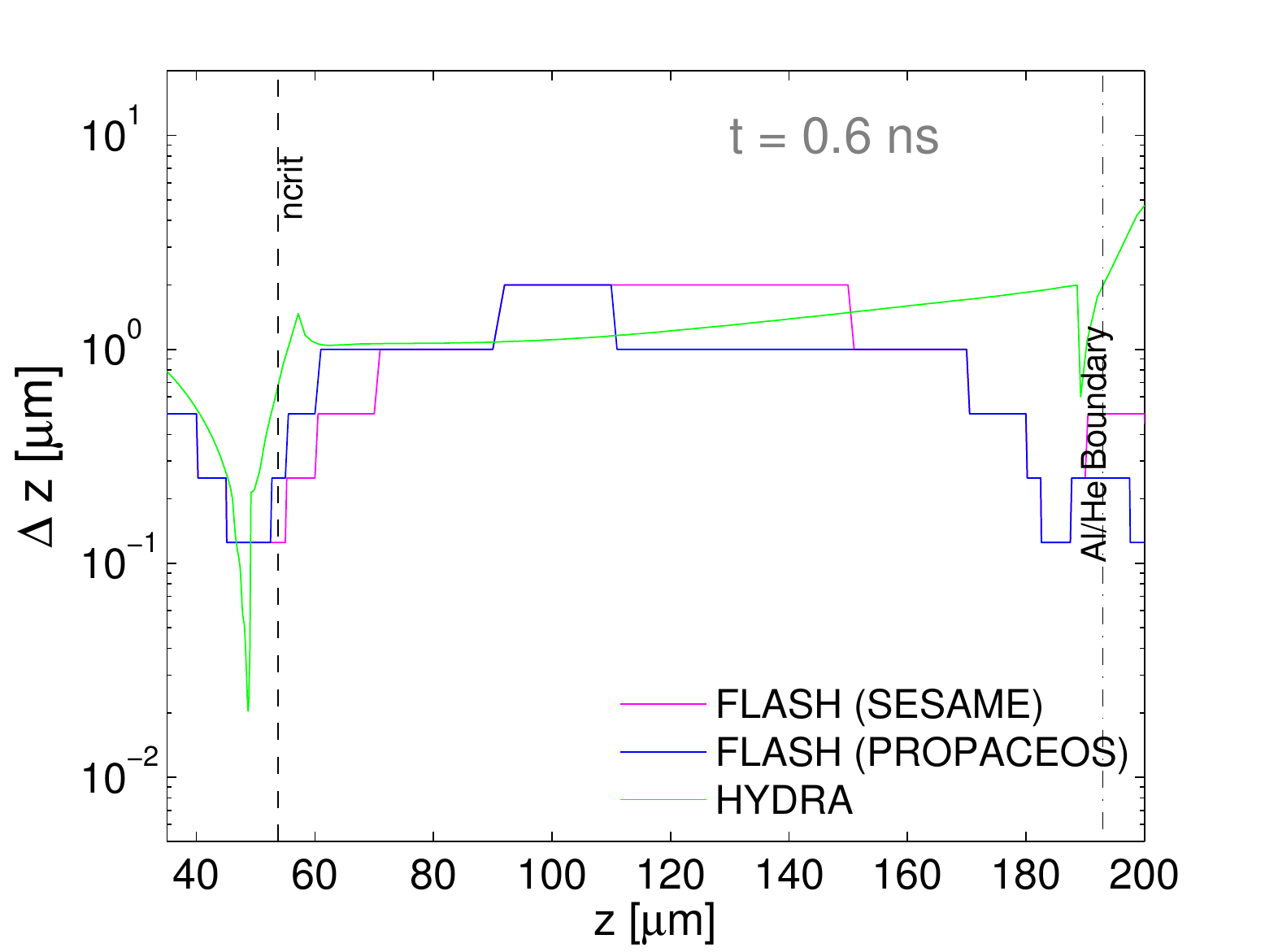}\includegraphics[angle=0,width=2.5in]{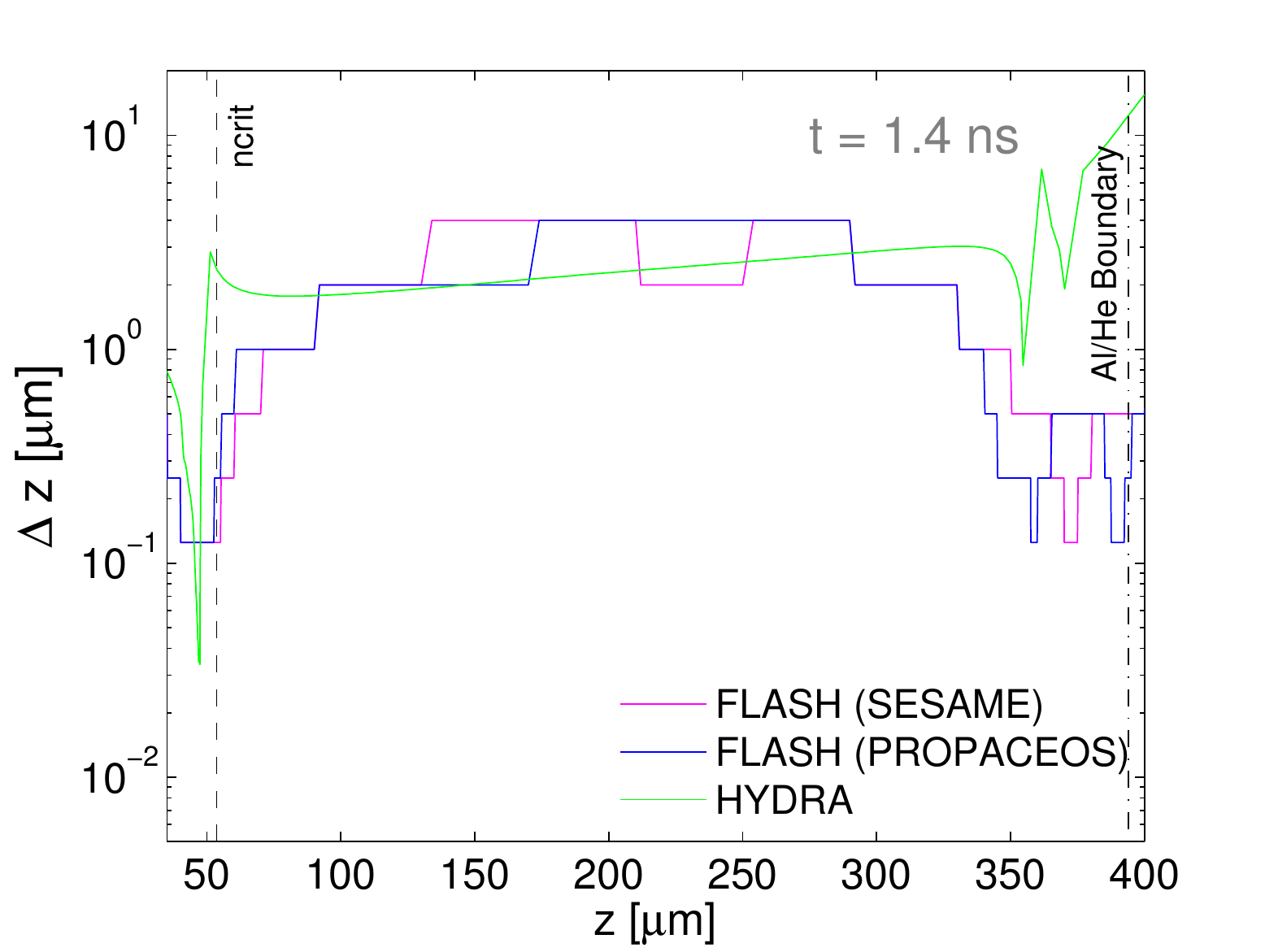}
\vspace{-0.4cm}
\caption{  
Results from simulations of 15.3 mJ pre-pulse irradiation of an Aluminum slab target. Panels show HYDRA results (green) 
and FLASH results using two different EOS models (magenta for SESAME, blue for PROPACEOS). Left-column panels are for 
$t = 0.2$~ns, middle-column panels are $t = 0.6$~ns, right-column panels are $t = 1.4$~ns. The top panels show
ion densities, the next-to-top row panels show electron temperatures in solid lines and ion temperatures in dashed lines,
the next-to-bottom row of panels show the mean ionization state of the plasma and the bottom row of panels compare the resolutions of
the computational meshes. All plots compare quantities along the laser axis ($r = 0, z$).
}\label{fig:20mJresults}
\end{figure*}

\begin{figure*}
\includegraphics[angle=0,width=2.5in]{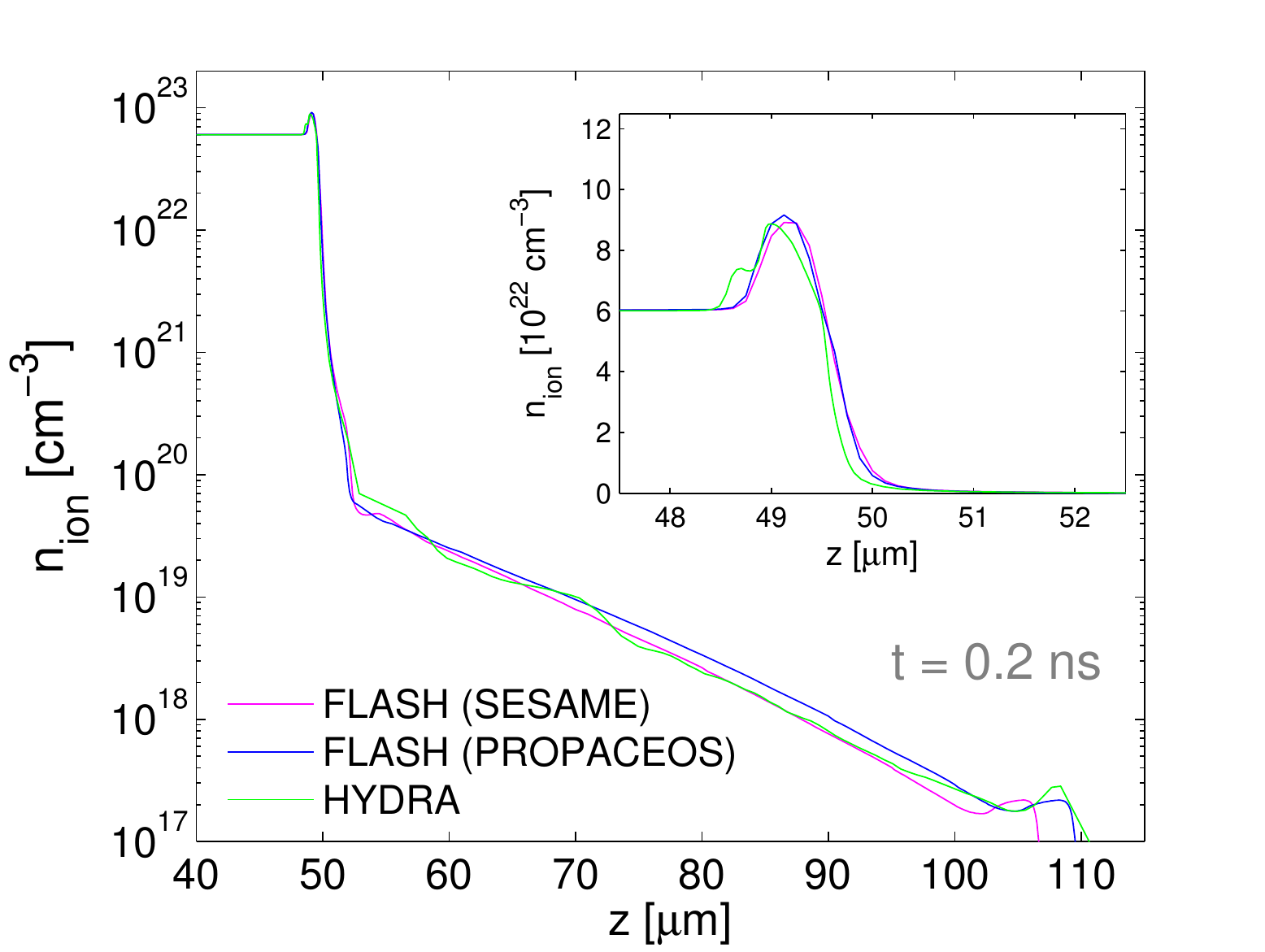}\includegraphics[angle=0,width=2.5in]{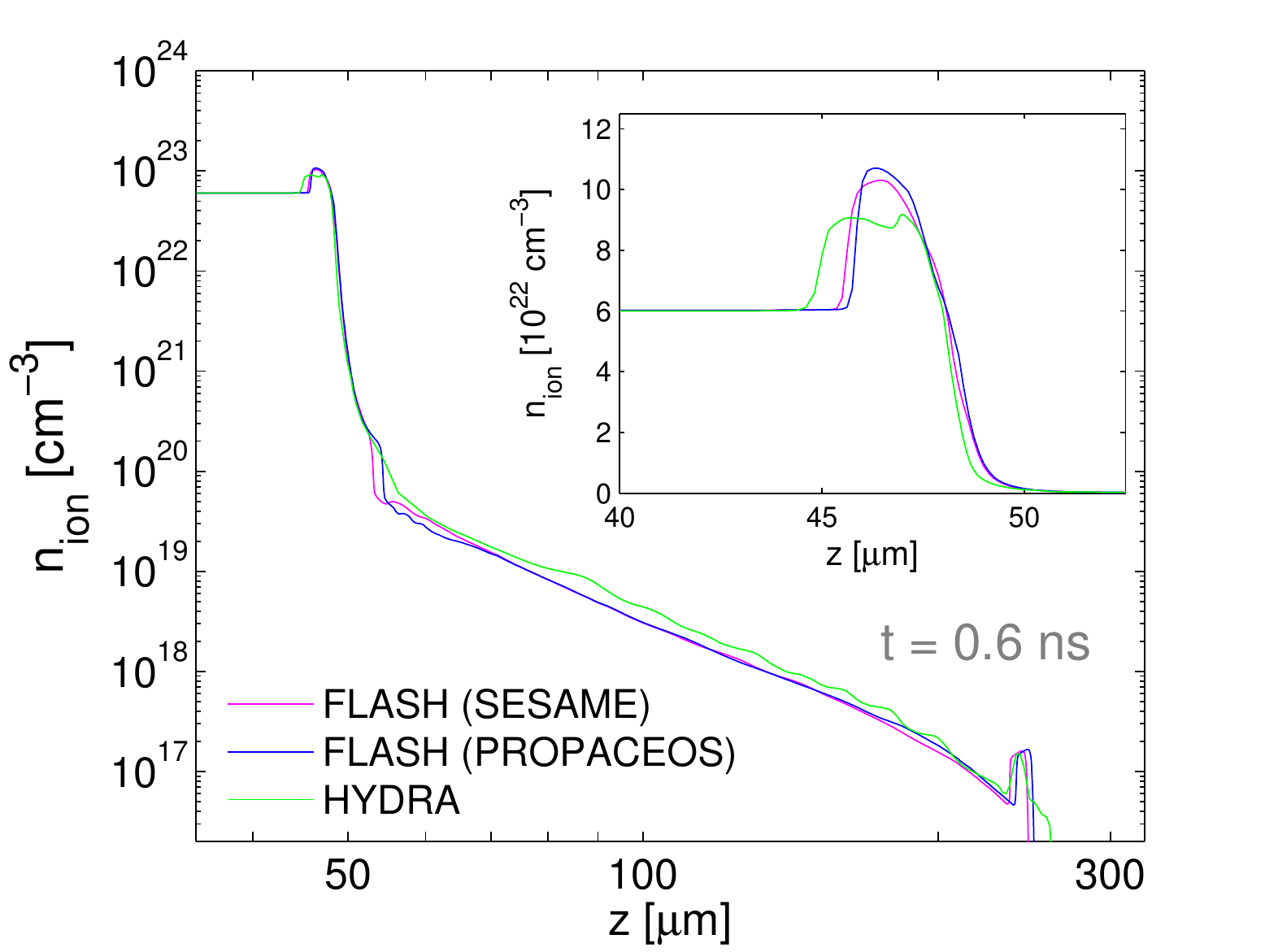}\includegraphics[angle=0,width=2.5in]{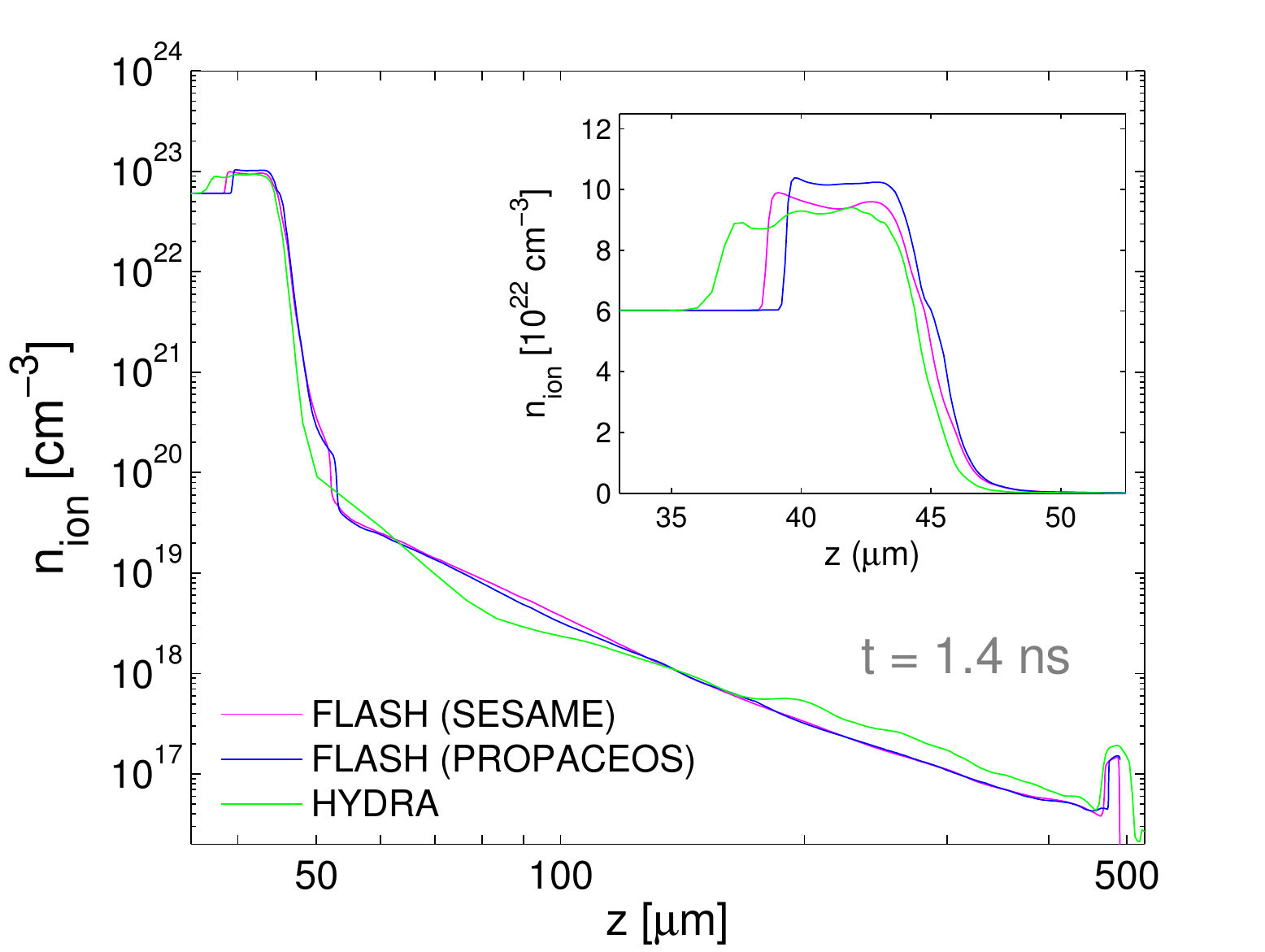}
\includegraphics[angle=0,width=2.5in]{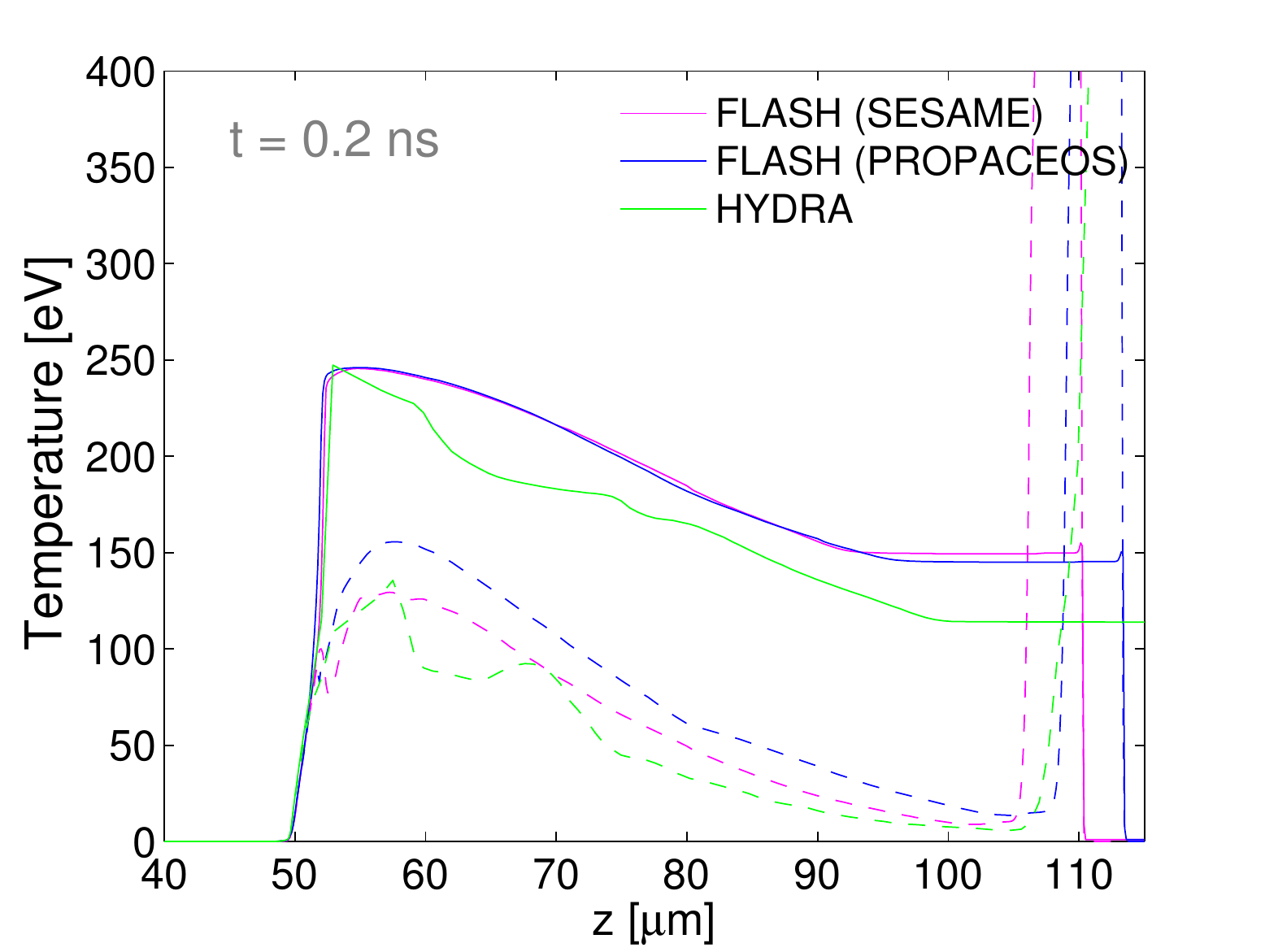}\includegraphics[angle=0,width=2.5in]{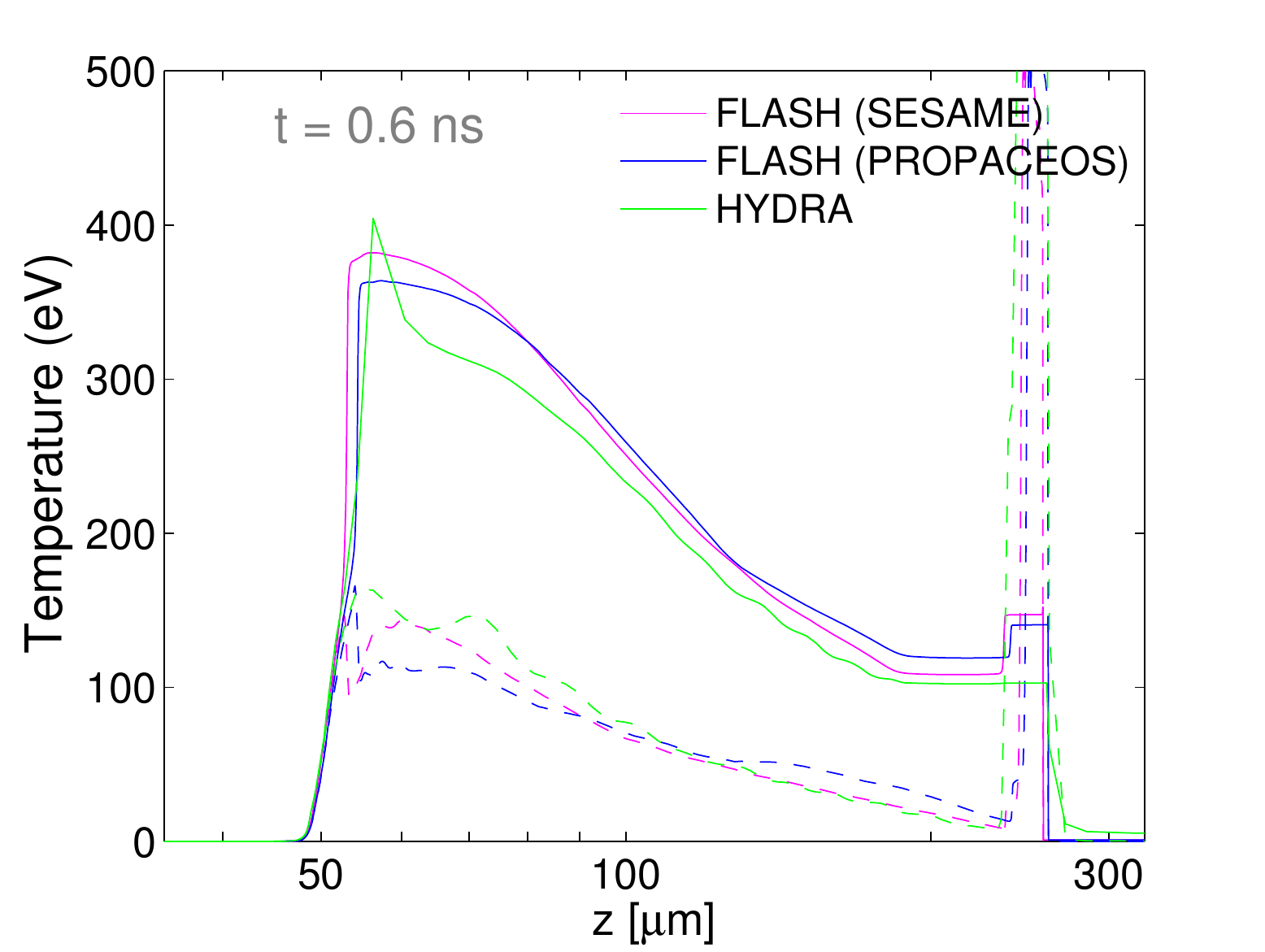}\includegraphics[angle=0,width=2.5in]{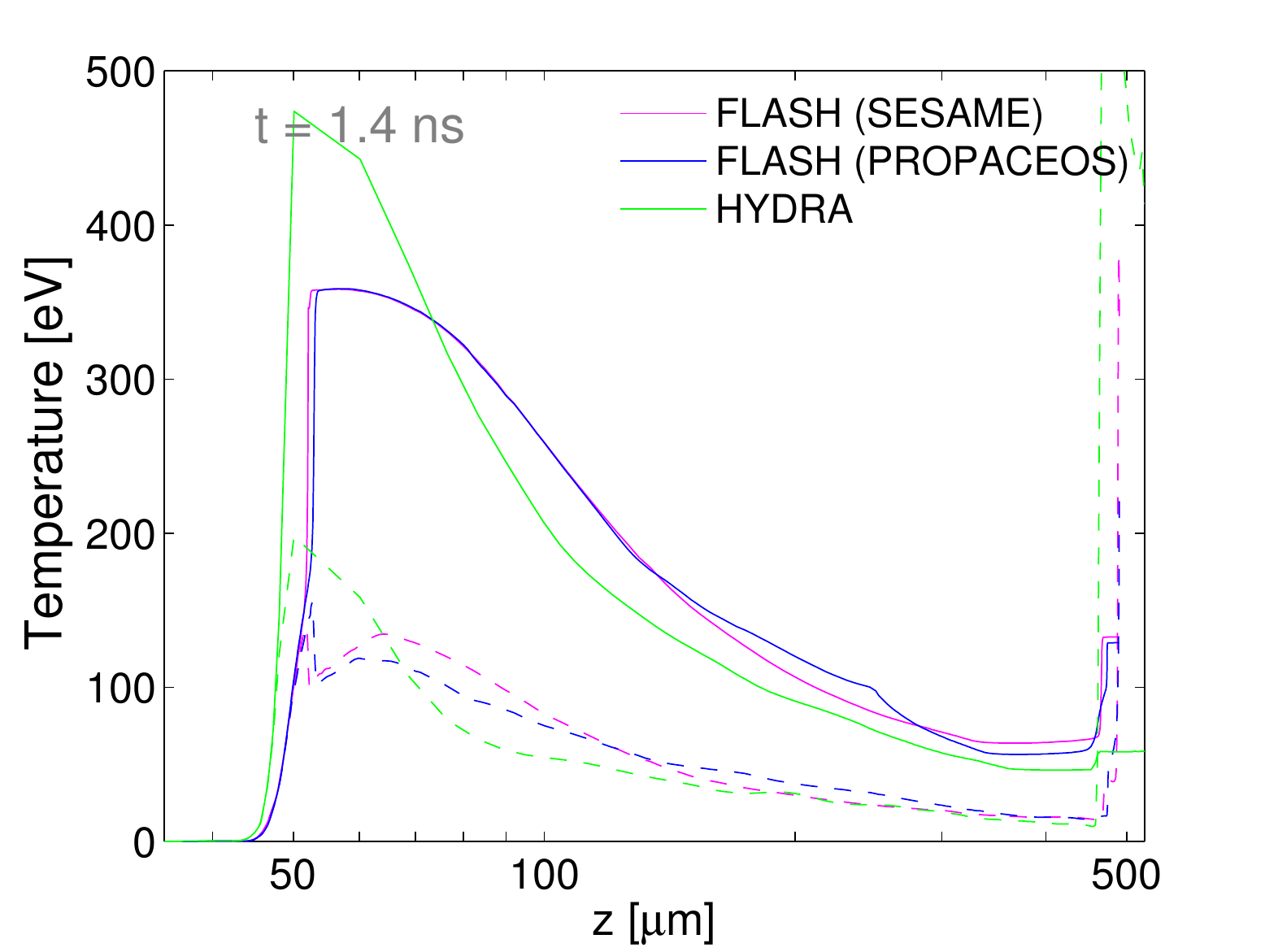}
\includegraphics[angle=0,width=2.5in]{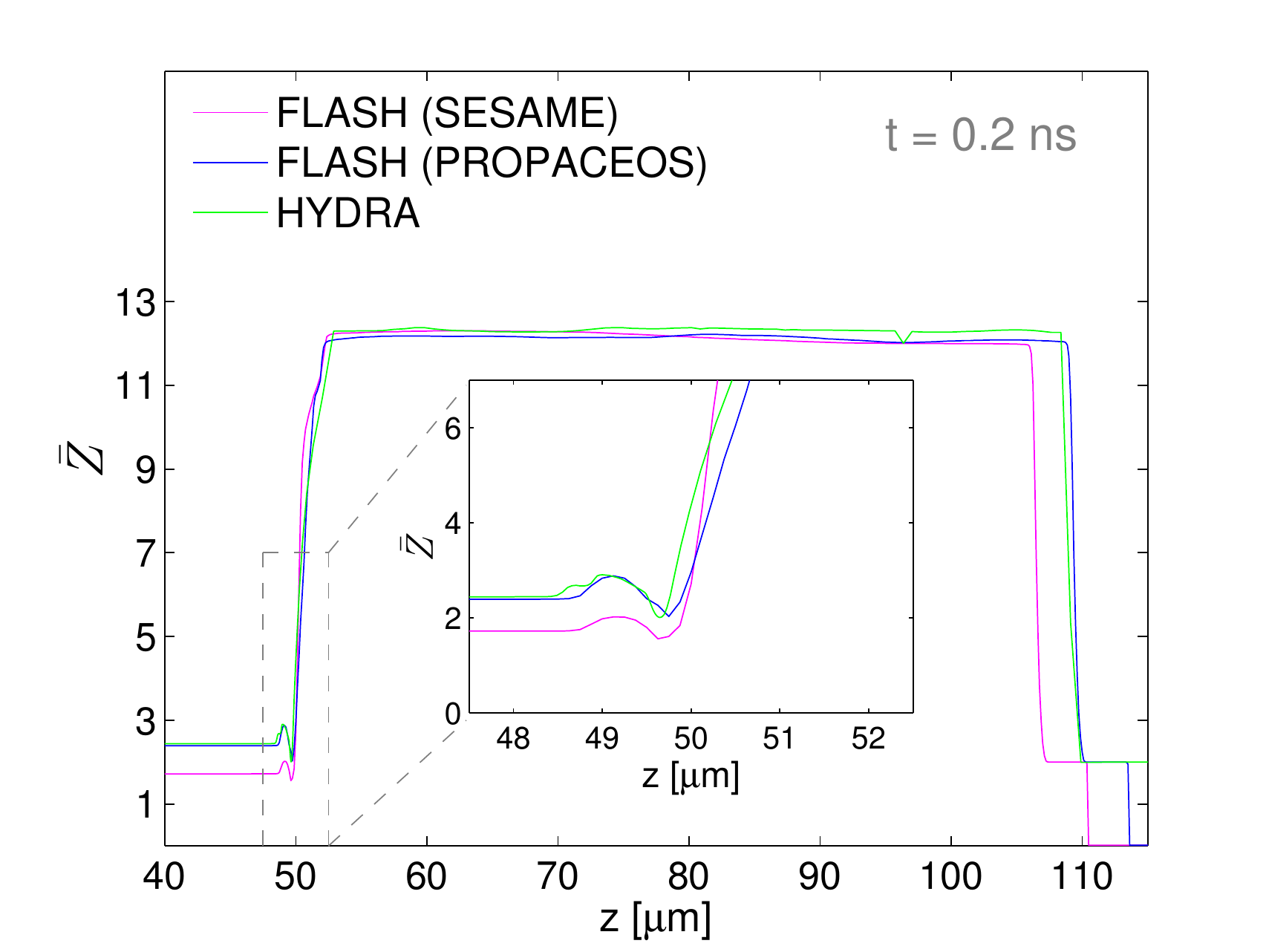}\includegraphics[angle=0,width=2.5in]{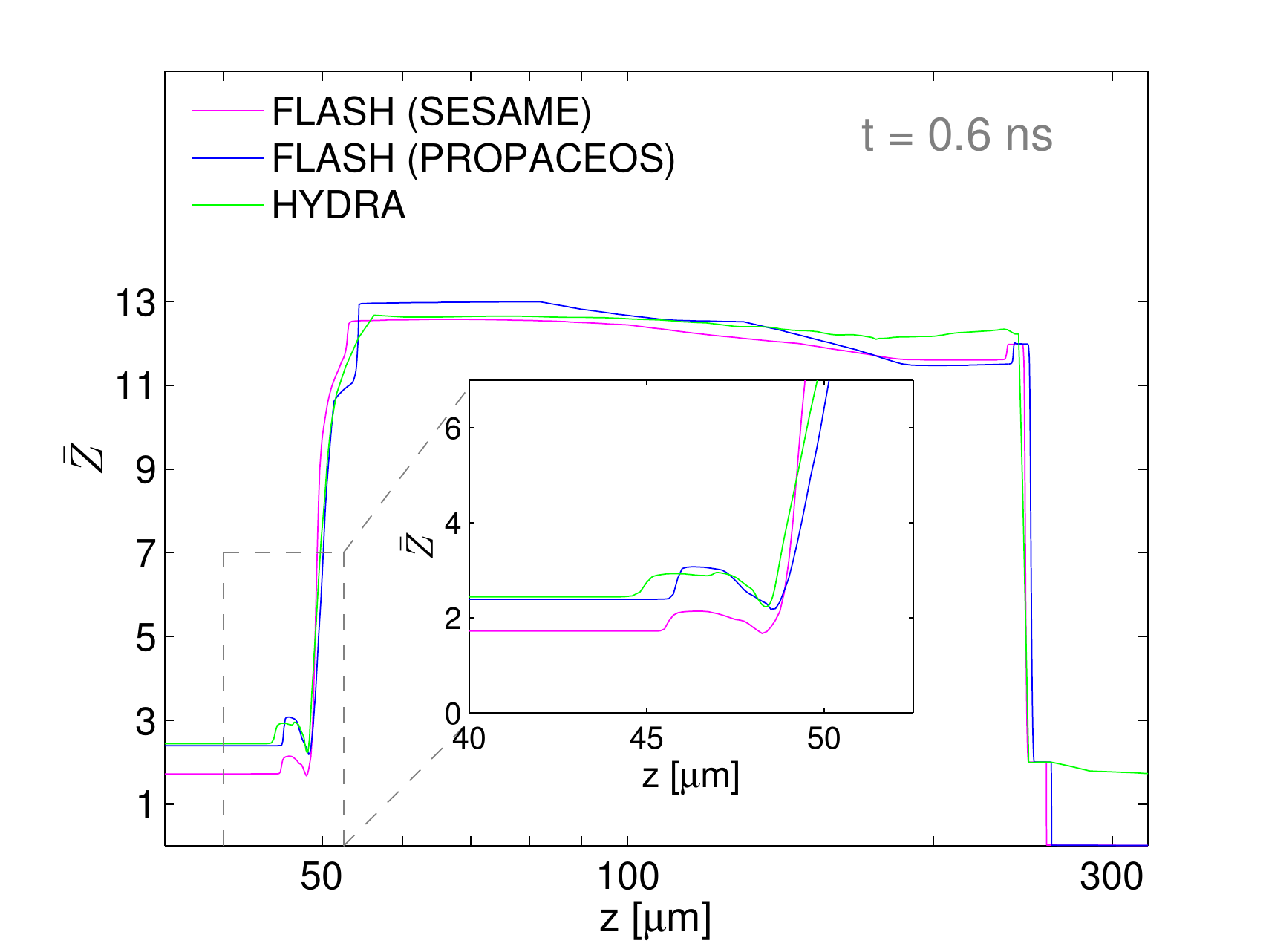}\includegraphics[angle=0,width=2.5in]{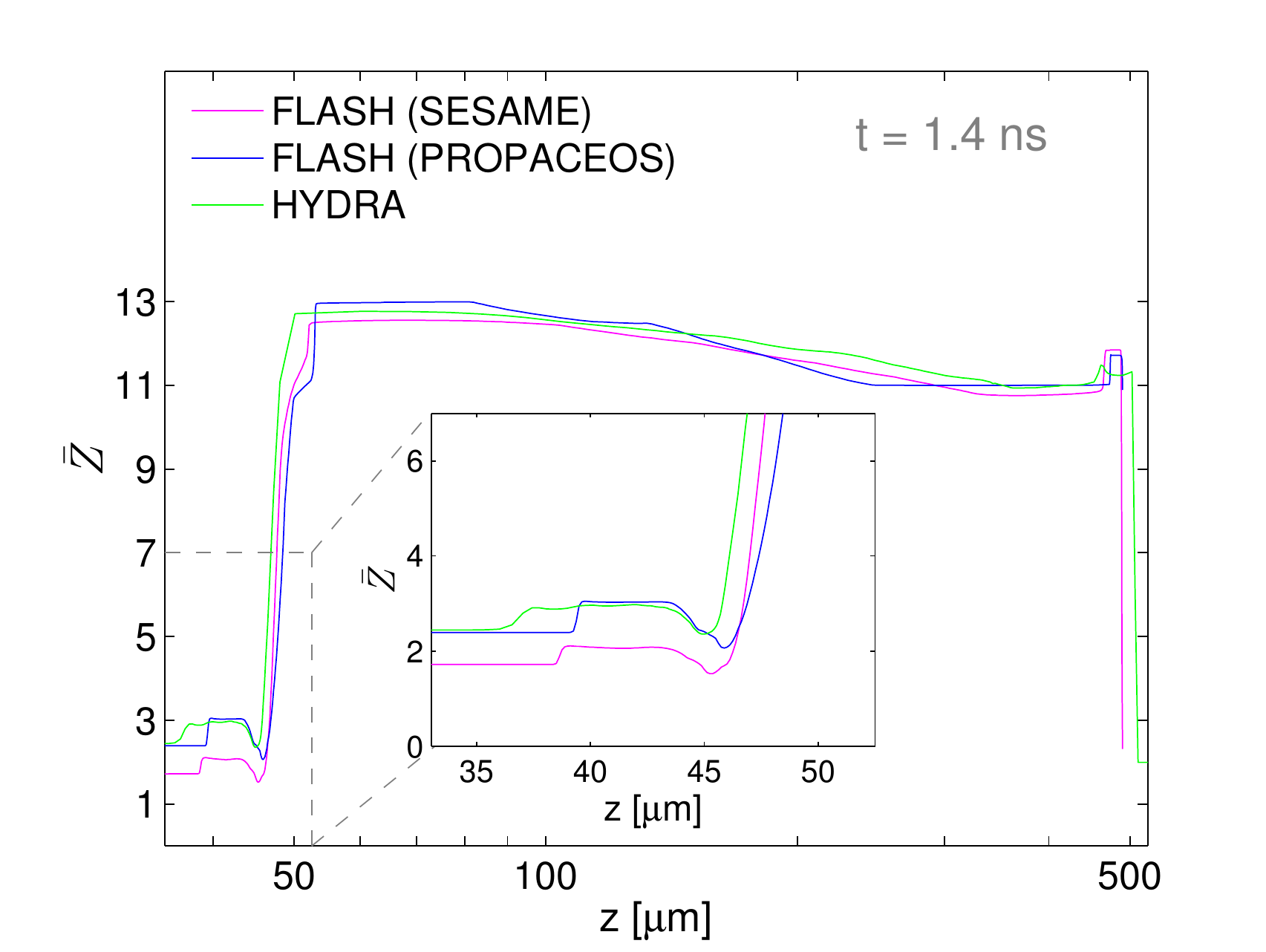}
\includegraphics[angle=0,width=2.5in]{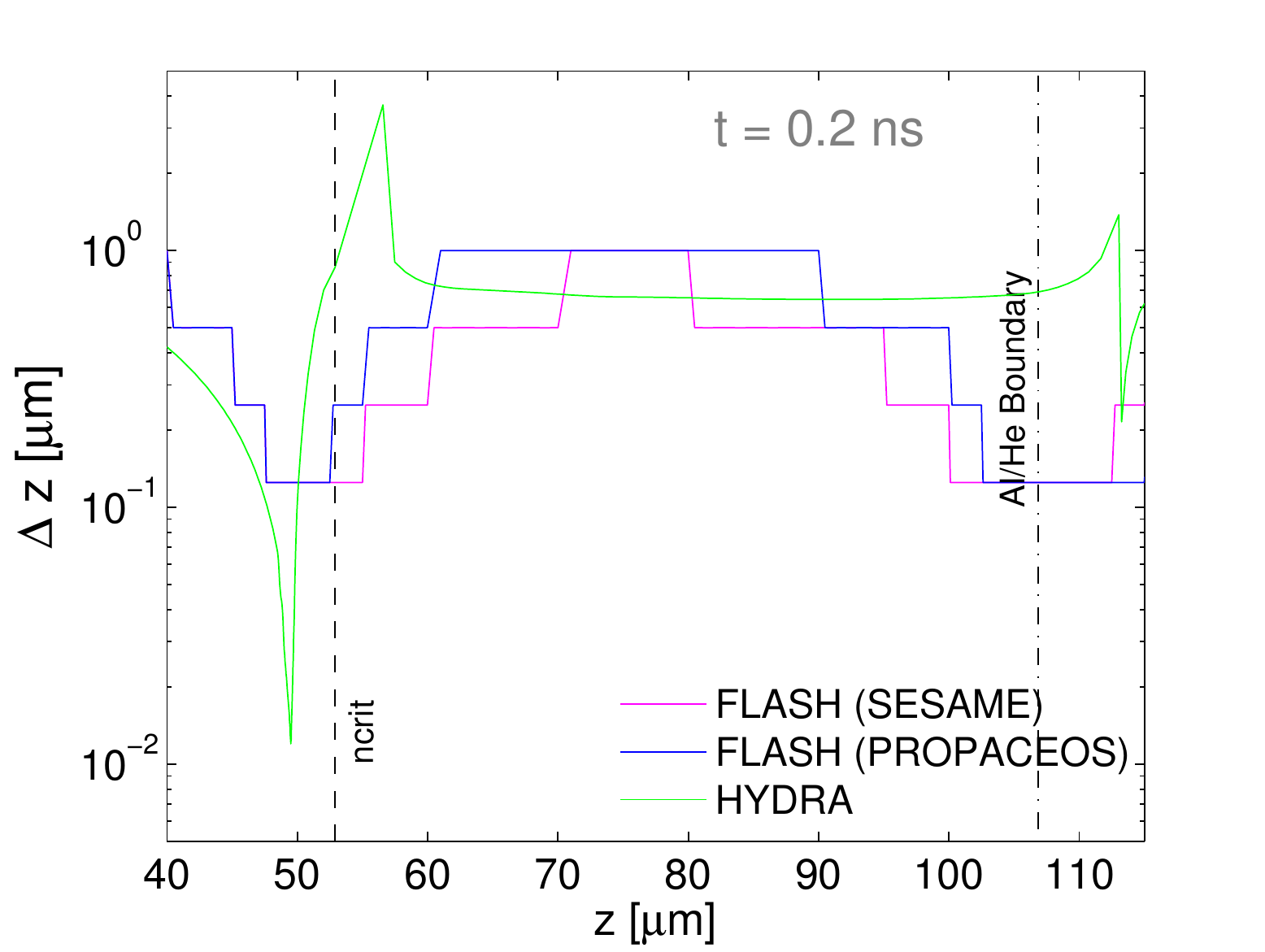}\includegraphics[angle=0,width=2.5in]{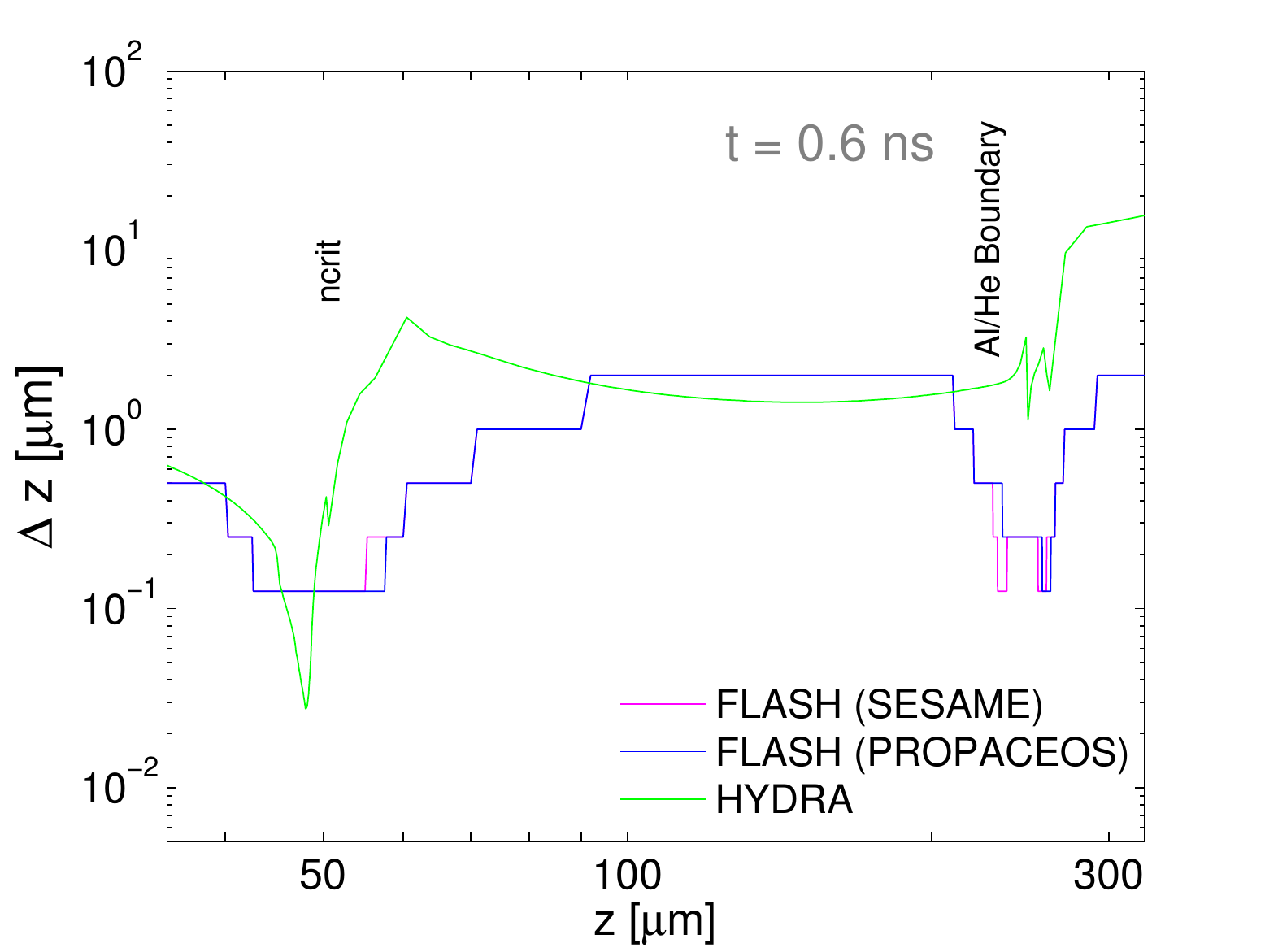}\includegraphics[angle=0,width=2.5in]{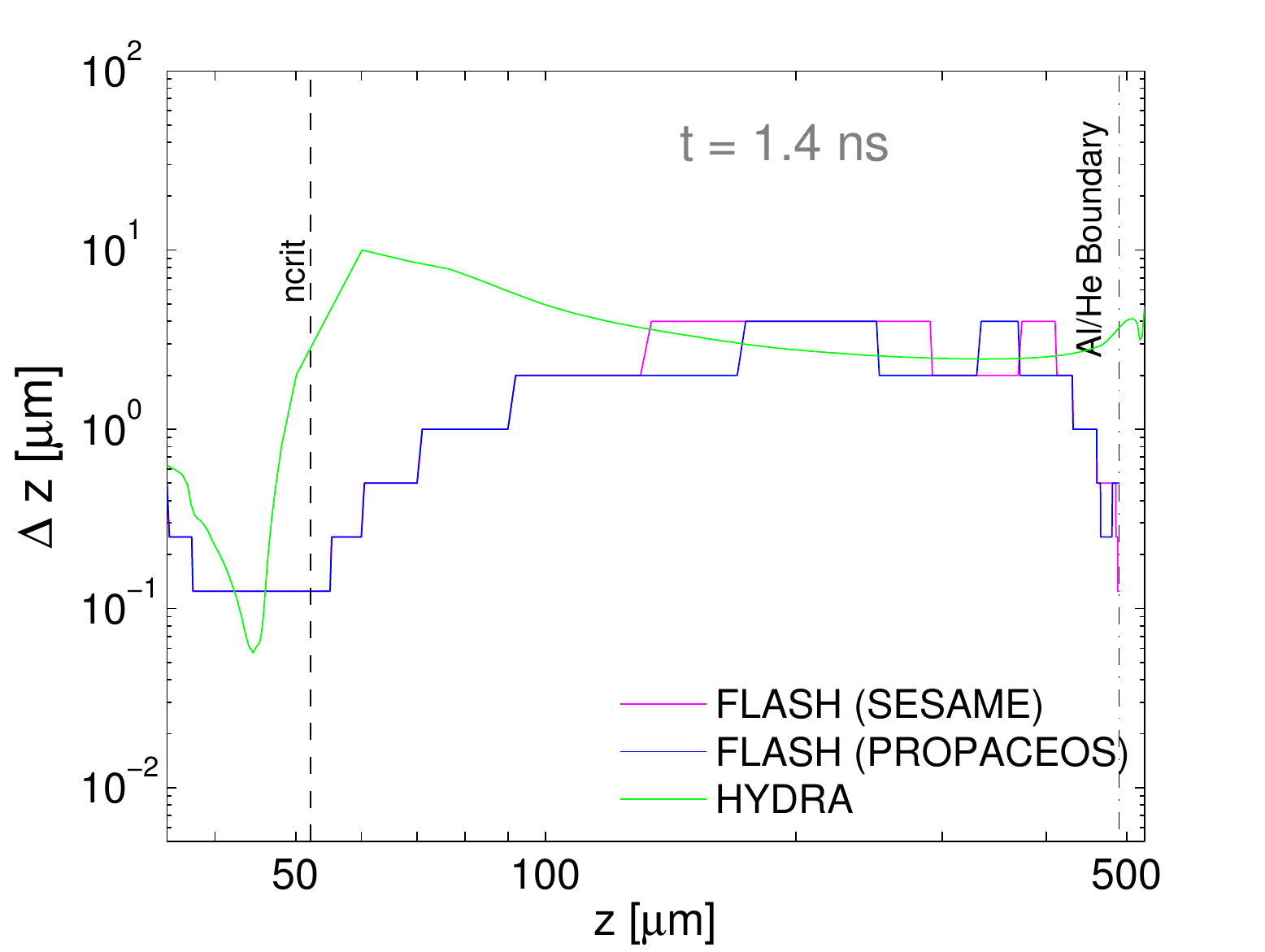}
\vspace{-0.4cm}
\caption{  
Results from simulations of 76.7 mJ pre-pulse irradiation of an Aluminum slab target. Panels are as described in Fig.~\ref{fig:20mJresults}.
}\label{fig:100mJresults}
\end{figure*}

\subsection{Results}

Fig.~\ref{fig:20mJresults} presents detailed profiles of the plasma properties through the laser axis at
$t = 0.2$~ns, 0.6~ns, and 1.4~ns for the 15.3~mJ pre-pulse simulations. Fig.~\ref{fig:100mJresults}
presents this same information for the 76.7~mJ pre-pulse. HYDRA results are shown with 
green lines while FLASH results are shown for with two different EOS models: 
magenta for SESAME and blue for PROPACEOS. All FLASH simulations use 
opacity data from PROPACEOS.

Surveying the top three panels in both figures, it is clear that the plasma expansion 
proceeds quite similarly in both codes and mostly independent of the choice of EOS 
in FLASH. For the purpose of using these pre-pulse simulations
as initial conditions for PIC simulations of short-pulse 
laser interactions, the plasma expansion with time and change
in density profile are the single most important aspects of 
 the resulting pre-plasma properties. While 
Figs.~\ref{fig:20mJresults} \& \ref{fig:100mJresults}
also show the electron and ion temperatures and mean ionization states,
the ion motion due to the pre-pulse alone will be negligible 
over the duration of a short-pulse ultra-intense laser-matter interaction
and hence the results of the PIC simulation should be insensitive
to the initial ion temperature. Likewise, the 
field and electron-impact ionization from the ultra-intense pulse
are likely to be strong enough that the sensitivity, at fixed ion density,
to the mean ionization state or the initial electron temperature 
will also be low. This being the case, the second-highest rows, which 
compare electron (solid lines) and ion
(dashed lines) temperatures show a great deal of similarity between
the FLASH and HYDRA results. Well away from the Al/He interface at low 
densities, FLASH gives slightly hotter results. This may stem 
from increased absorption or slightly lower heat capacity. The electron
temperatures in HYDRA deviate more from the FLASH results 
in the 76.7~mJ test. However the HYDRA resolution at the critical 
surface coarsens to $\Delta z \sim 10 \mu$m by the end of this simulation, 
and so the spike in the electron temperature at the $t = 1.4$~ns output
is unphysical.

At very low plasma densities and around the Al/He interface, the
ion temperatures skyrocket as the Al blowoff plasma shock heats the 
cold, low density helium initially at rest outside the Al target.
The relatively long timescale of electron-ion thermal 
equilibrium and the comparatively rapid timescale of electron heat conduction 
prevent the electrons from becoming super-heated as well, as discussed in 
\cite{MihalasMihalas1984}, and originally investigated in \cite{Shafranov1957}. 
This essential physics is in both codes and although there may
be differences in how this shock is handled, we focus our attention on the results 
at higher densities since He is merely a stand-in for vacuum 
conditions.\footnote{n.b. FLASH compares well and is tested daily against an 
exact solution for a 1D non-equilibrium shock \cite{Fatenejad_etal2011}.}

Regarding the mean ionization state results in the second from the bottom columns
in Figs.~\ref{fig:20mJresults} \& \ref{fig:100mJresults},
 the $\bar{Z}$ values are typically consistent to within
$\pm 0.5$, except at the Al/He interface. In the under-dense corona,
away from the target, the PROPACEOS
and SESAME models do about equally well in matching the $\bar{Z}$ from
HYDRA, while the near-or-slightly-above solid density $\bar{Z}$ results
highlighted by the inset figures strongly favor the PROPACEOS model
as the most HYDRA-like way of running FLASH (at least for Aluminum 
plasmas). 

\begin{figure}
\includegraphics[angle=0,width=3.0in]{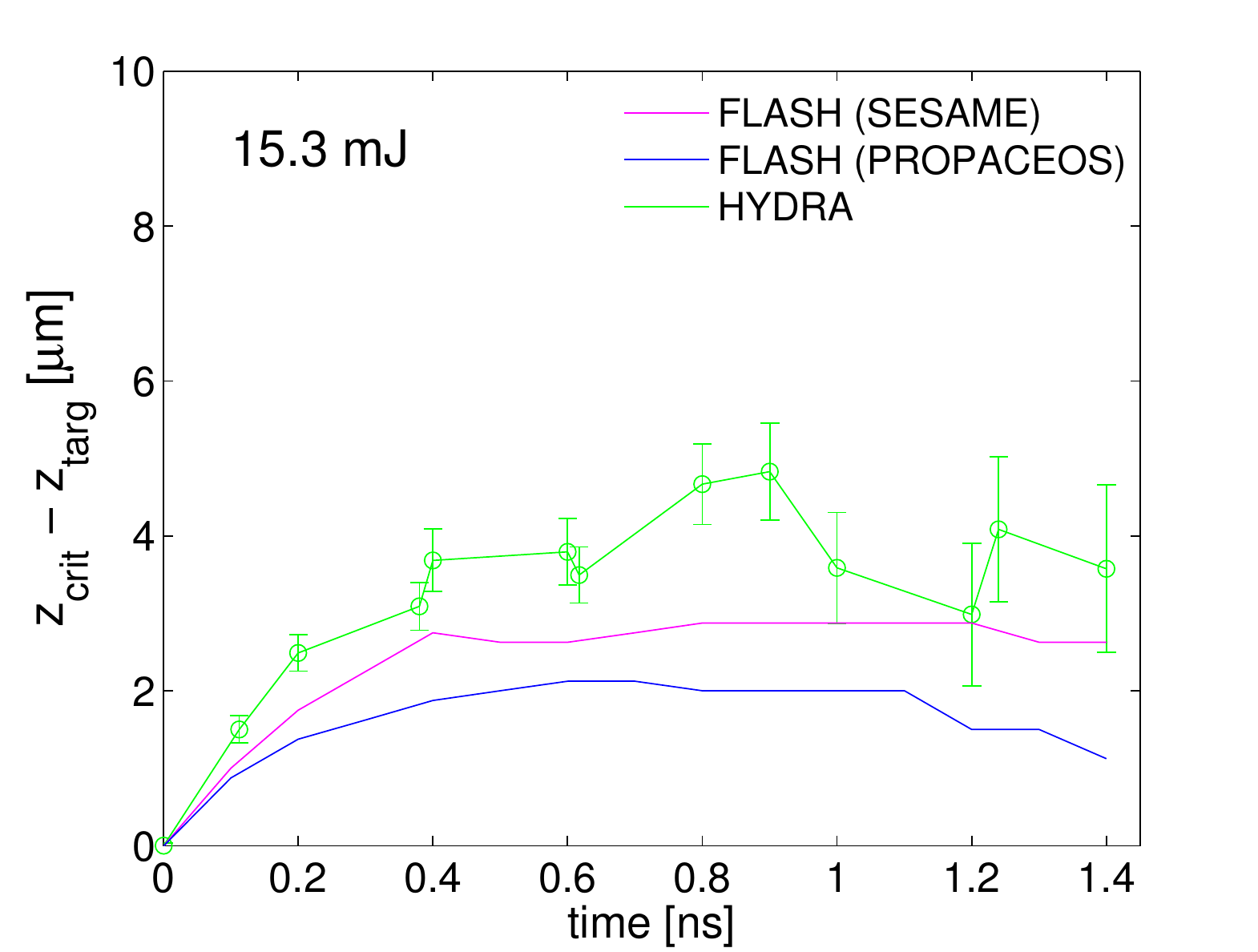}
\includegraphics[angle=0,width=3.0in]{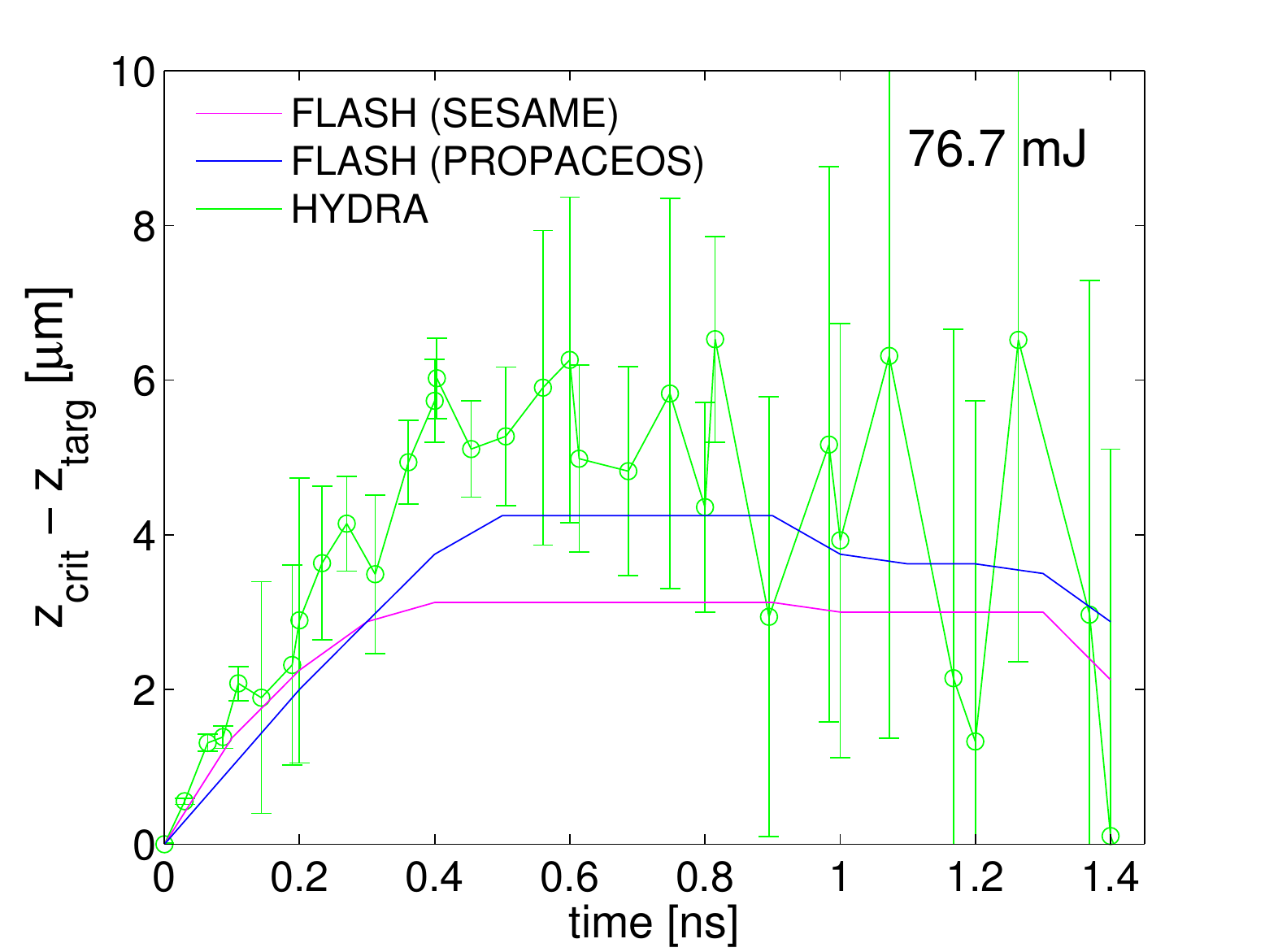}
\caption{ Position of critical density ($n_c = 1.1 \cdot 10^{21}$~cm$^{-3}$) along the laser axis versus time relative to its initial position at the target edge ($z_{\rm targ} = 50 \, \mu$m). Upper panel shows results for the 15.3~mJ pre-pulse while the lower panel shows results for the 76.7~mJ pre-pulse. Since the resolution at the critical density in the HYDRA results (green) is generally much coarser than in FLASH (magenta or blue, depending on the EOS model) we show the 
HYDRA results with error bars to indicate the cell size at critical density.
}\label{fig:ncrit}
\end{figure}

The bottom row of panels in Figs.~\ref{fig:20mJresults} \& \ref{fig:100mJresults}
compare the spatial resolutions in FLASH and HYDRA. The most striking 
result is that the resolution at the critical surface in HYDRA 
quickly becomes an order of magnitude coarser than in FLASH. The HYDRA
error bars in Fig.~\ref{fig:ncrit}, where we highlight the motion of 
the critical density along the laser axis, reflect this result with
the large size of the 76.7~mJ error bars especially apparent. 
As discussed earlier, the cell stretching in HYDRA is closely coupled to the divergence
of the fluid velocity whereas in FLASH the resolution is decided by other
means. Regardless, the motion of the critical density in FLASH and
HYDRA is consistent to within $\lesssim 2 \mu$m with neither the
use of PROPACEOS nor SESAME models for the EOS standing out as 
agreeing better with the HYDRA result.

Overall, FLASH and HYDRA agree reasonably well for densities significantly
away from where the Al/He interface occurs and in moderate-to-high
resolution regions. Despite important differences between the AMR
and ALE schemes for determining the computational mesh
and other details discussed in Sec.~\ref{sec:diff}, most of the remaining differences
seem attributable to the different EOS models employed
in the simulations.
A possible exception to this are the near-solid-density regions
shown by the inset figures in the upper panels of 
Figs.~\ref{fig:20mJresults} \& \ref{fig:100mJresults}. In both the 
15.3~mJ and 76.7~mJ cases the above-solid-density feature in 
HYDRA is lower in density and further into the target at all outputs.
This discrepancy is still to be adequately explained and we discuss
it further in Sec.~\ref{sec:disc}.








\section{Code-to-Code Comparisons and Validation for an Aluminum Slab Target With a V-Shaped Groove Irradiated by a Rectangular Laser Beam}
\label{sec:grava}

This section describes code-to-code comparisons between FLASH and HYDRA for an experiment that was carried out at Colorado State University and modeled using HYDRA in which an Al target with a V-shaped groove is irradiated by a rectangular laser beam \cite{Grava_etal2008}. 
The data from the experiment was used as a powerful validation test for HYDRA in \cite{Grava_etal2008}. In this section the data will be used as a validation test for FLASH for the first time.

The Grava et al. \cite{Grava_etal2008} study is unique in both the quality of the experimental data
collected and in the sophistication of the radiative-hydrodynamic modeling with HYDRA, 
which is described in some detail. In the experiment, the V-shaped groove target is irradiated
by a rectangular 360 $\mu$m FWHM laser pulse with peak intensity $\sim 10^{12}$ W/cm$^2$.  The temporal 
behavior of this pulse, approximately 120~ps FWHM, is shown in Fig~10 of \cite{Grava_etal2008}.

The geometry of the groove-shaped target, which is reminiscent of a well-known validation 
experiment on the NOVA laser \cite{Wan_etal1997,Stone_etal2000}, allows interferometric 
measurements of the electron density in the blowoff plasma
with a few-ns cadence. Importantly Grava et al. \cite{Grava_etal2008}  use soft x-ray 
(46.9 nm wavelength) interferometry to conduct this measurement,
which implies a critical density of $5 \cdot 10^{23}$~cm$^{-3}$. Taking into account 
instrumental resolution and other details puts the largest measurable electron density
at $5 \cdot 10^{20}$~cm$^{-3}$.

Grava et al. \cite{Grava_etal2008} pursued the experiment as a scaled version of astrophysically occurring 
radiative shocks, explaining that the radiative energy loss timescale in the problem, $\tau_{\rm rad}$, is 
comparable to hydrodynamic expansion timescale, $\tau_{\rm hydro}$. It was also significant that earlier NOVA experiments 
along these lines led to some puzzling results, raising the question whether collisionless Particle-In-Cell (PIC) 
codes might actually be more appropriate
to the experimental regime than rad-hydro codes \cite{Wan_etal1997}. Grava et al. and later work
by that collaboration \cite{Filevich_etal2009,Purvis_etal2010} showed that rad-hydro codes can indeed be trusted for
these experiments, and for a variety of different elements.


Grava et al. focused on an experiment with Aluminum, the same target material as
in the pre-pulse investigation in \S~\ref{sec:prepulse}. Restricting our investigation to 
Aluminum greatly simplifies the task of doing code-to-code comparisons and validating the HEDP extensions of FLASH. 
We hope to investigate other validation experiments with other target materials in the 
near future \cite{Filevich_etal2009,Purvis_etal2010}.

\subsection{\, \, \,  Non-Radiative Results: \newline Electron Number Density}

\begin{figure}
\centerline{\includegraphics[angle=0,width=3.9in]{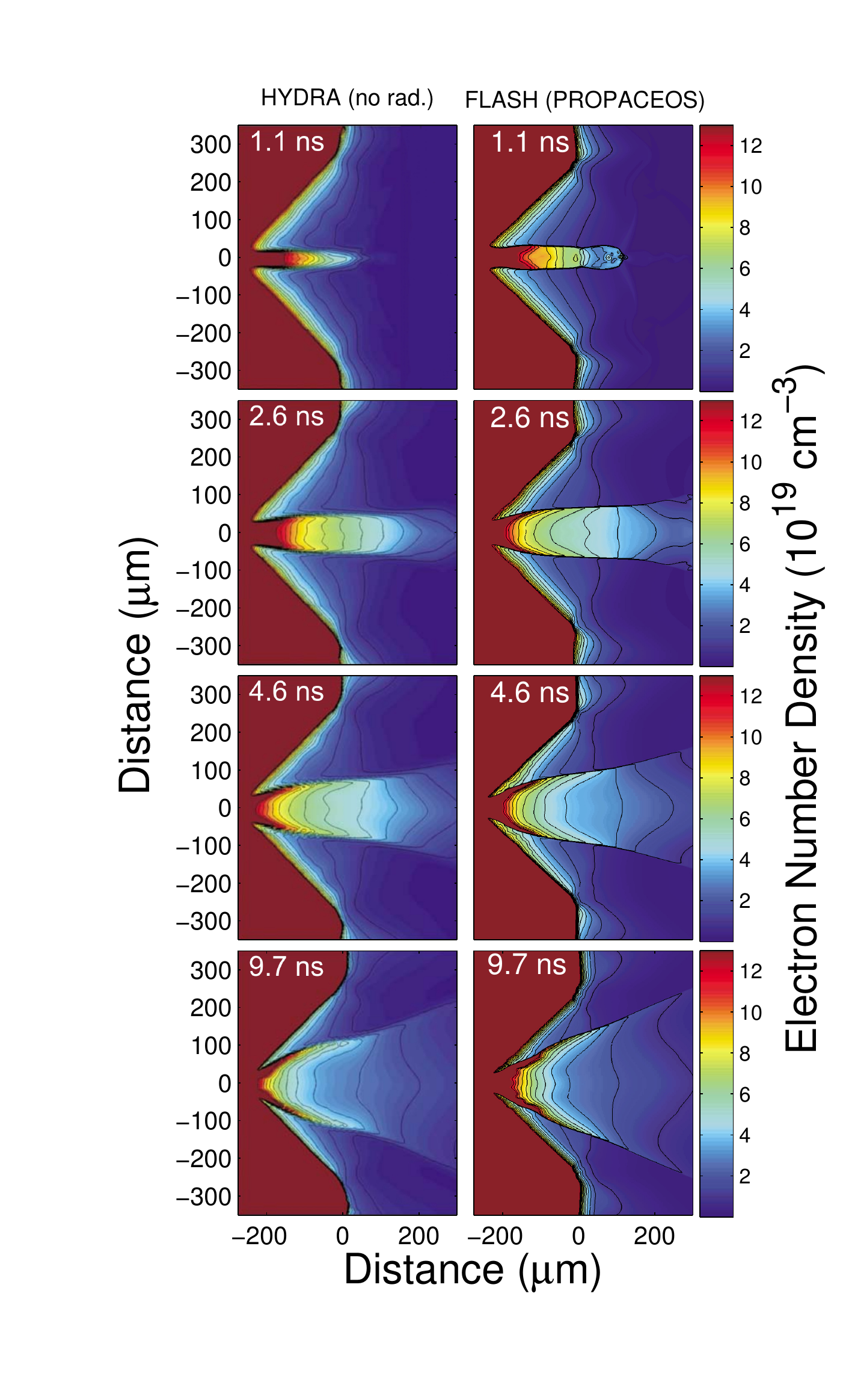}}
\vspace{-1.4cm}
\caption{Electron number densities at various times from a HYDRA simulation of the irradiation of a 
V-shaped groove {\it without} including radiation diffusion (left panels) and results from the 
same setup with FLASH using the PROPACEOS equation of state, also without radiation transport (right panels).
HYDRA panels are adapted with permission from Fig.~9 in Grava et al. \cite{Grava_etal2008} (copyrighted by the American Physical Society).
}\label{fig:norad}
\end{figure}

Grava et al. present HYDRA results with and without multi-group radiation diffusion to demonstrate the effect of 
radiation on their simulation results. We use the non-radiative HYDRA simulations as a starting 
point for our code-to-code comparison since it removes any dependence on the opacity model. 
Fig.~\ref{fig:norad} shows our primary result for simulations without radiation transport. 
For brevity, we only show FLASH results in this section with the 
PROPACEOS EOS, which produced better agreement with HYDRA in the previous section and came closest to resembling 
the HYDRA results in Fig.~\ref{fig:norad}.
The simulation results in Fig.~\ref{fig:norad} show that the two codes do qualitatively agree. 
There are no important features in the HYDRA
results that do not appear in the FLASH results and many of the contours from the simulations bear a remarkable resemblance to
each other. The FLASH results in the blowoff plasma, for reasons connected to earlier discussions
regarding the stretching of ALE cells, are somewhat higher resolution than the HYDRA results and therefore
there may be slight differences between the results merely because FLASH is resolving the plasma 
in more detail. Convergence with resolution could be explored in more depth; however, the purpose of the comparison in 
Fig.~\ref{fig:norad} is
merely to show that the FLASH simulation is realistic enough to pursue results including radiative effects. 
It is clear from Fig.~\ref{fig:norad} that FLASH passes this test.

\subsection{\, Results including Radiation: \newline Electron Number Density}

\begin{figure}
\centerline{\includegraphics[angle=0,width=3.5in]{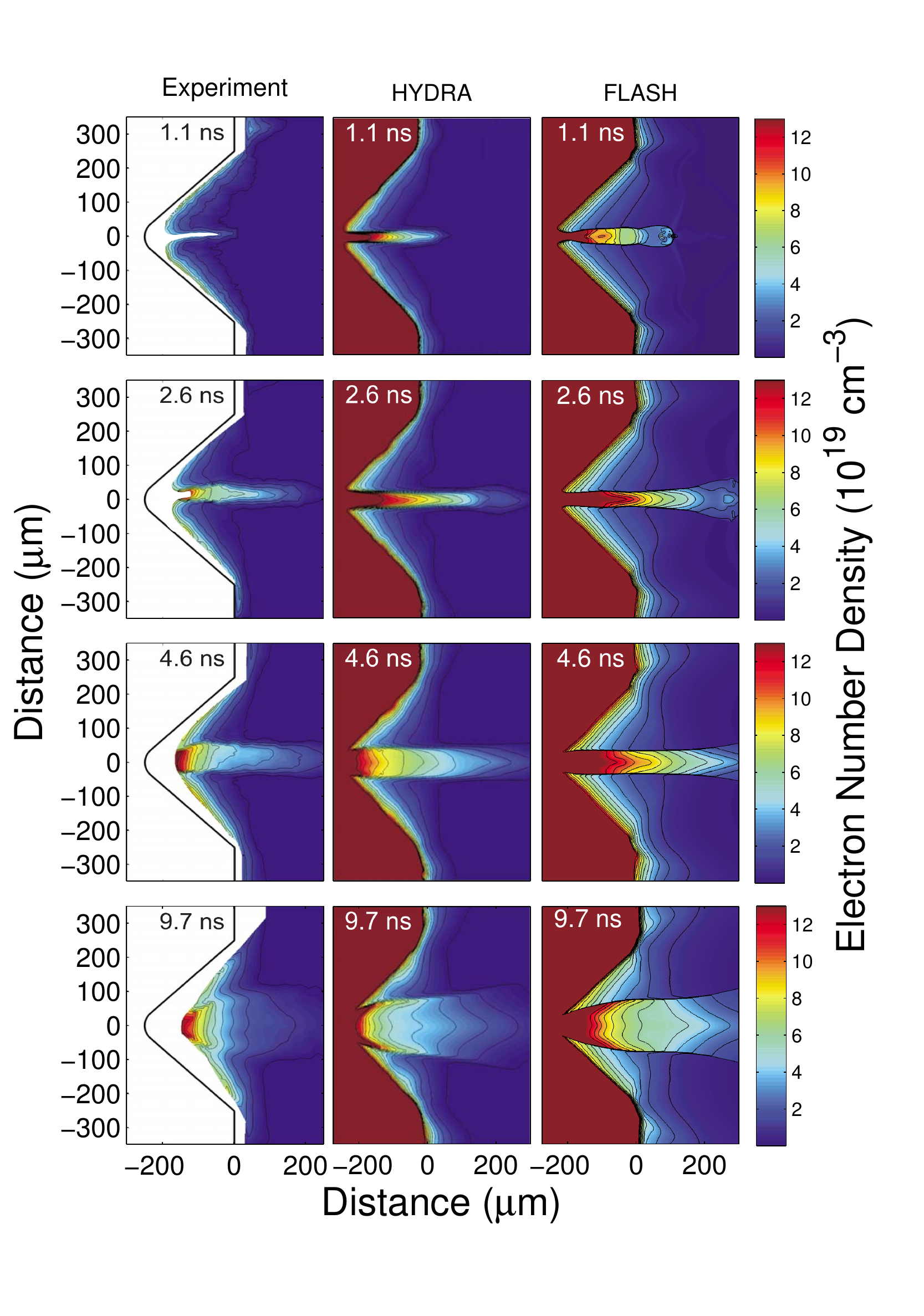}}
\vspace{-1.0cm}
\caption{
Comparing inferred electron number density from soft x-ray interferometry (left column) to HYDRA (center column) and FLASH (right column) simulations, both including multi-group radiation diffusion. The FLASH simulation uses PROPACEOS opacity 
and EOS data. Left and center columns are adapted with permission from Fig.~9 in Grava et al. \cite{Grava_etal2008} (copyrighted by the American Physical Society).
}\label{fig:rad}
\end{figure}

As discussed earlier, the simulations discussed here use a diffusion approximation to model the 
effect of radiation \cite{Castor2004}, which provides a channel for energy to escape from the hot 
plasma, thus lowering the temperature and consequently the pressure. As a result, plasmas with 
large radiative-energy fluxes can be compressed to higher densities than non-radiating plasmas.
This physics is evident in comparing the 
non-radiative results in Fig.~\ref{fig:norad} to the radiative results in Fig.~\ref{fig:rad}, which includes 
the experimental measurements of electron number density in the left hand column. The ablating plasma is colliding 
with itself at 1.1~ns (as in Fig.~\ref{fig:norad}), creating a relatively thin line of high density, high temperature 
Aluminum extending from the target. 
But instead of rebounding from the high pressure, thus creating the feature in Fig.~\ref{fig:norad} at 2.6~ns and 4.6~ns, 
the Aluminum stays compressed for longer so that at 4.6~ns in Fig.~\ref{fig:rad} an only slightly broadened version 
of this feature remains. Only between the 4.6~ns and 9.7~ns snapshots is 
there time enough for the pressure to broaden the feature beyond recognition. For a more extreme example
of the effect of radiation in this problem see the Cu or Mo results in \cite{Purvis_etal2010}.

Despite the possibility of important differences arising from either the implementation of multi-group
diffusion or in the opacity models used, the radiative results shown in Fig.~\ref{fig:rad} exhibit 
agreement between HYDRA and FLASH at essentially the same level as the non-radiative results shown
in Fig.~\ref{fig:norad}. Importantly, the agreement with the experimental data is qualitatively good as well.

\begin{figure}
\centerline{\includegraphics[angle=0,width=3.5in]{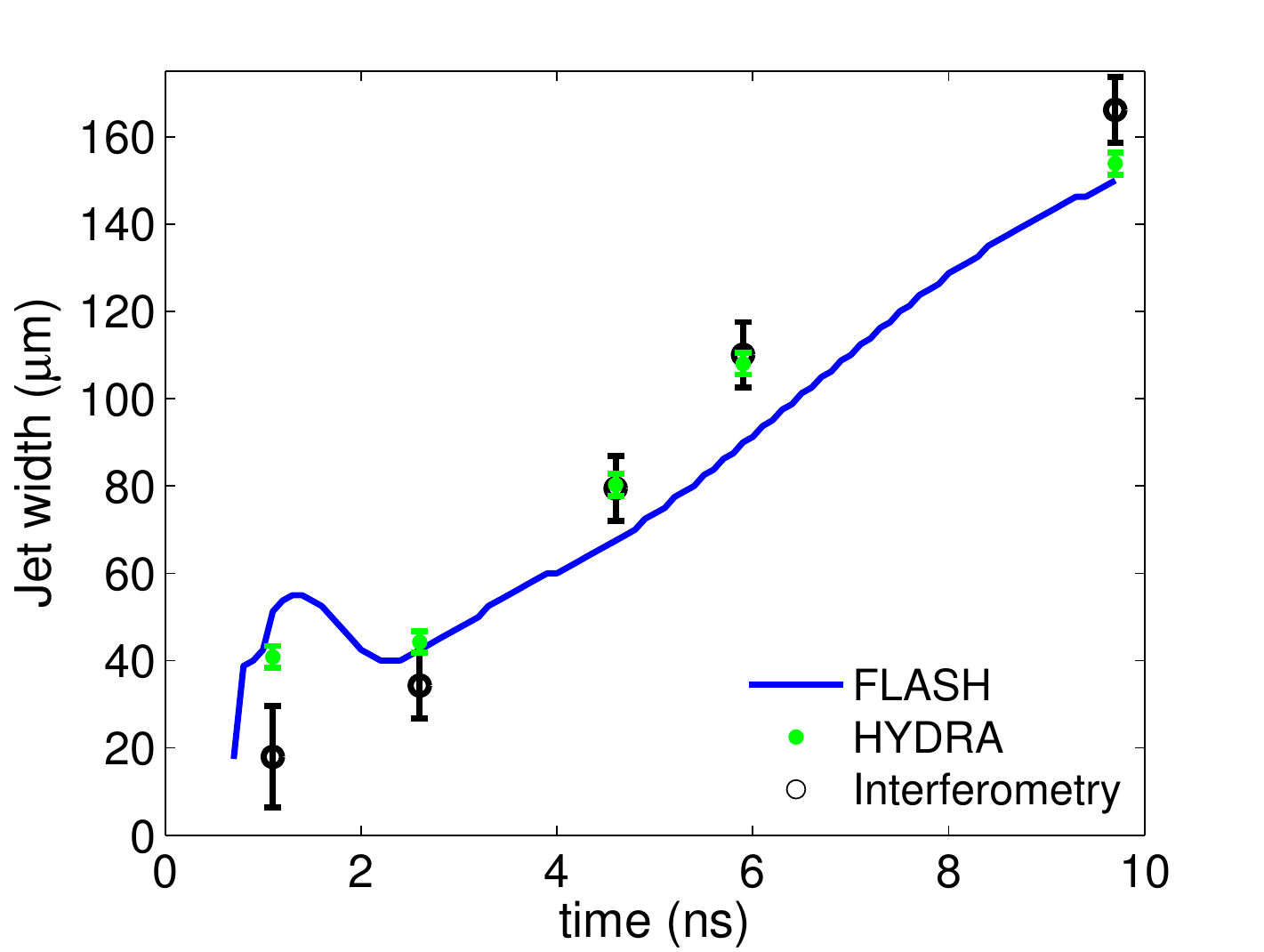}}
\vspace{-0.25cm}
\caption{A quantitative comparison of the jet width versus time as inferred from the electron number density (Fig.~\ref{fig:rad})
on a line perpendicular to the laser axis that is centered at (0,0). HYDRA results (green points) and interferometric measurements (black points)
from Grava et al. \cite{Grava_etal2008} are presented at 1.1, 2.6, 4.6, 5.9 and 9.7 ns. Measurements from the FLASH simulation (solid blue line) are more 
finely spaced in time. For brevity, the comparison at 5.9~ns is not shown in Fig.~\ref{fig:rad}.
}\label{fig:jetwidth}
\end{figure}

As a quantitative comparison, Fig.~\ref{fig:jetwidth} compares the width of the jet versus time.
The jet width is measured as the distance between the most steeply rising features 
in the electron density along a line perpendicular to the laser axis 
centered at (0,0). This definition is convenient for inferring the jet width from 
the figures in Grava et al. \cite{Grava_etal2008} since the point of steepest rise is 
simply where the electron number density contours are closest together.
The HYDRA results in Fig.~\ref{fig:jetwidth} are assigned an error bar
of $\pm$ 2.5~$\mu$m which is a typical distance between these two closest 
contours. Likewise, the jet width from the interferometric measurements is measured from the contours of electron number density 
in the same way. The error bars for the interferometric jet width
are estimated as roughly half the distance between the fringes ($\pm 7.5 \, \mu$m), except at 1.1~ns
when the jet is still forming. The true resolution of the interferometric data may be slightly smaller than
indicated in Fig.~\ref{fig:jetwidth}, although, in principle, a slight misalignment of the laser
on target would be an additional source of uncertainty \cite{Purvis_etal2010}. In the FLASH simulations the jet width can be inferred in a 
precise way according to the definition above and at a number of outputs.

Fig.~\ref{fig:jetwidth} shows that the HYDRA result is typically closer to 
the interferometry than the FLASH result which slightly underpredicts the jet width
at 4.6~ns and 5.9~ns. The reason for this is unclear. The lower resolution of the HYDRA 
simulation may have enlarged the jet to some degree. Interestingly, both FLASH and 
HYDRA overpredict the width of the jet at 1.1 ns.

\subsection{Results including Radiation: \newline Electron Temperature and Mean Ionization State}

\begin{figure}
\centerline{\includegraphics[angle=0,width=3.5in]{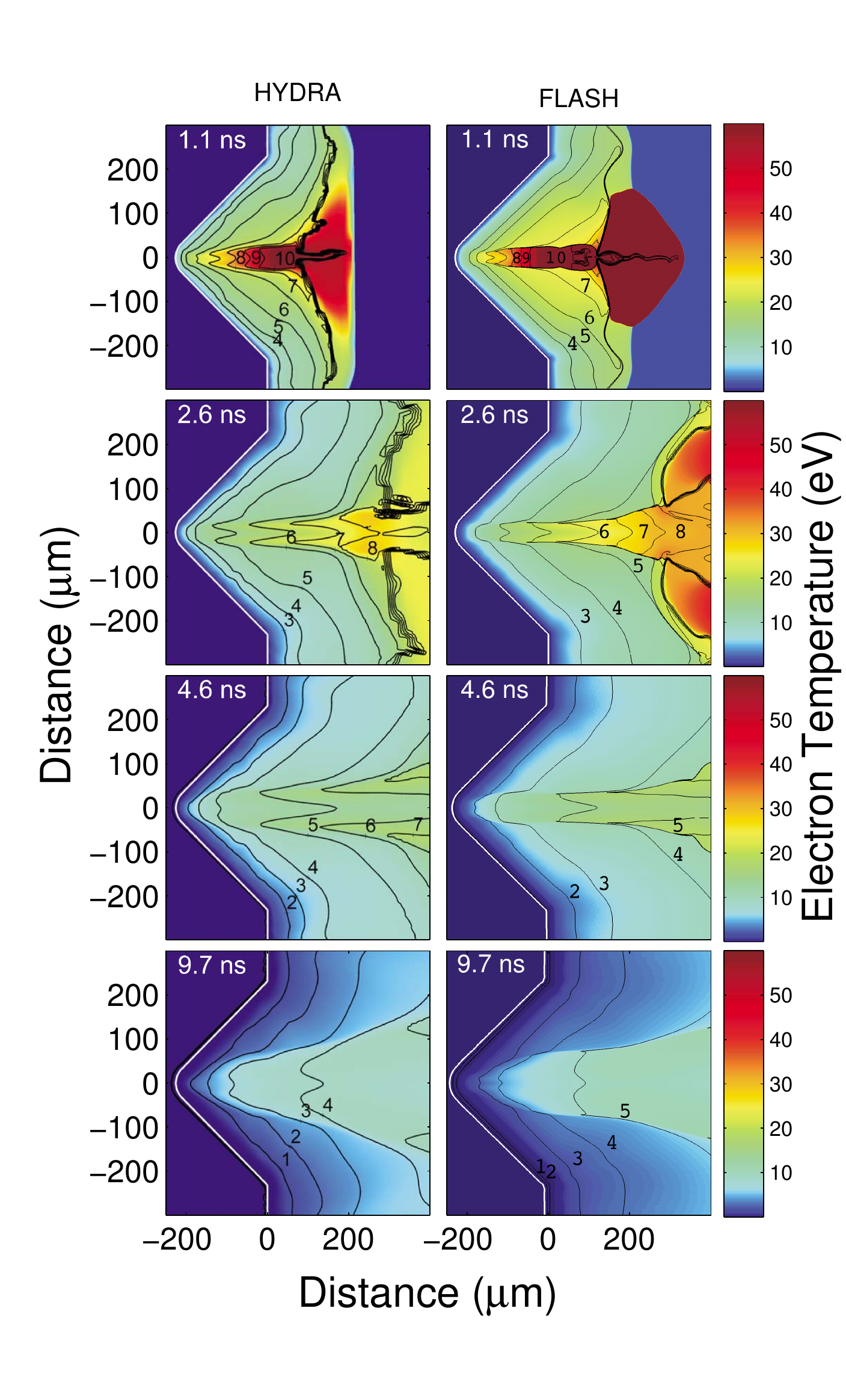}}
\vspace{-1.0cm}
\caption{
Comparing electron temperatures in radiation-hydrodynamics simulations from HYDRA (left column) to results from FLASH (right column); the FLASH simulation uses PROPACEOS opacity and EOS data. Contours for the mean ionization state, $\bar{Z}$, are overplotted. HYDRA panels are adapted with permission from Fig.~6 in Grava et al. \cite{Grava_etal2008} with permission (copyrighted by the American Physical Society).
}\label{fig:tele}
\end{figure}

Fig.~\ref{fig:tele} compares the electron temperatures in FLASH and HYDRA at the same times reported 
in Fig.~\ref{fig:rad}. In comparing these results, it is important to note that the interface between the 
Aluminum blowoff plasma and the 
vacuum density Helium is visible in the 1.1~ns and 2.6~ns results. The $\bar{Z}$ contours in Fig.~\ref{fig:tele} show where
this boundary occurs because the plasma goes from significantly ionized Al to a region where $\bar{Z}$ can at most be 
equal to two. As a result there is a kind of pileup of contours moving steadily away from the target; by 4.6~ns this 
transition is off the grid. Rightward of this interface at 1.1~ns, Fig.~\ref{fig:tele} indicates that the FLASH 
simulation has somewhat higher He temperatures than the HYDRA simulation. Without giving credence to the idea that 
this difference is especially meaningful, 
it could stem from a difference in handling the non-equilibrium (i.e. $T_{\rm ele} \neq T_{\rm ion}$) nature of the shock 
interface or, more prosaically, the He density in FLASH may simply be 
lower than what was assumed but not reported in Grava et al. \cite{Grava_etal2008}. 
The FLASH simulation assumes the same He density, $\rho = 5 \cdot 10^{-7}$ g/cm$^{3}$, as 
in the pre-pulse tests in Sec.~\ref{sec:prepulse}.

Leftward of this Al/He interface, in the Al blowoff plasma, the results
are again qualitatively similar with the FLASH results being slightly hotter
than HYDRA at 1.1~ns. Some combination of this slightly higher temperature
and differences in the tabulated data for the mean ionization state
 make the FLASH blowoff plasma slightly more ionized
than the HYDRA result. The $\bar{Z}$ contours in Fig.~\ref{fig:tele} are 
consistent with an overall $\Delta \bar{Z} \sim 1$ difference between the 
two simulation results. A closer look at the FLASH output for 1.1~ns reveals
that this is true for the highest ionization state as well, and we
find some regions where $\bar{Z} \sim 11$. Grava et al. state that the 
highest mean ionization state in their simulations is $\bar{Z} \sim 10$
and that this result is confirmed by the absence of signatures of more-highly-ionized charge states in extreme UV spectroscopy. Nevertheless,
both FLASH and HYDRA simulations agree that the mean ionization state never reaches
$\bar{Z} \sim 12$, which would require much higher temperatures.

\begin{figure}
\centerline{\includegraphics[angle=0,width=3.5in]{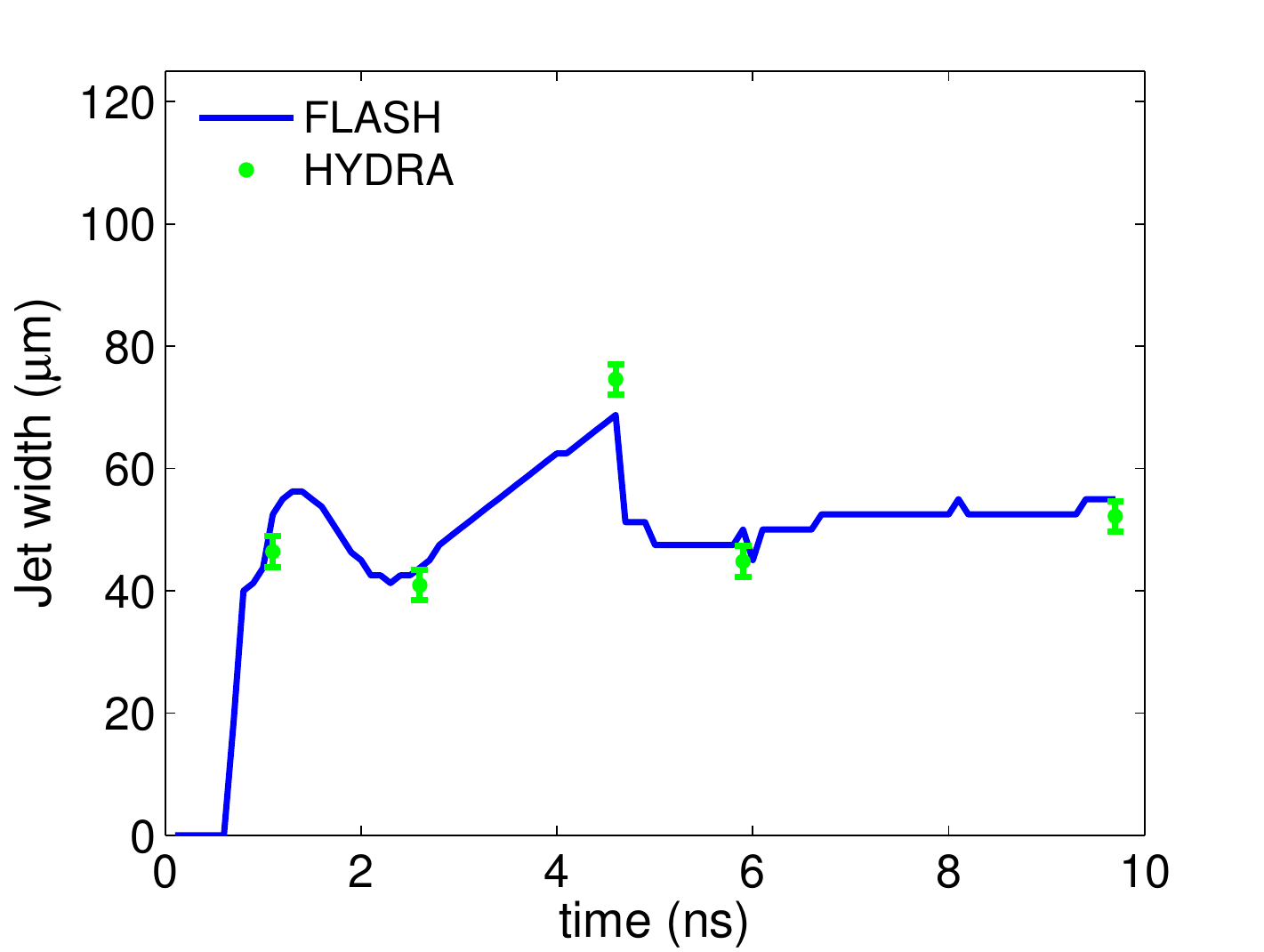}}
\vspace{-0.25cm}
\caption{A quantitative comparison of jet width versus time as measured from the distance between the peaks in 
the mean ionization state ($\bar{Z}$) on a line perpendicular to the laser axis that is centered on (0,0). 
The HYDRA measurements (green points) come from the $\bar{Z}$ contours in Fig.~\ref{fig:tele}, which are derived 
from Grava et al. \cite{Grava_etal2008}. Measurements from the FLASH simulation (solid blue line) are presented
at finely spaced intervals in time.}\label{fig:jetwidth_zbar}
\end{figure}

To quantify the level of agreement in Fig.~\ref{fig:tele}, the mean ionization state contours can be used 
as another measure of the width of the jet versus time. Material ablated from
the walls will expand and collide with the plasma on axis which, up to and before 5~ns, creates a 
unique feature where the temperature and mean ionization state peaks just above and below the laser axis
instead of on the laser axis itself. The two-finger-like appearance of the mean ionization state contours 
 in Fig.~\ref{fig:tele} arises from this interaction. By drawing a line between the ``fingers'' in 
the contours, the distance between these peaks in the mean ionization state on the line perpendicular to
the laser axis centered at (0,0) can be inferred for each output. Fig.~\ref{fig:jetwidth_zbar} compares
this measurement from the HYDRA simulations to a very precise measurement of distance between these
peaks in $\bar{Z}$ from the FLASH simulation for a number of outputs. At 4.6~ns and earlier this definition
of the jet width gives similar results as Fig.~\ref{fig:jetwidth}, however at around 5~ns and later
the $\bar{Z}$ feature is much less pronounced because a transition is taking place 
in which the plasma arriving at the center is no longer plasma that was directly heated by the
laser (c.f. Fig.~10 in Grava et al. \cite{Grava_etal2008}). The speed and momenta of the impact from 
plasma arriving from the walls at the center has decreases significantly and a new maxima 
in $\bar{Z}$ forms closer to the laser axis. The abrupt transition to smaller jet width in 
Fig.~\ref{fig:jetwidth_zbar} around 5~ns comes from transitioning to this new maxima.

Having explained the features and subtleties of Fig.~\ref{fig:jetwidth_zbar}, it is clear that
the two codes agree well on the jet width according to the peak in the mean ionization 
state. This is in spite of the overall $\Delta \bar{Z} \sim 1$ difference in the 
mean ionization state mentioned earlier. Since there is no experimental measurement of the mean ionization
state of the plasma, Fig.~\ref{fig:jetwidth_zbar} is strictly a comparison of the 
two codes.


\subsection{Results including Radiation: Total Pressure}

\begin{figure}
\centerline{\includegraphics[angle=0,width=3.5in]{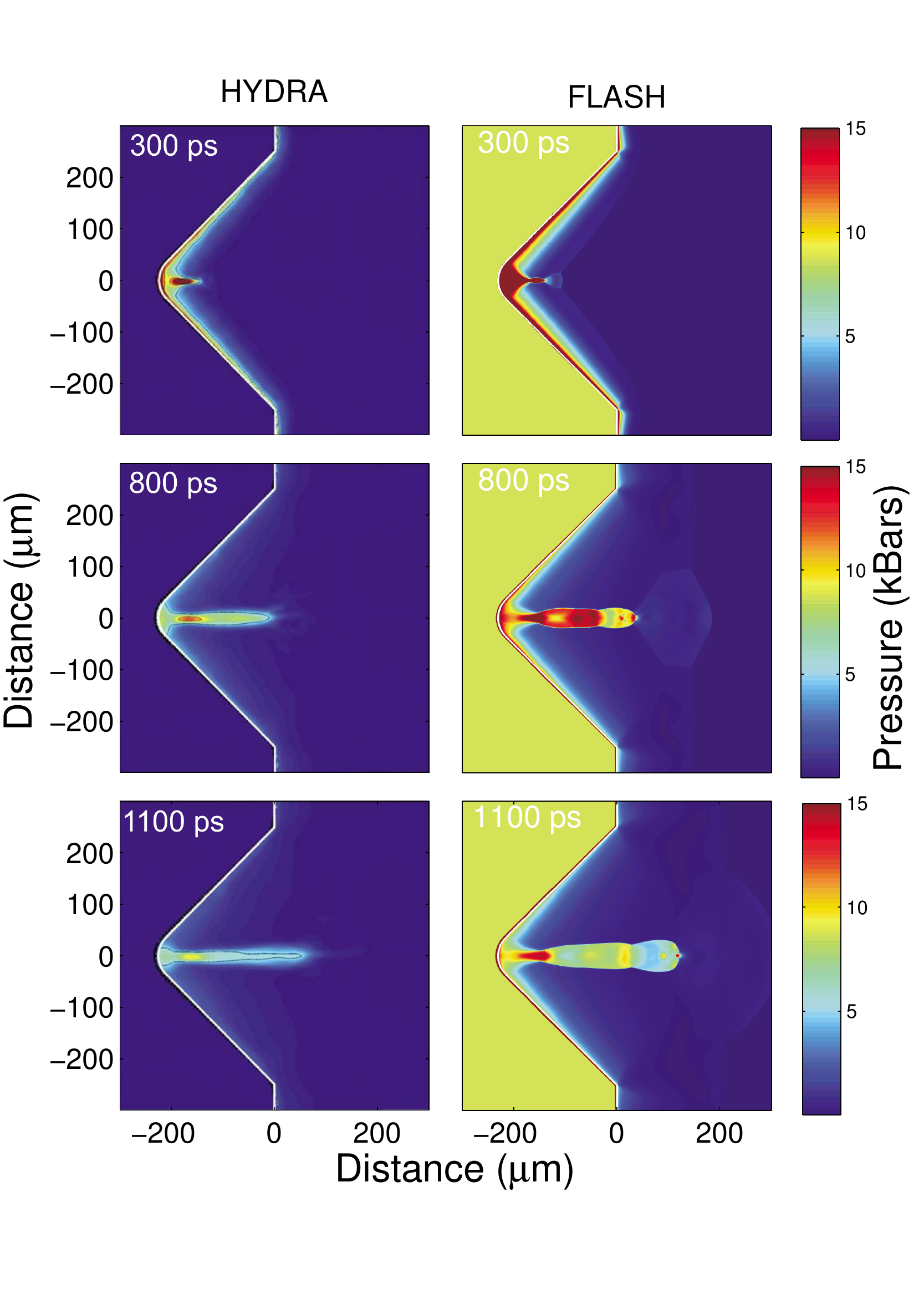}}
\vspace{-1cm}
\caption{
Comparing total plasma pressure at early times from radiation-hydrodynamics simulations
from HYDRA (left column) and FLASH (right column). The FLASH simulation was performed using PROPACEOS opacity and 
EOS data. HYDRA panels are adapted from Fig.~8 in Grava et al. \cite{Grava_etal2008} with permission
(copyrighted by the American Physical Society).
}\label{fig:pres}
\end{figure}

Fig.~\ref{fig:pres} presents a comparison of the total plasma pressure at early times in
the radiation-hydrodynamics simulations. While both FLASH and HYDRA use EOS data that
are the same as or related to the QEOS model constructed by More et al. \cite{QEOS}, the implementations
are clearly somewhat different. In HYDRA the total pressure at solid density
(i.e. well into the target) is vanishingly small, as one would expect for a cold solid.
FLASH uses the (QEOS-based) PROPACEOS equation of state which achieves vanishingly small 
total pressures for cold solids, in part, by allowing the electron pressures
to be negative. Currently, FLASH requires positive electron and ion pressures in
solving the momentum equation and in interpolating EOS data. So we set to a 
very small, positive value the few negative electron pressures that exist in the 
PROPACEOS table. The $\sim 7$ kBar (olive-colored) pressures well into the target in
the FLASH results thus reflect only the {\it ion} pressure reported from 
PROPACEOS\footnote{Both SESAME and BADGER EOS models gave even higher 
total pressures for cold solids, which is part of our reasoning for highlighting the PROPACEOS
results in this section.}.
Bearing in mind that, from a hydrodynamic point of view, only pressure 
{\it differences} matter, and that laser-heated Aluminum will quickly enter a regime
where the total pressure is naturally well above zero, it is understandable
why this detail seems not to have affected the agreement between the two codes,
given the level of agreement that we have seen in this section and in
the earlier pre-pulse tests.

Morphologically, the total pressures in the blowoff plasma are qualitatively similar
in the FLASH and HYDRA simulations. Since the resolution of the FLASH simulation
is very likely somewhat higher than the HYDRA simulation in this region it is reasonable that the FLASH
results show more detailed features. Overall, the total pressures in FLASH
are slightly higher than HYDRA, probably due to slightly higher temperatures, 
but, again, only the pressure differences matter to the hydrodynamics and 
therefore the plasma expansion is very similar.

\begin{figure*}
\centerline{\includegraphics[angle=0,width=5in]{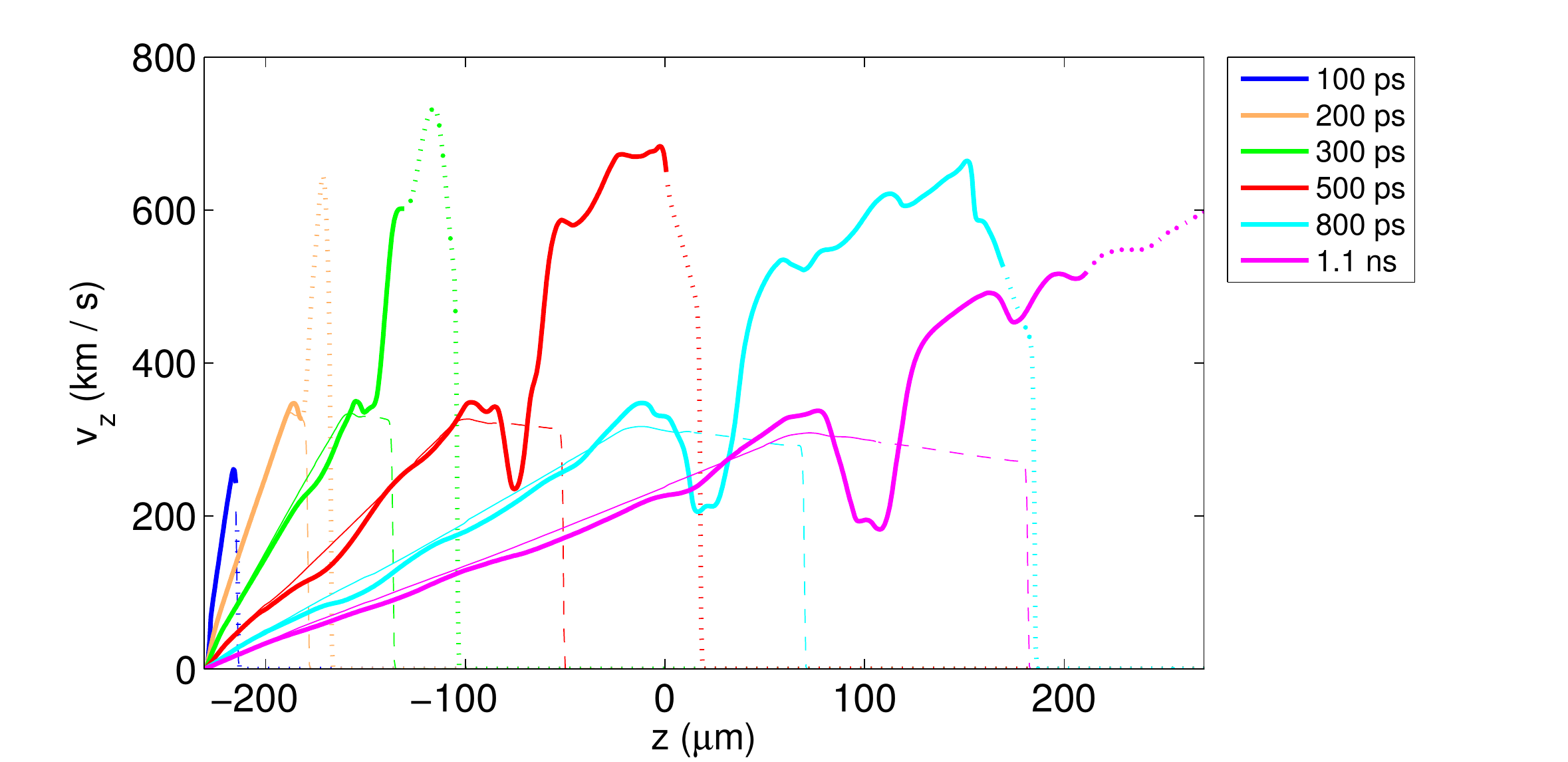}}
\vspace{-0.3cm}
\caption{
Lineouts along the laser axis of the component of the velocity parallel to the laser axis at various times. The thick solid lines show measurements from the FLASH simulation of a V-shaped groove target described in \S~\ref{sec:grava}. These thick solid lines become dotted lines at the point where the cell material is mostly very low density Helium instead of Aluminum. Thin solid lines (which become dashed lines at the Al/He transition) show the same measurements from a simulation where a flat target of the same material is irradiated with the Grava et al. laser pulse.
}\label{fig:vely1}
\end{figure*}

\begin{figure*}
\centerline{\includegraphics[angle=0,width=5in]{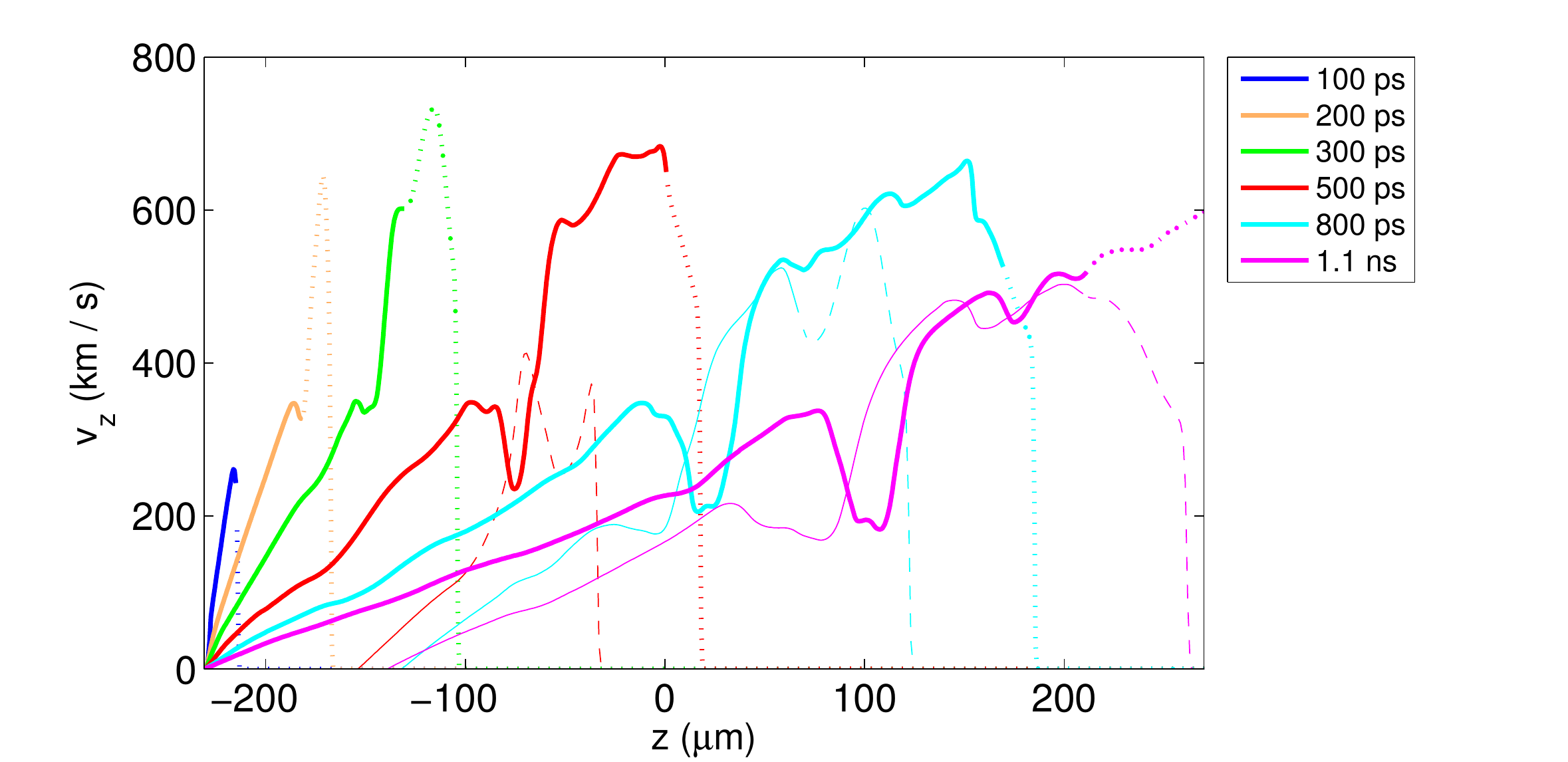}}
\vspace{-0.3cm}
\caption{
Lineouts along the laser axis of the component of the velocity parallel to the laser axis at various times. The thick solid lines show measurements from the FLASH simulation of a V-shaped groove target described in \S~\ref{sec:grava}. These thick solid lines become dotted lines at the Al/He transition. Thin solid lines (which likewise become dashed lines at the Al/He transition) show the same measurements from a simulation where, similar to the target geometry of Wan et al. \cite{Wan_etal1997}, a large gap of material is missing from the center of the V-shaped groove. This target is irradiated by the Grava et al. laser pulse. 
}\label{fig:vely2}
\end{figure*}

\section{Formation and Properties of the Jet in the Grava et al. Experiment}
\label{sec:jet}

The Grava et al. \cite{Grava_etal2008} experiment had two objectives:  (1) to obtain data that would make possible an important additional validation test of radiative hydrodynamics codes in the wake of the apparent failure of LASNEX \cite{LASNEX} simulations to reproduce a similar experiment done using the Nova laser \cite{Wan_etal1997}; and (2) to create a jet analogous to astrophysics jets, following Stone et al. \cite{Stone_etal2000}.  So far we have used the data from Grava et al. \cite{Grava_etal2008} to validate the FLASH code.  Here we consider the Grava et al. experiment with the second objective in mind.

Having validated FLASH for the Grava et al. experiment, we now use the FLASH simulations of the experiment to better understand the formation and properties of the jet seen in it.  We focus on early times ($\leq$~1.1~ns) and to inform our discussion we use comparisons between (1) the FLASH simulations of this experiment presented in \S~\ref{sec:grava},  (2) FLASH simulations of a flat Aluminum target irradiated by the same rectangular laser beam used in the Grava et al. experiment, and (3) FLASH simulations of the Grava et al. experiment but with the inner $\pm 75 \mu$m section of the target removed. This latter configuration resembles an earlier experiment done by Wan et al. \cite{Wan_etal1997} in which two slabs with a gap between them were oriented perpendicular to each other and irradiated by beams of the NOVA laser. Grava et al. \cite{Grava_etal2008} cite this experiment as an important motivation for their work.

A number of questions naturally arise in drawing parallels between the self-colliding blowoff plasma in Grava et al. and astrophysical jets.  While the radiative cooling timescale may be similar in magnitude to the hydrodynamic expansion timescale (and therefore the adiabatic cooling timescale) in both problems, how similar are they in other 
respects?  Here we address three specific questions about the formation and properties of the jets seen in the laser experiments whose answers enable us to compare them with astrophysical jets:

\begin{enumerate}

\item
What determines the physical conditions (e.g., the density, temperature, and velocity) in the core of the jet?

\item
Is the collimating effect of the the plasma ablating from the angular sides of the groove due to thermal pressure (i.e., the internal energy of the ablating plasma) or ram pressure (i.e., the component of the momentum of the ablating plasma perpendicular to the mid-plane of the experiment)?

\item
Does the collimation of the flow by the plasma ablating from the angular sides of the groove and the entrainment of this plasma in the resulting jet increase or decrease the velocity of the jet?

\end{enumerate}

\subsection{What determines the physical conditions in the core of the jet?}

Figures \ref{fig:jetwidth} \& \ref{fig:jetwidth_zbar} show that the width of the jet at early times ($\leq$~1.1~ns) is similar to the width of the rounded region of the groove in the Al slab at the mid-plane, which is relatively flat and so relatively perpendicular to the laser beam illuminating the target.  The incident laser beam has a Gaussian cross section with a full width such that most of the laser energy is deposited within $\pm 100 \mu$m of the laser axis. Accordingly, the intensity of the laser beam will be much greater on the relatively flat portion of the target than on the sloping sides of the V-shaped groove.  This suggests that the velocity in the core of the jet might be produced by the same physical mechanism that produces the velocity of the flow when an Al slab target is illuminated by a laser (e.g. as in \S~\ref{sec:prepulse}); namely, heating of the Al target by the laser followed by free expansion of the heated target material due to the thermal pressure of the hot electrons.

To investigate this possibility, we have compared lineouts of the velocity along the mid-plane for the two problems at five relatively early times ($\leq$ 1.1 ns).  Figure \ref{fig:vely1} shows the results.  Ignoring the very high ($v_z~\gtrsim~350$~km/s) velocities of very low density gas and {\it focusing on} $v_z \lesssim 350$~km/s, {\it the velocity profiles for the two problems agree closely at all five times, supporting our conjecture that the mechanism producing them is the same}. The differences seen between the two simulations at very high velocities ($v_z \gtrsim 350$~km/s) are at very low densities, as already mentioned, and near or approaching the transition from cells that are mostly Aluminum to cells that are mostly very low density Helium. This transition is marked by a change in line type from thick solid to dotted lines for the V-shaped target or from thin solid to dashed lines for the flat target. Because the Helium serves only as an approximation to vacuum conditions, the results at this very low density interface are both questionable and irrelevant to the questions we are concerned with in this section.


\begin{figure*}
\centerline{\includegraphics[angle=0,width=5in]{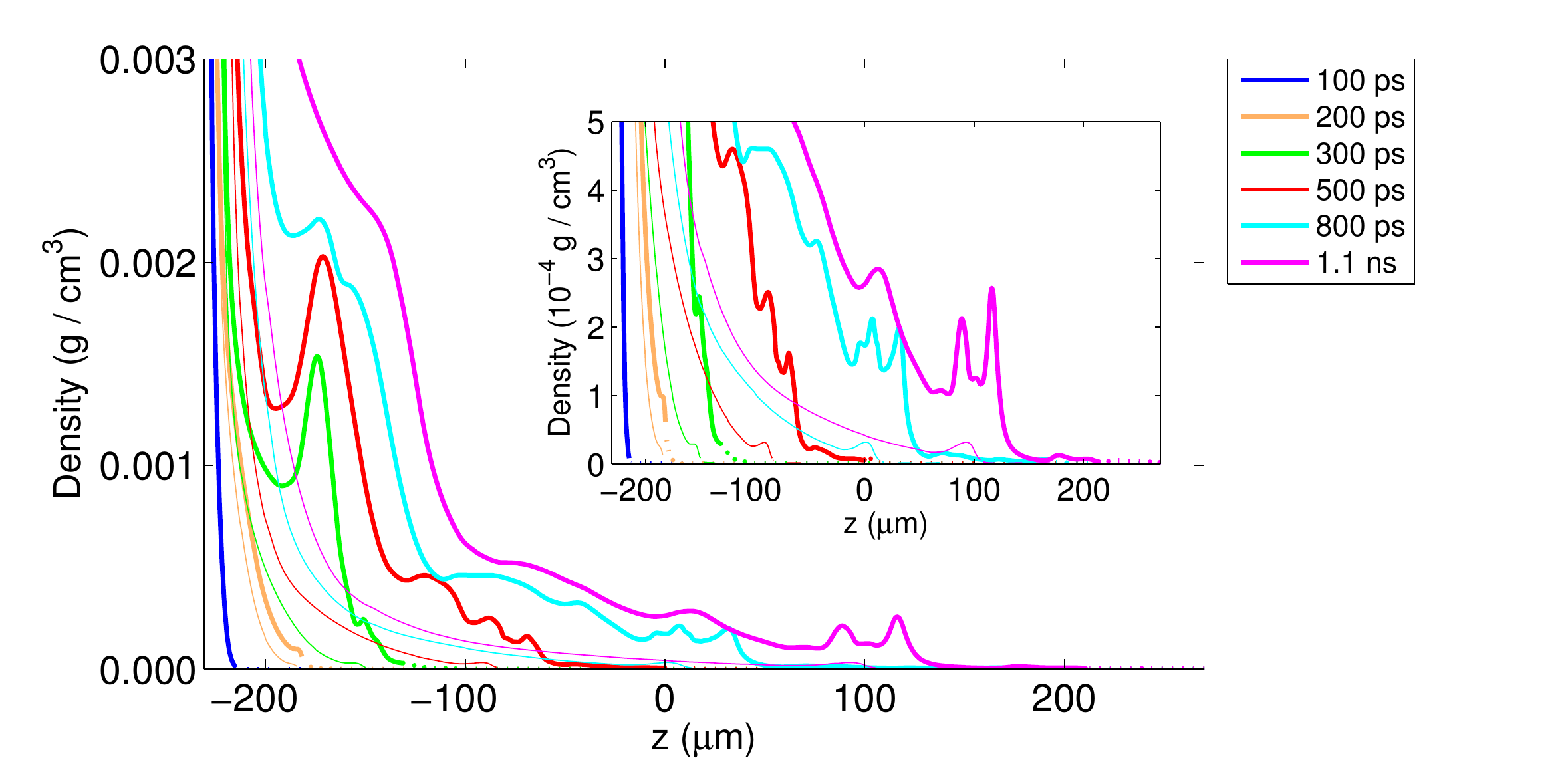}}
\vspace{-0.3cm}
\caption{
Lineouts of the mass density along the laser axis at various times. The thick solid lines show measurements from the FLASH simulation of a V-shaped groove target described in \S~\ref{sec:grava}. Thin solid lines show the same measurements from a simulation where a flat target of the same material is irradiated with the Grava et al. laser pulse. As in Fig.~\ref{fig:vely1}, a change in line type to dotted or dashed indicates the Al/He transition. 
}\label{fig:dens1}
\end{figure*}

\begin{figure*}
\centerline{\includegraphics[angle=0,width=5in]{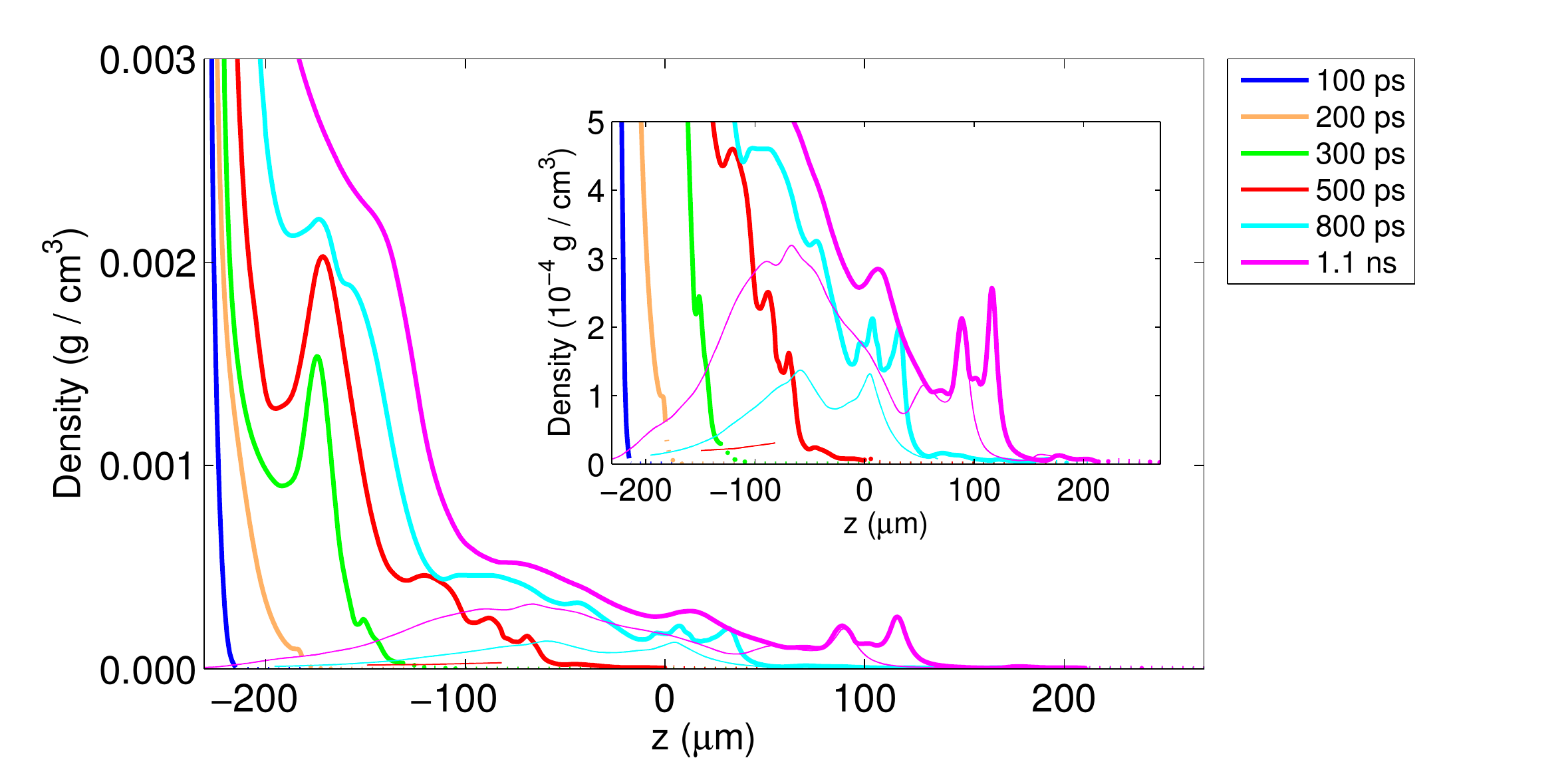}}
\vspace{-0.3cm}
\caption{
Lineouts of the mass density along the laser axis at various times. The thick solid lines show measurements from the FLASH simulations of a V-shaped groove target described in \S~\ref{sec:grava}. Dashed lines show the same measurements from a simulation where, similar to the target geometry of Wan et al. \cite{Wan_etal1997}, a large gap of material is missing from the center of the V-shaped groove. This target is irradiated by the Grava et al. laser pulse. As in Fig.~\ref{fig:vely2}, a change in line type to dotted or dashed indicates the Al/He transition. 
}\label{fig:dens2}
\end{figure*}

We have also compared these same lineouts of the velocity along the mid-plane for an Al target with a V-shaped groove, which was used in the Grava et al. \cite{Grava_etal2008} experiment, and for a target consisting of two Al slabs oriented perpendicular to each other with a gap between them, similar to the Wan et al. \cite{Wan_etal1997} experiment. This comparison allows us to contrast the properties of the jet produced by an Al target with a V-shaped groove that has a relatively flat portion near the mid-plane and one that does not.  Figure \ref{fig:vely2} shows the results with the V-shaped groove simulation with thick solid lines of various colors depending on the output and the Wan et al.-like simulation shown likewise with thin solid lines. As in Fig.~\ref{fig:vely1} a change of line type indicates the transition from mostly Aluminum to mostly Helium. Again ignoring the high velocities ($v_z \gtrsim 350$~km/s) that occur at low densities, the two velocity profiles differ greatly.  The absence of a relatively flat portion of the target near the mid-plane means that formation of the jet is delayed until the plasma ablating from the sloping sides of the target has had time to meet at the mid-plane.  Furthermore, the velocity profile of the jet is much shallower and its maximum velocity is much smaller.  The results provide further support for the hypothesis that the physical mechanism producing the jet in the Grava et al. experiment is the same as in the slab problem.

Figure \ref{fig:dens1} compares lineouts of the density along the mid-plane for the V-shaped groove target used in the Grava et al. experiment (thick solid lines) and an Al slab target (thin solid lines), while Fig.~\ref{fig:dens2} compares the same quantity from the V-shaped groove target (thick solid lines) and for a target consisting of two Al slabs oriented perpendicular to each other with a gap between them (thin solid lines), similar to the Wan et al. (1997) experiment. As in Figs.~\ref{fig:vely1} \& \ref{fig:vely2} these lines become dotted or dashed when the cells are mostly He instead of Al.

Clearly, ablation from the sloping sides of the V-shaped groove confines the flow, as discussed by Wan et al. \cite{Wan_etal1997} and Grava et al. \cite{Grava_etal2008}, greatly increasing the density in the jet relative to the density in the case of the Al slab target, where the flow can expand laterally as well as away from the surface of the target.  Collimation of the flow by the plasma ablating from the sloping sides of the V-shaped groove in the Grava et al. experiment raises the question of whether the collimation is due primarily to thermal pressure or to ram pressure.  We now address this question.

\subsection{Is the collimating effect of the plasma ablating from the angular sides of the groove due to thermal pressure or ram pressure?}

\begin{figure}
\centerline{\includegraphics[angle=0,width=3.5in]{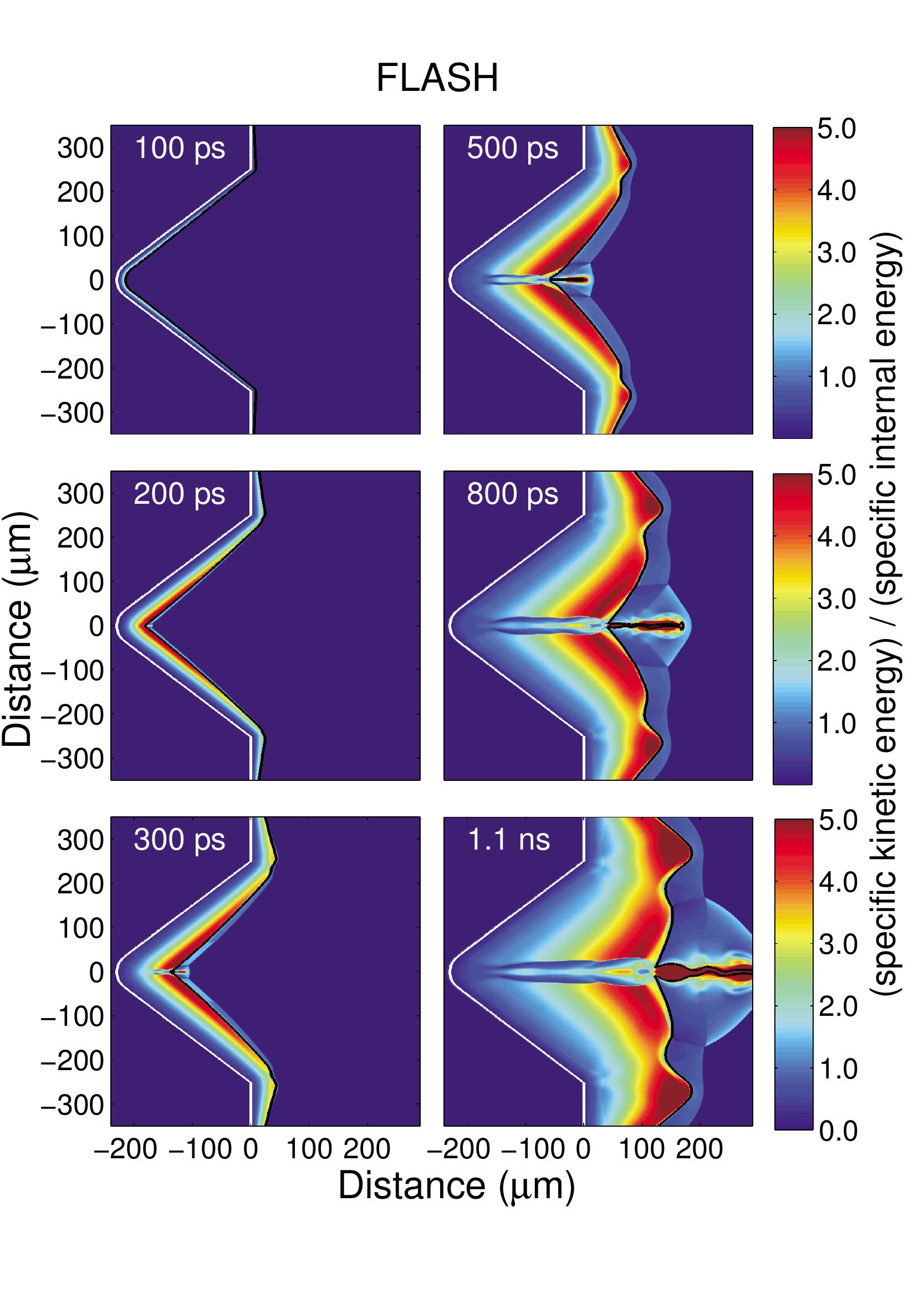}}
\vspace{-1cm}
\caption{
Plotting the ratio of $e_{\rm kin} / e_{\rm int}$ at various times for the FLASH simulation of the V-shaped groove described in \S~\ref{sec:grava}. The original target location is shown with a white line. The transition from mostly Al to mostly He cells is indicated with a solid black line in each panel.
}\label{fig:ekin}
\end{figure}

\begin{figure}
\centerline{\includegraphics[angle=0,width=3.5in]{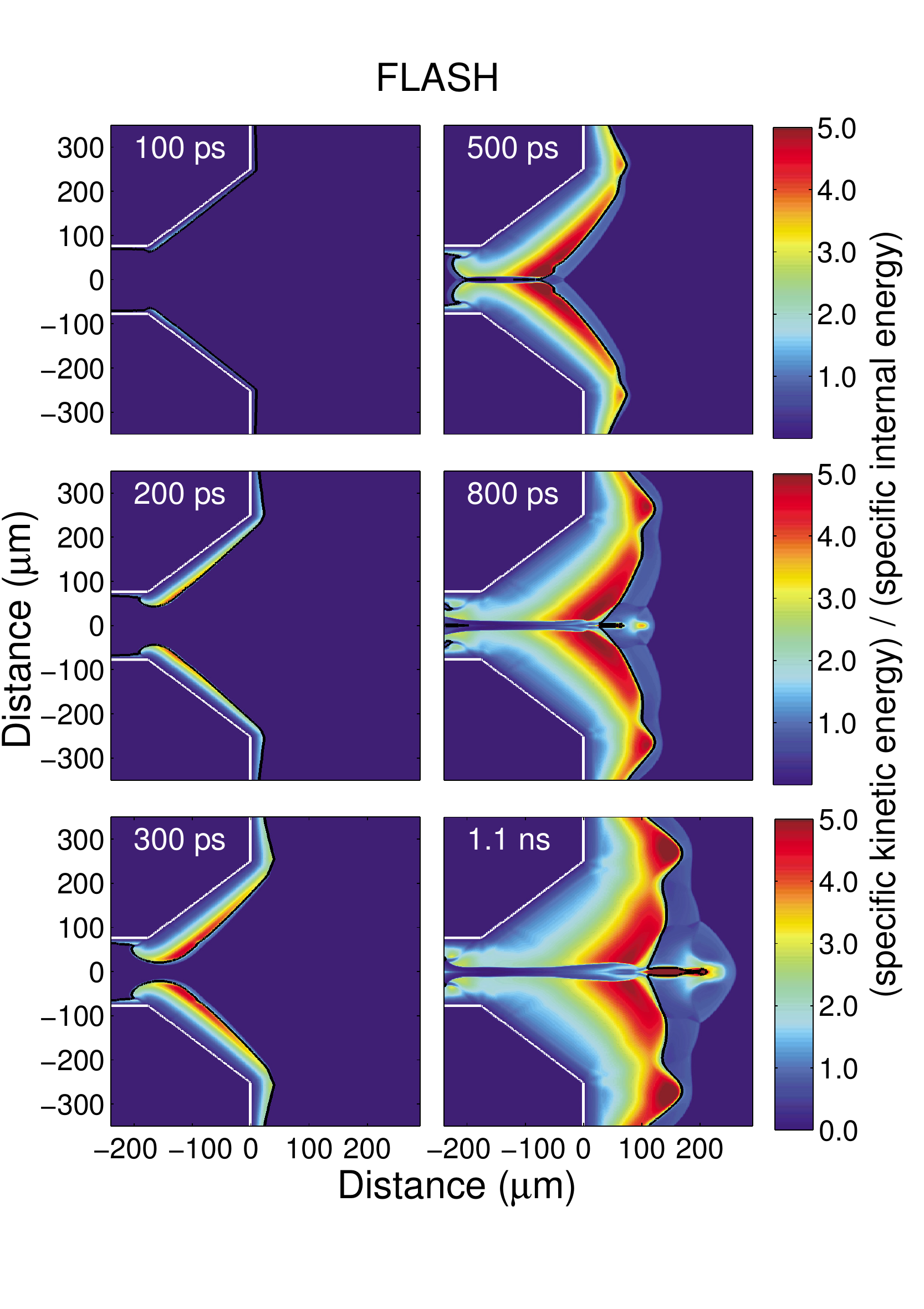}}
\vspace{-1cm}
\caption{
Plotting the ratio of $e_{\rm kin} / e_{\rm int}$ at various times for a FLASH simulation of a target with a gap in the center (similar to Wan et al. 1997) that is irradiated by the Grava et al. laser pulse. The original target location is shown with a thick white line. The transition from mostly Al to mostly He cells is indicated with a solid black line in each panel.
}\label{fig:ekin_nomid}
\end{figure}

\begin{figure}
\centerline{\includegraphics[angle=0,width=3.5in]{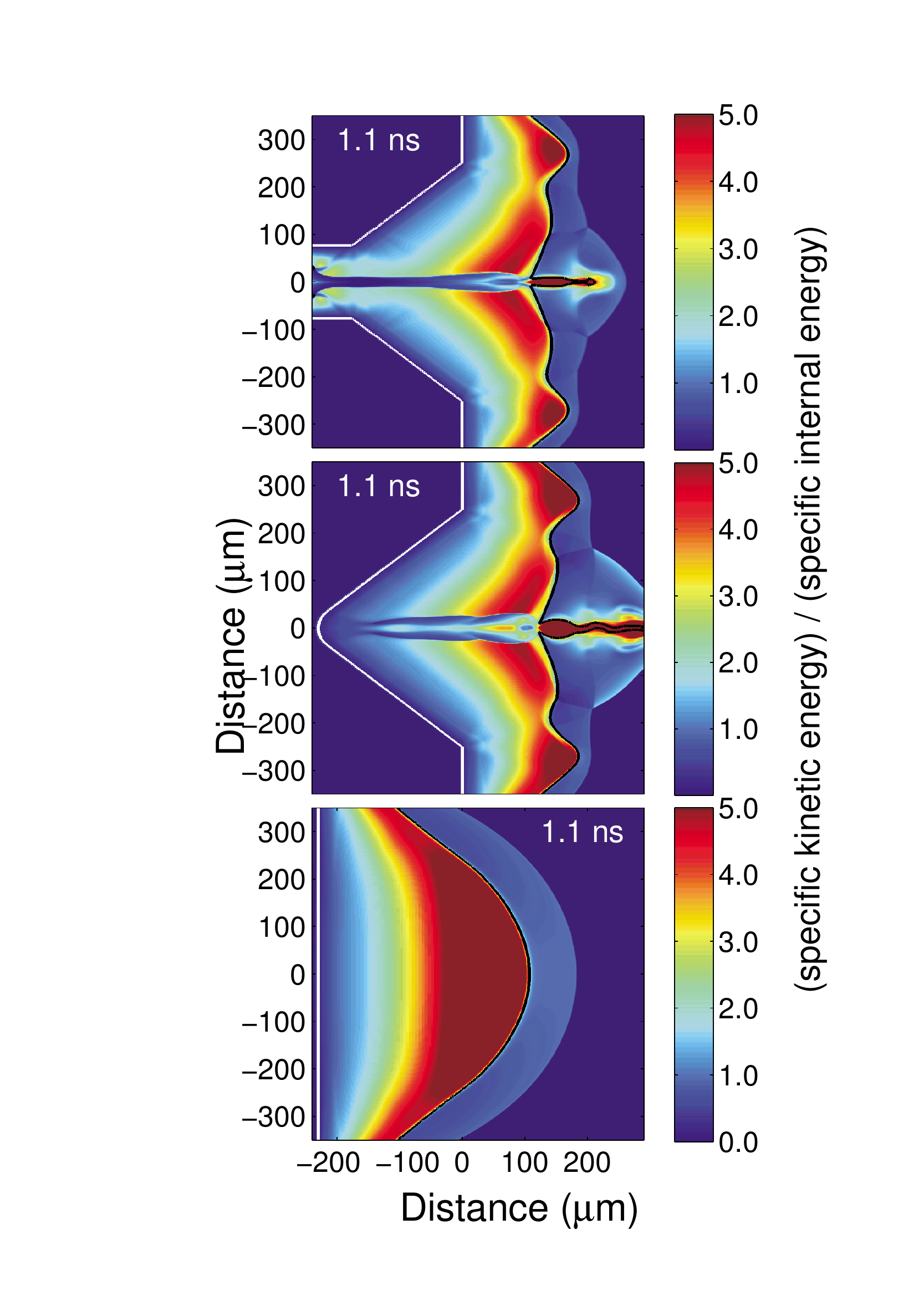}}
\vspace{-1cm}
\caption{
Plotting the ratio of $e_{\rm kin} / e_{\rm int}$ at 1.1~ns for three different FLASH simulations: (Top) the V-shaped groove simulation described in \S~\ref{sec:grava}, (middle) results from a V-shaped groove with a gap as in Wan et al. (bottom) results from a FLASH simulation of a flat target that is likewise irradiated by the Grava et al. laser pulse. In each plot the original target location is shown with a thick white line. The transition from mostly Al to mostly He cells is indicated with a solid black line in each panel.
}\label{fig:ekin_nomid_nc_flat}
\end{figure}

To address this question, we calculate the ratio of the specific kinetic energy,
\begin{equation}
e_{\rm kin} = \frac{1}{2} |\vec{v}|^2
\end{equation}
to the total specific internal energy,
\begin{equation}
e_{\rm int} = e_{\rm ele} + e_{\rm ion}.
\end{equation}
If $e_{\rm kin} / e_{\rm int} \gg 1$, the kinetic energy of the ablation flow is dominant;  if instead, $e_{\rm kin} / e_{\rm int} \lesssim 1$ the internal energy due to the temperature of the plasma is dominant.

Figure~\ref{fig:ekin} shows the ratio $e_{\rm kin}/e_{\rm int}$ throughout the computational domain at six different times for the Al slab target with a V-shaped groove, while Figure~\ref{fig:ekin_nomid} shows the same quantity at the same times for the Al target comprised of two slabs oriented perpendicularly to each other with a gap in between, similar to the Wan et al. experiment.  Figure~\ref{fig:ekin_nomid_nc_flat} compares the same quantity for an Al slab target, and the Grava et al. and Wan et al.-like targets at 1.1~ns.  In all three figures, we see a similar behavior:  even at very early times, in the plasma very near the target, the internal energy of the plasma dominates the kinetic energy of the bulk flow away from the target, but further out the kinetic energy of the bulk flow dominates the internal energy.  Once the laser turns off, the region where the kinetic energy of the bulk flow dominates the gas internal energy begins to grow substantially larger.

Examining the properties of the ablation flow as it approaches the mid-plane, we see that the internal energy of the plasma dominates at distances $< ~50~\mu$m from the target, but the kinetic energy of the bulk flow dominates at all larger distances.  This indicates that the collimation of the jet is due primarily  to ram pressure except very near the target, where it is due primarily to thermal pressure of the hot plasma.  

The plasma in the jet is heated when the plasma ablating from the sloping sides of the target in both the Grava et al. and the Wan et al.-like experiments collides with it and the kinetic energy in the bulk flow of the ablating plasma is released.  The ratio $e_{\rm kin}/e_{\rm int}$ is small where this happens.  However, the ratio is large in the mid-plane because the outward velocity in the jet is so large.  The result is a complex lateral structure within the jet in which $e_{\rm kin}/e_{\rm int}$ is large in the core of the jet and small at its edges, and large again in the ablating plasma above and below the jet.  This is the origin of the double-horn structure evident at very late times in the electron density, which can be seen in Figures \ref{fig:norad}-\ref{fig:jetwidth} and in the ionization state, which can be seen in Figures \ref{fig:tele}-\ref{fig:jetwidth_zbar}.

While the complex lateral structure of the jet in these experiments is qualitatively similar to astrophysical jets, it differs in that the ratio $e_{\rm kin}/e_{\rm int}$ in astrophysical jets is expected to be large in the core of the jet and progressively smaller values further away from the jet axis with $e_{\rm kin}/e_{\rm int} \rightarrow 0$ in the ambient medium \cite[e.g.][]{Stone_Norman1993}.

\subsection{Does the ablation from the angular sides of the groove increase or decrease the velocity of the jet?}

We are now in a position to address whether the ablating plasma from the angular sides of the groove in the target increases or decreases the velocity of the jet.   A key piece of information is that the velocity of the jet is much smaller in the Wan et al.-like experiment in which the target is two Al slabs oriented perpendicular to each other with a gap in between than in the Grava et al. experiment in which the target is an Al slab with a V-shaped groove it it.  We can now understand the reason why from the answer we obtained to the previous question.  The energy density of the ablating plasma is dominated by its bulk kinetic energy by the time it approaches the mid-plane, except at very small distances ($< 50~\mu$m) from the target.  The component of the momentum of the ablating plasma that is perpendicular to the mid-plane will go into heating the jet, while the component parallel to the mid-plane will add to the velocity of the jet.  However, because the laser intensity is much lower away from the mid-plane, due both to the profile of the laser beam and the slanted angle of the surface of the groove, the {\it specific} internal energy (i.e., the internal energy per gram) generated by the component of the momentum of the ablating plasma when the ablating plasma collides with the jet, and the {\it specific} component of the momentum of the accreting plasma parallel to the mid-plane (i.e., the momentum per unit mass) are both smaller than in the jet flow itself, which is generated by the most intense part of the laser beam illuminating the nearly flat part of the groove near the mid-plane.  This suggests that the entrainment in the jet of the plasma ablating from the sloping sides of the groove will {\it decrease slightly} the velocity of the jet compared to the velocity along the mid-plane of the freely expanding plasma in the case of a slab target.  This expectation is consistent with the results shown in Figure~\ref{fig:vely1}.

\section{Discussion of Code-to-Code Comparisons}
\label{sec:disc}

We find excellent agreement between the FLASH and HYDRA simulations for both the pre-pulse problem described in Sec.~\ref{sec:prepulse} and the V-shaped groove-target 
problem investigated in Sec.~\ref{sec:grava}.  In this section, we consider the differences between the two codes that are described in Sec.~\ref{sec:diff}
and comment on the remaining discrepancies or differences found in both 
suites of tests generally rather than focusing on any particular result.

As already noted, the two codes were {\it not} run with the same EOS 
and opacity data, and therefore the small discrepancies that do exist 
can reasonably be attributed to this fact, rather than to anything more
fundamental, such as the different computational mesh schemes they use,
which we described in Sec.~\ref{sec:aleamr}. We consider that differences in the EOS
models are likely to be a larger source of uncertainty than the opacity data:
comparing Figs.~\ref{fig:norad} and \ref{fig:rad}, the effect of
including radiative-energy loss is still relatively subtle, and FLASH 
simulations of the pre-pulse tests without radiation (not shown)
gave very similar results to the simulations that include 
radiation transport that we described in Sec.~\ref{sec:prepulse}. 

To explore the extent to which different EOS models affect the results,
we presented FLASH simulations from both the PROPACEOS and 
SESAME EOS models for the pre-pulse problem.  For the V-shaped
groove targets, we presented only the PROPACEOS EOS for
brevity, but we did perform FLASH simulations with  
the SESAME EOS.  The results did not agree nearly as well
with the experimental measurements of electron number density and
the HYDRA simulations. This result, taken together with
the finding in Figs.~\ref{fig:20mJresults} and \ref{fig:100mJresults}
that the PROPACEOS $\bar{Z}(\rho,T)$ model near solid density
agrees much better with HYDRA than does the SESAME model, 
indicates that using the PROPACEOS model is the most HYDRA-like
way of running FLASH and, of course, validation using the interferometry 
data from Grava et al (2008) implies that it is also the most accurate 
approach, at least for Aluminum plasmas at these 
densities and temperatures.

The only systematic difference between the FLASH and HYDRA results
that cannot be attributed to the EOS models is the near-solid-density
results in the pre-pulse problem, highlighted by the inset figures
in the upper panels of Figs.~\ref{fig:20mJresults} \& \ref{fig:100mJresults}.
Regardless of the EOS, the above-solid-density features are typically somewhat
more pronounced in FLASH and are not as far into the target as with HYDRA,\footnote{Pre-pulse tests with FLASH using the BADGER EOS \cite{BADGER} (not shown) confirm this as well.}
although the resolution in this region is reasonably high for both codes.

Of all the differences between FLASH and HYDRA discussed in Sec.~\ref{sec:diff},
the lack of the ponderomotive force in FLASH would seem to be a  likely culprit, since by $t = 0.2$~ns the above-solid-density feature
is already further into the target in HYDRA, and it is at 
very early times when the plasma pressure is low that the ponderomotive
force may become a non-negligible effect. However, HYDRA pre-pulse 
simulations without the ponderomotive force (not shown) gave identical 
results for the early-time plasma properties.  The lack of the ponderomotive force in FLASH is therefore unlikely to be the explanation of the discrepancy.
Conceivably, as discussed in Sec.~\ref{sec:laser},
the difference could arise from the convergence of the laser
rays on the target in HYDRA as opposed to the initially parallel rays
in FLASH. However, Fig.~\ref{fig:ncrit} indicates $\lesssim 5 \mu$m
movement of the critical surface over the duration of the simulation,
and the discrepancy appears at early times (even before $t = 0.2$~ns),
so this hypothesis seems unlikely, and the differences remain unexplained.

Unfortunately, the simulations of the Al target with a V-shaped groove, which involve electron densities
below the $\sim 5 \cdot 10^{20}$ cm$^{-3}$ experimental threshold of detection,
 add little to the discussion. Despite this lingering issue, the 
properties of the Aluminum blowoff plasma are in reasonably
good agreement between the codes and with the experimental data. Since PIC
simulations of short-pulse laser interactions with pre-formed plasmas 
depend crucially on the plasma properties near or below the critical 
density ($n_c \sim 10^{21}$ cm$^{-3}$), this level of agreement is sufficient 
to use the output from either FLASH or HYDRA radiation-hydrodynamics simulations as initial 
conditions for these codes. Equivalently, the predictive accuracy of the pre-plasma
properties seems to be limited only by the quality of the EOS and opacity
data, and the accuracy of the {\it measurements} of the pre-pulse properties.








\section{Conclusions}
\label{sec:concl}

We carried out a series of studies comparing the FLASH code to previously-published and new
simulations from the HYDRA code used extensively at LLNL. These tests include a 
comparison of results at pre-pulse intensities and energies (15.3~mJ and 76.7~mJ)
for an Aluminum slab target in 2D cylindrical geometry over 1.4~ns of evolution.
We also compared the results of FLASH to previously-published experimental results and HYDRA 
modeling for the irradiation of a mm-long V-shaped groove cut into an Aluminum target
as a test of the code in a 2D cartesian geometry \cite{Grava_etal2008}. 
Importantly, these experiments, conducted at Colorado State University, included 
soft x-ray interferometric measurements of the electron density in the Al blowoff plasma
as a powerful validation test.

In all cases the FLASH results bore a remarkable resemblance to results from HYDRA.
This is especially true for the properties of the underdense Al blowoff plasma, which matters most
for the use of these codes in calculating pre-plasma properties as initial conditions
for PIC simulations of ultra-intense, short-pulse laser-matter interactions.

These code comparisons were performed {\it without} using the same EOS and opacity models.
To assess the extent to which differences in EOS models can change the results, we
presented simulations with FLASH using both the PROPACEOS and SESAME EOS models.
While these simulations gave similar results, the PROPACEOS model gave the most
HYDRA-like way of running the code and also agreed better with the interferometric
measurements of the electron number density in the V-shaped groove experiment.

One difference between the FLASH and HYDRA results could not be attributed to
differences in EOS: the above-solid-density ablation front moving
into the target in the pre-pulse tests typically peaked at slightly higher density in FLASH,
and typically had moved slightly further into the target in HYDRA. We were able to 
rule out the lack of ponderomotive force (a.k.a. light pressure) in FLASH as the explanation
for this by looking closely at HYDRA pre-pulse simulations with and without the ponderomotive force
which yielded identical results. This confirmed our intuition that the ponderomotive force
is negligible for pre-pulse intensities and energies, and therefore the FLASH treatment of 
the laser as a source of energy deposition only should be an exceedingly good
approximation in the pre-pulse regime. 

Having validated the FLASH simulations for the Al target with a V-shaped groove, we have also used these FLASH simulations to better understand the formation and properties of the jet in the experiment.  We show that the velocity of the jet is produced primarily by the heating of the target in the relatively flat region of the V-shaped groove at the mid-plane, as in standard slab  targets.  We show that the jet is collimated primarily by the ram pressure of the plasma that ablates from the sloping sides of the groove.  We find that the interaction of the plasma ablating from the sloping sides of the groove with the jet produces the observed complex lateral structure in it, a structure that is qualitatively similar to astrophysical jets but differs significantly from them, quantitatively.  Finally, we show that the entrainment in the jet of the plasma ablating from the sloping sides of the groove slightly decreases the velocity in the jet compared to the velocity along the mid-plane of the freely expanding plasma in the case of a slab target.

For the FLASH code, which has historically been used primarily for astrophysics, this 
full-physics code comparison and validation of a number of new HEDP-relevant algorithms, and investigation into
the physics of the laboratory-produced jet
represents a milestone in the realization of FLASH as a well-tested open-source
tool for radiation-hydrodynamics modeling of laser-produced plasmas. More testing
can and should be done. We focused on laser-irradiated Aluminum plasmas. 
High-quality interferometric data also exists for C, Cu and Mo 
\cite{Filevich_etal2009,Purvis_etal2010}. Any future FLASH investigations using these
materials would do well to start with these validation tests.


\section*{Acknowledgements}

CO thanks the Flash Center for Computational Science for its hospitality and eager support throughout
the project. CO additionally thanks Douglass Schumacher and Richard Freeman for insightful conversations
and encouragement. Much thanks also to Marty Marinak for a close reading of an early manuscript, 
to Mike Purvis for helpful correspondence and to Pravesh Patel for pointing us to the code validation 
papers referenced extensively in this paper. Supercomputer allocations for this project included 
generous amounts of time from the Ohio Supercomputer Center as well as resources at Livermore Computing. 
The FLASH code used in this work was developed in part by the Flash Center for Computational Science at the 
University of Chicago through funding by DOE NNSA ASC, DOE Office of Science ASCR, and NSF Physics.
CO was supported by the U.S. Department of Energy contract DE-FG02-05ER54834 (ACE). 
SC \& SW performed their work under the auspices of the US DOE by Lawrence Livermore National Security, LLC, 
Lawrence Livermore National Laboratory under Contract DE-AC52-07NA27344. DL \& MF were supported in 
part at the University of Chicago by the U.S. Department of Energy NNSA ASC through the Argonne Institute for Computing
in Science under field work proposal 57789. We further acknowledge Los Alamos National Laboratory for the use of the 
SESAME EOS.

\bibliography{ms}
\bibliographystyle{apsrev}

\end{document}